\documentclass[article,11pt]{article}

\pdfoutput=1

\usepackage{jheppub}
\usepackage{caption}
\usepackage{subcaption}
\usepackage{amssymb}
\usepackage{amsbsy}
\usepackage{amsfonts}
\usepackage{amsmath}
\usepackage{epsfig}
\usepackage{cancel}
\usepackage{slashed}
\usepackage{array}
\usepackage{mathtools}

\allowdisplaybreaks

\newcolumntype{L}[1]{>{\raggedright\let\newline\\\arraybackslash\hspace{0pt}}m{#1}}
\newcolumntype{C}[1]{>{\centering\let\newline\\\arraybackslash\hspace{0pt}}m{#1}}
\newcolumntype{R}[1]{>{\raggedleft\let\newline\\\arraybackslash\hspace{0pt}}m{#1}}

\title{LHC phenomenology and baryogenesis in supersymmetric models with a $U(1)_R$ baryon number}

\author[a]{Hugues Beauchesne}
\author[b]{Kevin Earl}
\author[b]{Thomas~Gr\'egoire}

\emailAdd{hubea44@if.usp.br}
\emailAdd{KevinEarl@cmail.carleton.ca}
\emailAdd{gregoire@physics.carleton.ca}

\affiliation[a]{Instituto de F\'isica, Universidade de S$\tilde{a}$o Paulo, C.P. 66.318, 05315-970 S$\tilde{a}$o Paulo, Brazil.}
\affiliation[b]{Ottawa-Carleton Institute for Physics, Department of Physics, Carleton University 1125 Colonel By Drive, Ottawa, K1S 5B6 Canada}
\abstract{
We study the phenomenology of a supersymmetric extension of the Standard Model with an $R$-symmetry under which $R$-charges correspond to the baryon number. This identification allows for the presence in the superpotential of the $R$-parity violating term $\lambda''U^c D^c D^c$ without breaking baryon number, which loosens several bounds on this operator while changing considerably the phenomenology. However, the $R$-symmetry cannot remain exact as it is at least broken by anomaly mediation. Under these conditions, we investigate the constraints coming from baryon number violating processes and flavour physics and find that, in general, they are lessened. Additionally, we examine recent ATLAS and CMS experimental searches and use these to place limits on the parameter space of the model. This is done for both stop production, which now features both pair and resonant production, and pair production of the first two generations of squarks. Finally, we study the implications this model has on baryogenesis. We find that successful baryogenesis can potentially be achieved, but only at the cost of breaking the $R$-symmetry by a significant amount.
}

\begin{document}

\maketitle

\section{Introduction}
Supersymmetric (SUSY) extensions of the Standard Model (SM) have long been considered attractive candidates for physics beyond the Standard Model (BSM). In their simplest realization they solve the hierarchy problem, have a dark matter candidate and predict gauge coupling unification. As such, superpartners have been the focus of a very large number of searches  by collider experiments. Despite these intensive efforts, they have not been seen, putting the limit on their mass above the TeV scale in many cases. For many versions of the Minimal Supersymmetric Standard Model (MSSM), this means introducing large fine-tuning, therefore weakening one of the main motivations for these models. This has led to a renewed interest in supersymmetric models which depart from the MSSM in some way and lead to different collider phenomenology. This includes models with Dirac gauginos \cite{Fayet:1978qc, Hall:1990hq, Fox:2002bu, Nelson:2002ca, Kribs:2007ac, Amigo:2008rc, Benakli:2008pg, Benakli:2010gi, Kribs:2010md, Abel:2011dc, Csaki:2013fla, Frugiuele:2011mh, Davies:2011mp, Fok:2012fb, Frugiuele:2012kp, Frugiuele:2012pe, Beauchesne:2014pra, Bertuzzo:2014bwa, Carpenter:2015mna, Itoyama:2011zi, Itoyama:2013sn, Itoyama:2013vxa} which can exhibit supersoft supersymmetry breaking \cite{Fox:2002bu} and  lead to reduced cross section for the production of squarks \cite{Heikinheimo:2011fk,Kribs:2012gx,Kribs:2013eua}. The Dirac nature of the gauginos also enables the building of models that possess a $U(1)_R$ symmetry, which were shown to have weaker flavour constraints \cite{Kribs:2007ac,Fok:2010vk}. Furthermore, the $U(1)_R$ symmetry can be identified with a lepton or baryon number leading to models where the superpartners have non-standard charges under these symmetries. Such identification can lead to models with  unusual structure and phenomenology. For example, if the $U(1)_R$ symmetry is identified with a lepton number  the sneutrino can acquire a significant vacuum expectation value (vev) and play the role of the down type Higgs \cite{Gherghetta:2003he, Frugiuele:2011mh, Fok:2012fb, Frugiuele:2012kp, Frugiuele:2012pe, Riva:2012hz, Beauchesne:2014pra, Biggio:2016sdu}.

In this work we examine the phenomenology of models where the $U(1)_R$ is instead identified with baryon number \cite{Brust:2011tb, Frugiuele:2012pe}.  Because this symmetry does not commute with supersymmetry, superpartners have different baryon numbers than their corresponding Standard Model particles which themselves retain their standard baryon number.  Under this charge assignment, the standard $R$-parity violating superpotential term of the form $ \lambda'' U^c D^c D^c$ is now baryon number conserving. The bound on such a term is therefore weakened significantly which can modify the LHC phenomenology. For example, superpartners can decay promptly, making displaced vertices signatures, which are very constraining, less prevalent. Furthermore, an exact $U(1)_R$ would forbid stop decays containing two same sign leptons, which is also a very constraining signature.
 
Models with a $U(1)_R$ baryon number also have all the  necessary components  to generate successful baryogenesis: baryon number violation through unavoidable $U(1)_R$ breaking, the possibility of CP violation and out of equilibrium processes through the late decay of a gaugino \cite{Sakharov:1967dj}.
 
In this paper we first look at how the bounds on the $\lambda''$ couplings are modified by the presence of the approximate $U(1)_R$ symmetry. This is presented in section \ref{Sec:Model}. In section \ref{Sec:collider_constraints} we examine the collider constraints on the model when a single coupling of the form $\lambda''_{3ij}$ is important. This phenomenology is in many cases very similar to the one studied in \cite{Monteux:2016gag} (see also for example \cite{Dreiner:1991pe,Allanach:2012vj,Evans:2012bf,Bhattacherjee:2013gr,Graham:2014vya}). In section \ref{Sec:Baryogenesis}, we study how our model can lead to successful baryogenesis. The mechanism is similar to the one studied in \cite{Cui:2013bta, Cui:2012jh, Arcadi:2015ffa, Arcadi:2013jza} and rely on the out of equilibrium decays of gauginos through a baryon number violating interaction. This requires a split spectrum with gauginos much lighter than the scalars. As we will see, the (pseudo-)Dirac nature of the gauginos leads to new diagrams contributing to the decay process and as a result new portions of the parameter space can have successful baryogenesis. 

\section{The model}\label{Sec:Model}
The  model we consider is an extension of the minimal $R$-symmetric Supersymmetric Standard Model (MRSSM) \cite{Kribs:2007ac}. It has an approximate $U(1)_R$ symmetry and Dirac gauginos whose mass terms can be written as:
\begin{equation}
\label{gauginomass}
\sqrt{2} \int d^2 \theta \frac{{W'}^\alpha}{M_*}\left[c_1 W^{(1)}_\alpha S +c_2 W^{(2)i}_\alpha T^i+ c_3{W^{(3)a}_\alpha O^a} \right]+\text{h.c.},
\end{equation}
where ${W'}^\alpha = \theta^\alpha D'$ is a spurion vector superfield with a non-zero $D$-term. $S$, $T^i$ and $O^a$ are chiral superfields in the adjoint representation  of $U(1)_Y$, $SU(2)_L$ and $SU(3)_c$ respectively, $W^{(k)}_\alpha$ are the Standard Model superfield strengths and $M_*$ is the supersymmetry breaking mediation scale. The gaugino masses then take the form:
\begin{equation}
M_i^D = c_i \frac{D'}{M_*}.
\end{equation}

The standard $\mu$ term being forbidden by $U(1)_R$, the chiral superfields $R_u$ and $R_d$ are added to provide mass to the higgsinos. These new fields have the same gauge numbers as the higgsinos but different $U(1)_R$ charges.  They have bilinear $\mu$-like terms with the Higgs superfields but their scalar components do not acquire vevs. The $U(1)_R$ symmetry can then be identified with baryon number by assigning the right-handed quark superfields $R$-charge $2/3$ and the left-handed quark superfields $R$-charge $4/3$. The charge assignments of the remaining superfields are shown in table \ref{table:Rcharge}. 
\begin{table}[t]
\begin{center}
\begin{tabular}{|c|c|}
\hline
Fields & $R$-charge \\ \hline
$H_{u,d}$ & $0$ \\ \hline
$R_{u,d}$ & $2$ \\ \hline
$U^c$, $D^c$ & 2/3 \\ \hline
$S$, $T$, $O$ & 0  \\ \hline
$Q$ & $4/3$ \\ \hline
$L$, $E^c$ & 1\\ \hline
\end{tabular}
\end{center}
\caption{$R$-charge assignment of chiral superfields of the model.}
\label{table:Rcharge}
\end{table}
Under this symmetry  all the Standard Model particles have their usual baryon number. However superpartners have non-standard baryon numbers. For example, the right-handed squarks have baryon number $2/3$ and thus  are diquarks, while the left-handed squarks have baryon number $4/3$ and the gauginos baryon number 1. Gauge symmetries and the $U(1)_R$ symmetry lead to the following superpotential:
\begin{equation}
  \begin{aligned}
W = \; &y_u Q H_u  U^c - y_d Q H_d D^c - y_e L H_d E^c + \mu_u H_u R_d + \mu_d R_u H_d \\  &+ \lambda_u^t H_u T R_d + \lambda_d^t R_u T H_d +\lambda_u^s S H_u  R_d + \lambda_d^s S R_u H_d + \frac{1}{2} \lambda''_{i j k} U^c_i D^c_j D^c_k,
\end{aligned}
\end{equation}
where $T=T^i \sigma^i/2$. This superpotential is equal to the superpotential of the MRSSM to which the standard $R$-parity violating term of the form $U^c D^c D^c$ has been added.\footnote{This term violates the standard $R$-parity but not the $U(1)_R$ symmetry defined in table \ref{table:Rcharge}.} Beside gaugino masses, the soft SUSY breaking terms include non-holomorphic scalar masses, $B_{\mu}$ like terms and a linear term for $S$:
\begin{equation}
V_\text{soft} = \sum_\Phi M_\Phi \left| \Phi \right|^2 + \left[B_\mu H_u H_d + \frac{1}{2} b_S S^2 + \frac{1}{2} b_T T^2 + \frac{1}{2} b_O O^2 + f_S S+\text{h.c.}\right].
\end{equation}
Various tri-linear terms are also allowed by the symmetries of the model but can be suppressed \cite{Frugiuele:2012pe}. In addition, the $f_S$ term needs to be small to avoid destabilizing the hierarchy. On general ground, the $U(1)_R$ symmetry cannot remain an exact symmetry of the theory. The breaking will manifest itself at least through the gravitino mass. This breaking will then unavoidably be communicated to  the Standard Model sector through anomaly mediation \cite{Randall:1998uk, Giudice:1998xp}. Majorana gaugino mass terms and tri-linear $A$-terms will in this case be generated with size of order:
\begin{equation}
M \sim A \sim  \frac{1}{16 \pi^2} m_{3/2} .
\end{equation}

\subsection{Bounds on \texorpdfstring{$\lambda''$}{lambda''}}\label{sSec:BoundsOnLambdapp}
The bounds on the  $\lambda''$ couplings come in our model from the same sources as in the $R$-parity violating Supersymmetric Standard Model (RPVMSSM), namely flavour violating processes and baryon number violating processes.

The situation in the RPVMSSM goes as follow. The flavour violating processes put severe constraints on products of $\lambda''$s with different flavour structures while baryon number violating processes can impose strong constraints on the $\lambda''$ individually \cite{Barbier:2004ez}. The baryon number violating processes that put the most stringent bounds are proton decay, neutron antineutron oscillation and double nucleon decay. The proton decay constraint can be avoided if we assume that lepton number is conserved and that the gravitino is heavier than the proton, leaving neutron antineutron oscillation and double nucleon decay which are still very constraining for many of the $\lambda''$s, with the constraint on $\lambda''_{112}$ being the strongest. One approach to satisfy both the flavour violating and baryon number violating constraints is to assume a minimal flavour violating (MFV)  structure for the $\lambda''_{ijk}$ \cite{Csaki:2011ge}. This leads to very small couplings and the LHC phenomenology is then characterized by displaced vertices. Another approach to avoid the bounds is to assume that in the mass eigenstate basis only one coupling of the form  $\lambda_{3 i j}''$ is large  while the $\lambda''$s with different flavour structures are very suppressed. The bounds are then easily satisfied. Single stop production  becomes relevant at the LHC, and neutralinos can decay promptly via an off-shell stop to a top and two jets. This phenomenology was explored in \cite{Monteux:2016gag}. The difficulty in such a scenario is to build a flavour model that  leads in the mass basis to a large $\lambda''_{3 i j}$ coupling but a very small $\lambda''_{1 1 2}$ coupling. 

\subsubsection{Bounds from baryon number violating processes}
\begin{figure}[t!]
  \centering
  \begin{subfigure}{0.41\textwidth}
    \centering
    \includegraphics[width=\textwidth, bb = 0 0 469 339]{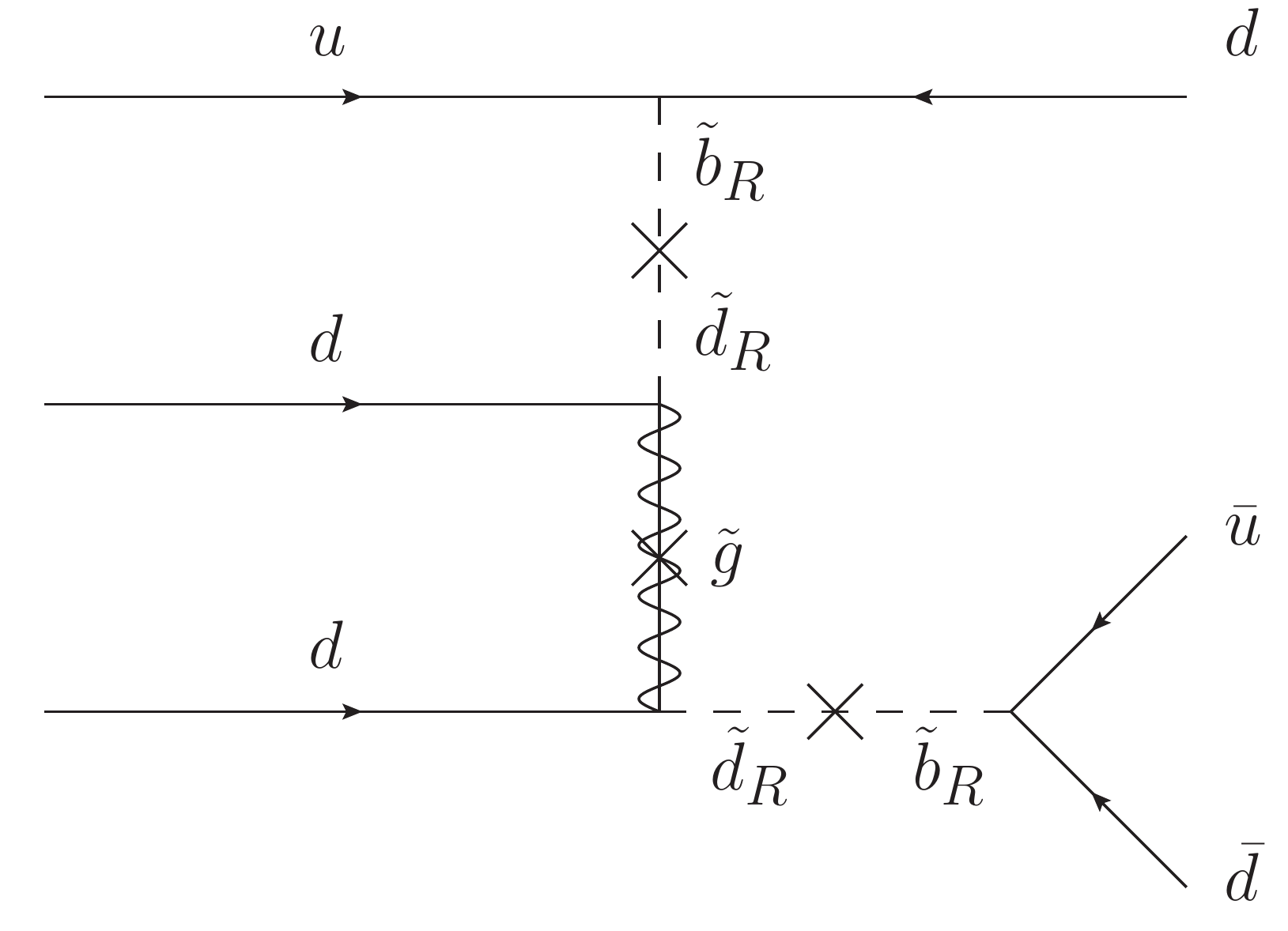}
    \caption{}
    \label{fig:nnbarRR}
  \end{subfigure}
  ~ ~ ~ ~ ~
    \begin{subfigure}{0.41\textwidth}
    \centering
    \includegraphics[width=\textwidth, bb = 0 0 469 339]{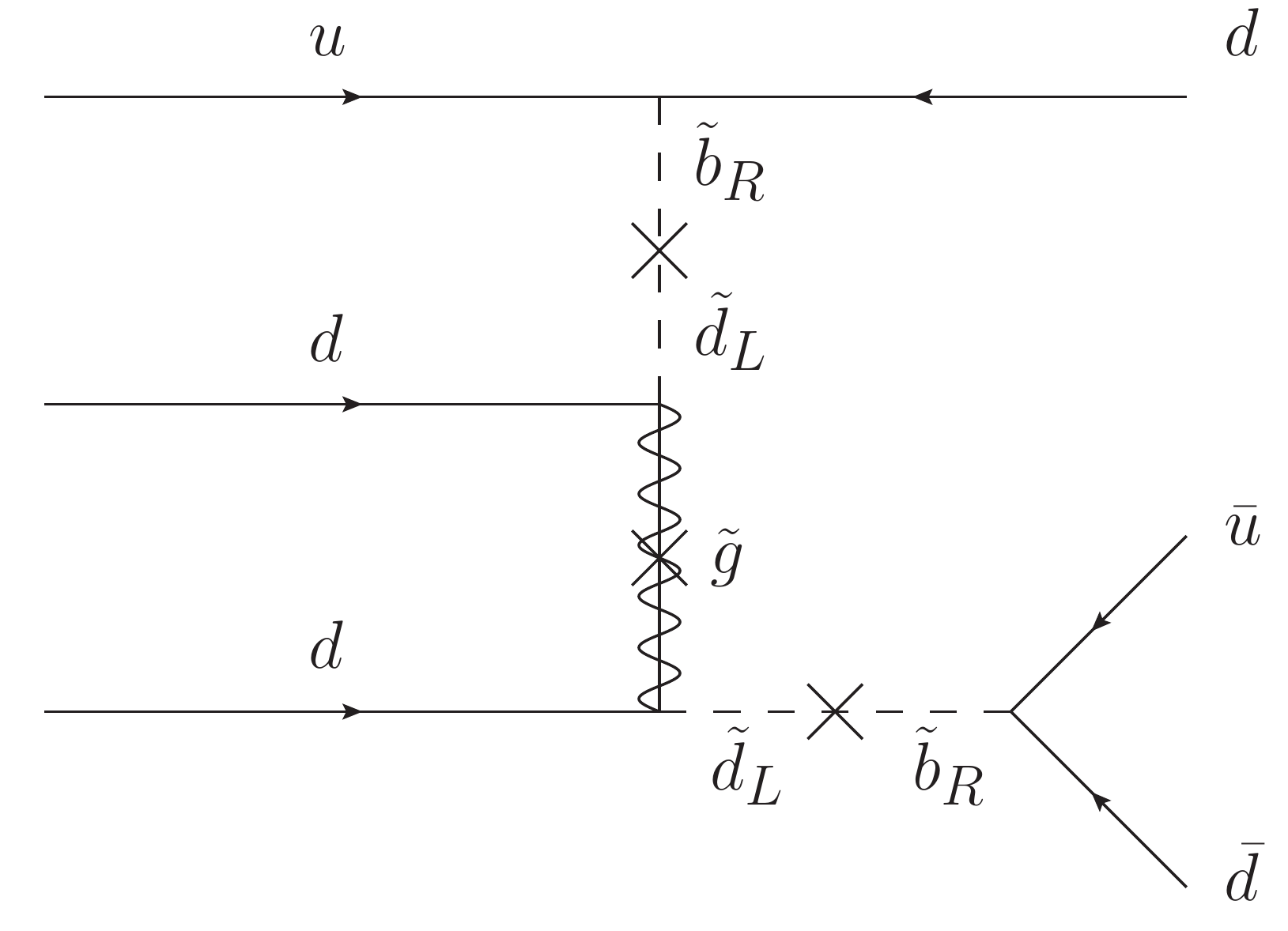}
    \caption{}
    \label{fig:nnbarLR}
  \end{subfigure}
\caption{Diagrams leading to neutron antineutron oscillation. Flavour changing insertions are needed on the squark lines and a Majorana mass insertion is needed on the gluino line. (a) shows a diagram with flavour changing insertions of the right-handed squarks. (b) shows a diagram requiring a left-right squark mixing which is further suppressed by the gravitino mass.}\label{fig:nnbar}
\end{figure}

In the model we consider, baryon number is  violated only by the small $U(1)_R$ breaking terms coming from anomaly mediation which are proportional to the gravitino mass. Constraints from baryon number violating processes are then potentially weaker than in the RPVMSSM. However, if the gravitino is lighter than the proton, the proton can decay to a gravitino and a kaon. This process proceeds through a $\lambda''_{112}$ coupling and is the same as in the RPVMSSM, leading to a bound of \cite{Barbier:2004ez}:
\[
\lambda''_{112} \lesssim 6 \times 10^{-15} \left(\frac{m_{\tilde{q}}}{1 \text{TeV}} \right)^2 \left(\frac{m_{3/2}}{1 \text{eV}}\right).
\]
When the gravitino is heavier than the proton, the bounds from neutron antineutron oscillation and double nucleon decay still apply. The best experimental limit on neutron antineutron oscillation comes from the non-observation of $^{16}\text{O}$ decay to various final states with multiple pions and omega particles at SuperKamiokande \cite{Abe:2011ky}. This process receives tree-level contributions from diagrams of the form shown in figure \ref{fig:nnbarRR}. 
This leads to a bound on $\lambda_{11i}''$ which is somewhat model dependent as the diagram requires flavour mixing mass insertions on the squark lines. It also requires the insertion of a Majorana mass term for the gluino which we take to be given by anomaly mediation: $M_3  =3 \alpha_s m_{3/2}/4 \pi $. The amplitude for this process can be estimated to be \cite{Csaki:2011ge}:
\[
M_{n-\bar{n}} \sim 4\pi \alpha_s   \left(\lambda''_{11i}\right)^2 \frac{\left(\delta_{i1}^{RR}\right)^2 }{m_{\tilde{q}}^4} \frac{M_3}{\left(M^D_3\right)^2}  \Lambda^6,
\]
where $\delta_{ij}^{RR}$ is the ratio of the flavour non-diagonal elements of the right-handed down-type squark mass matrix to the flavour diagonal ones and  $\Lambda$  is the characteristic scale for the neutron matrix elements which is expected to be close to the QCD scale. Taking $\alpha_s=0.12$, we find a bound of the form \cite{Csaki:2011ge}:
\begin{equation}
\lambda''_{11i} \lesssim 2 \times 10^{-5} \left(\frac{1}{\delta_{i1}^{RR}}\right)\left(\frac{M_3^D}{1 \text{TeV}}\right) \left(\frac{1 \text{GeV}}{m_{3/2}}\right)^{1/2}\left(\frac{m_{\tilde{q}}}{1 \text{TeV}}\right)^2 \left(\frac{250 \text{MeV}}{\Lambda}\right)^3.
\end{equation}
If for some reason the effect of the flavour mixing in the right-handed squark mass matrix is small, which could happen if, for example, this matrix follows an MFV pattern,\footnote{With an MFV structure, there exists a basis where both the right-handed squark matrix  and the gauge Yukawa interactions involving down-type quarks are flavour diagonal.} then the process needs to involve left-right squark mixing (see figure \ref{fig:nnbarLR}). In the limit of an exact $U(1)_R$, these mixings, which come from $A$-terms, are forbidden. They are however expected to be generated with a size proportional to the gravitino mass once $U(1)_R$ breaking effects are taken into account. Taking the anomaly mediation value for the $A$-terms, the bound  becomes:
\begin{equation}
\lambda''_{1 1 i} \lesssim 2 \left(\frac{1}{y^d_{1i}} \right) \left(\frac{M_3^D}{1 \text{TeV}}\right) \left(\frac{1 \text{GeV}}{m_{3/2}}\right)^{3/2} \left(\frac{m_{\tilde{q}}}{1 \text{TeV}}\right)^4 \left(\frac{250 \text{MeV}}{\Lambda}\right)^3,
\end{equation}
where $y^d_{ij}$ is the down-type Yukawa matrix, and we see that order one $\lambda''$ become easily allowed. 

\begin{figure}
\begin{center}
\includegraphics[width=0.41\textwidth, bb = 0 0 406 406]{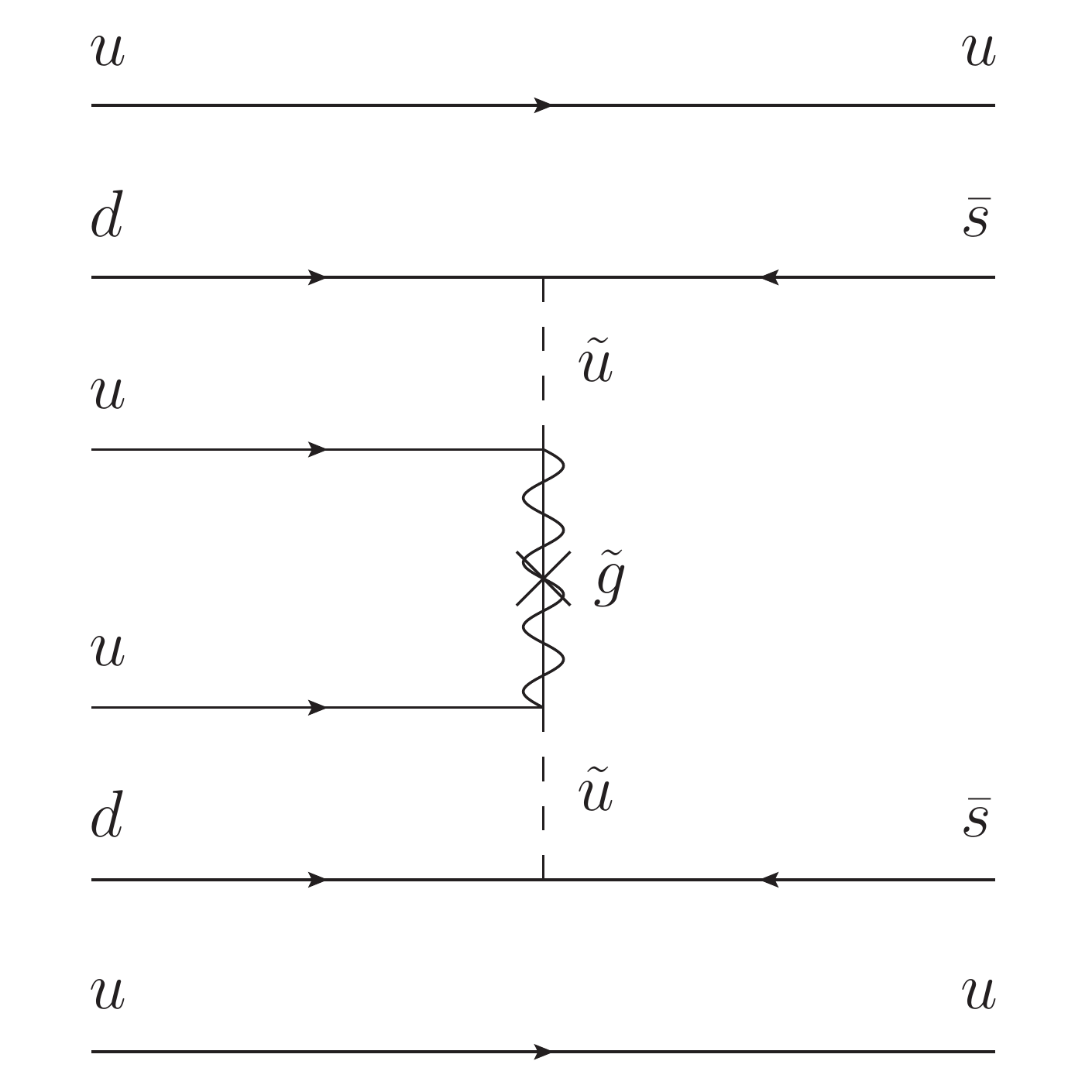}
\caption{Diagram mediating the $ p p \rightarrow K^+ K^+$ process.}
\label{fig:ppKK}
\end{center}
\end{figure}

The bound coming from double nucleon decay is more  independent from flavour physics as it can proceed through a diagram such as the one showed in figure \ref{fig:ppKK} which does not require flavour mixing on the squark lines. The diagram on the other hand still  requires the insertion of a gluino Majorana mass term. 
The best limit on this process also comes from the non-observation of $^{16}\text{O}$ decay to $^{14}\text{C} K^+ K^+$ at Superkamiokande \cite{Litos:2014fxa}. The bound on the partial lifetime is found to be  $1.7 \times 10^{32}$ years. A rough estimate for the amplitude can be obtained in a similar way to the $n-\bar{n}$ process \cite{Goity:1994dq,Csaki:2011ge}.  It leads to a bound on $\lambda''_{1 1 2}$ of the form:
\begin{equation}
\label{eq:boundppKK}
\lambda''_{112} \lesssim 2 \times 10^{-4} \left(\frac{M_3^D}{1 \text{TeV}}\right) \left( \frac{1 \text{GeV}}{m_{3/2}}\right)^{1/2} \left(\frac{m_{\tilde{q}}}{1 \text{TeV}}\right)^2 \left(\frac{150 \text{MeV}}{\tilde{\Lambda}}\right)^{5/2},
\end{equation}
where $\tilde{\Lambda}$ is the hadronic scale which is hard to estimate and introduces significant uncertainty on the bound. It is  expected to be suppressed compared to $\Lambda_{\text{QCD}}$ due to nucleon repulsion \cite{Goity:1994dq,Csaki:2011ge}.

\subsubsection{Bounds from flavour physics}
Flavour physics also puts strong bounds on the $\lambda''$ parameters. The bounds are on products of two $\lambda''$s with different flavour structures \cite{Barbier:2004ez,Giudice:2011ak}. For example, there are loop diagrams that contribute to $\epsilon_K$ and $\Delta m_K$, leading to a bound of the form \cite{Giudice:2011ak}:
\begin{equation}
\sqrt{|\text{Im}(\lambda''_{i 23} {\lambda''^*_{i13}})^2|} \lesssim 2.8 \times 10^{-3} \left(\frac{m_{\tilde{u}_i}}{1 \text{TeV}}\right),
\end{equation}
from $\epsilon_K$ while $\Delta m_K$ gives:
\begin{equation}
\sqrt{|\text{Re}(\lambda''_{i 23} \lambda''^*_{i13})^2|} \lesssim 4.6 \times 10^{-2} \left(\frac{m_{\tilde{u}_i}}{1 \text{TeV}}\right).
\end{equation}
There are also strong bounds on $\lambda''_{i 23} \lambda''^*_{i 12}$ from $B$-mixing and from bounds on $BR(B^{\pm} \rightarrow \phi \pi^{\pm})$. These bounds can be satisfied by having only one of the $\lambda''$s sizable in the mass eigenstate basis. Whether or not this can be easily achieved depends on the structure of the flavour physics. For example it might be possible to arrange for one of the $\lambda''$ to be dominant in the gauge basis, but when rotating to the mass basis other flavour structures will be generated. If the rotation has the same structure as the CKM matrix  a $\lambda''_{312}$ coupling of order one in the gauge basis is allowed by flavour physics constraints provided the squarks have masses in the TeV range \cite{Giudice:2011ak}. However, in order to satisfy the bound on $\lambda''_{112}$ from eq$.$ (\ref{eq:boundppKK}), the rotation of the right-handed up squarks from the gauge to the mass basis must induce a suppression of $\sim 10^{-4}$, and a CKM like structure will be insufficient. If $\lambda''_{313}$ is dominant in the gauge basis, $K-\bar{K}$ mixing constrains this coupling to be $\lesssim 0.1$, and in this case a CKM-like rotation structure for the up squarks will put $\lambda''_{112}$ close to the bound of eq$.$ (\ref{eq:boundppKK}).

\subsection{Spectrum and parameter space}\label{subsec:spectrum}
In view of the strong constraints on the $\lambda''$ couplings, we focus from now on models where only a single coupling of the form $\lambda''_{3ij}$ is important.  The relevant features of the phenomenology will crucially depend on the size of this coupling and on the spectrum. For example, for large $\lambda''_{3ij}$ single stop production can be important while a smaller  coupling  leads to pair production being dominant. Large $\lambda''_{3ij}$ couplings will also lead to the prompt decay of neutralinos to a top quark and two jets, leading to a distinct phenomenology from the displaced vertices characteristic of the small RPV coupling case.

\begin{figure}
 \begin{center}
 \includegraphics[width=0.41\textwidth, bb = 0 0 471 217]{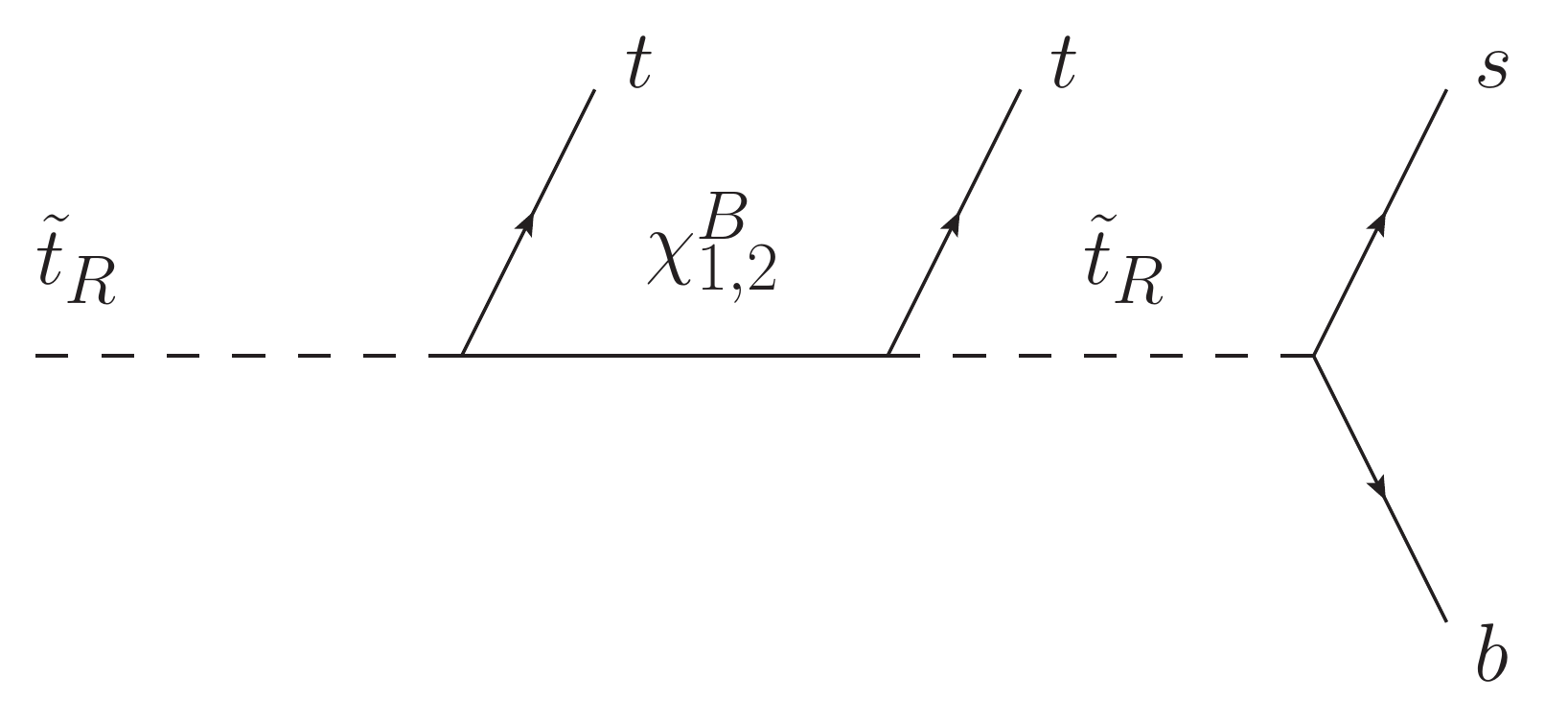}
 \caption{Decay of the stop through  the two on-shell pseudo-Dirac states $\chi^B_1$ and $\chi^B_2$. When the mass difference between the two states is smaller than the width, the diagram with $\chi^B_1$ cancels the one with $\chi^B_2$.}
 \label{fig:stopdecay}
 \end{center}
 \end{figure}

In the limit where the $U(1)_R$ symmetry is exact there is a distinction between neutralinos and antineutralinos. One of them has baryon number $1$ and decays to $t j j$,  while the other has baryon number $-1$ and decays to $\bar{t} j j $. In this case, the decay of a stop will always involve opposite sign tops: $\tilde{t} \rightarrow  t \bar{\chi}_0 \rightarrow t \bar{t}  j j$. However, in the presence of Majorana mass terms for the gauginos, the Dirac neutralinos split into two Majorana states which can both decay to either $t j j$ or $\bar{t} j j $.  This is important for the phenomenology as in this case there will be a signature with two same-sign leptons. We can see how this works by looking at a bino LSP interacting with the stop through the following potential: 
\begin{equation}
M^D_1 \tilde{S} \tilde{B}  + \frac{1}{2} M_1 \tilde{B} \tilde{B}-\frac{2 \sqrt{2}}{3} g' \tilde{t}_R^\dagger (t_R \tilde{B}) + \lambda''_{323} \tilde{t}_R(b_R s_R) + \text{h.c.},
\end{equation}
where $M_1$ is a small Majorana mass term for the bino. 
 The mass eigenstates are two pseudo-Dirac states \cite{DeSimone:2010tf} given by:
 \begin{eqnarray*}
 \chi^B_1 &=& i \frac{1}{\sqrt{2}} (\tilde{B} - \tilde{S})\\
 \chi^B_2 &=& \frac{1}{\sqrt{2}}(\tilde{B} + \tilde{S})
 \end{eqnarray*}
with corrections of order $M_1/M^D_1$. The masses of the two eigenstates are given by $m^B_1= M^D_1 - M_1/2$ and $m^B_2=M^D_1+M_1/2$ to leading order.  In term of the mass eigenstates, the potential can then be written as:
 \begin{equation}
\frac{m^B_1}{2} \chi^B_1 \chi^B_1 + \frac{m_2^B}{2}  \chi^B_2 \chi^B_2 + i \frac{2 g'}{3} \tilde{t}_R^\dagger t_R \chi^B_1 -\frac{2 g'}{3} \tilde{t}_R^\dagger t_R \chi^B_2 + \lambda''_{323} \tilde{t}_R(b_R s_R)+ \text{h.c..}
\end{equation}
The decay of the stop can proceed via an on-shell $\chi_1^B$ or $\chi^B_2$ as shown in figure \ref{fig:stopdecay}. In the case of a decay to $t t j j$, the two amplitudes have opposite sign and interfere destructively, while for the decay  to $t \bar{t} j j$, the amplitudes have the same sign and add. Therefore for a mass splitting smaller than the width of $\chi^B_1$ and $\chi^B_2$, decay chains with  same sign tops are suppressed whereas for a larger mass splitting they occur as often as opposite sign tops.
In our study of the LHC phenomenology we will compute the bound on squark masses as a function of the $\lambda''_{3 ij }$ for bino and Higgsino-up LSP. For gravitino masses slightly above $\sim 1$ GeV, for which the bound from proton decay to gravitino does not apply, the mass splitting between the pseudo-Dirac neutralino states is small enough to be ignored for most processes, except for the stop decay. A mass splitting of order 1 GeV is much larger than the typical neutralino width and as a consequence decay chains with same-sign tops will occur.  We will also show bounds for a case where the $U(1)_R$ symmetry is nearly exact  with no same-sign top signatures. This last case requires a very low $m_{3/2}$, which might be difficult to achieve, but could have interesting consequences for cosmology \cite{Ipek:2016bpf}.  

For our study of baryogenesis, we need to consider a different region of parameter space. In order to have a gaugino that decays out of equilibrium and generate a baryon antibaryon asymmetry through its decay, we will be led to consider a split spectrum with very heavy scalar masses. The bounds on $\lambda''$ are then considerably relaxed. Also, as explained in more details in section \ref{Sec:Baryogenesis} we will need to consider significantly larger mass splitting between the gauginos which means larger  $U(1)_R$ breaking.

\section{Collider constraints}\label{Sec:collider_constraints}
In this section we constrain the parameter space of the model by using a variety of LHC searches. We focus on two different scenarios. The first scenario is resonant stop production together with stop pair production. The second scenario is pair production of the first and second  generations of squarks. From here on out we simply refer to this scenario as squark production. For both scenarios, we consider the cases in which the $U(1)_R$ symmetry is either strictly preserved or, alternatively, broken. As mentioned above, whether or not the $U(1)_R$ symmetry is broken changes the phenomenology. 

\subsection{Placing limits on stops}\label{sSec:limits_stops} 

\subsubsection{Stop production}\label{ssSec:stop_production}
The main phenomenological novelty of the model is the presence in the superpotential of the term:
\begin{align}
\frac{1}{2}\lambda''_{3ij}U_3^c D_i^c D_j^c ,
\end{align}
which can only contain stops that are right-handed. Consequently, we concentrate on the production of right-handed stops, which we simply refer to as stops from now on. The left-handed stop, which does not mix with the right-handed one as it possesses a different $R$-charge,  is assumed to be decoupled.

If any of $\lambda''_{312}$, $\lambda''_{313}$ or $\lambda''_{323}$ is non-zero, resonant stop production can potentially take place at the LHC. For example, turning on $\lambda''_{312}$ will result in the partonic level processes $d s \rightarrow \tilde{t}^*$ and $\bar{d} \bar{s} \rightarrow \tilde{t}$, provided that the stop is not too heavy. Precisely, the partonic level cross section for $d_i d_j \rightarrow \tilde{t}^*$ is \cite{Berger:1999zt}:
\begin{align}
\hat{\sigma}(d_i d_j \rightarrow \tilde{t}^*) = \frac{\pi}{6}\frac{|\lambda''_{3ij}|^2}{m_{\tilde{t}}^2}\delta(1 - m^2_{\tilde{t}}/\hat{s}),
\end{align}
where $\hat{s}$ is the partonic centre of mass energy. Due to the valance down quark, the cross section to produce $\tilde{t}^*$ is generally much larger than that to produce $\tilde{t}$ (although if only $\lambda''_{323}$ is non-zero than $\tilde{t}^*$ and $\tilde{t}$ are produced in roughly equal amounts). Additionally, due to the small content of strange and bottom  within the proton, stop production through $\lambda''_{312}$ is larger than that through $\lambda''_{313}$, which is itself larger than that through $\lambda''_{323}$, assuming equal values for $\lambda''_{312}$, $\lambda''_{313}$ and $\lambda''_{323}$. We use MadGraph5\_aMC@NLO \cite{Alwall:2014hca} to calculate the leading order (LO) cross section at centre of mass energies of 8 and 13 TeV for resonant stop production (summing both $\tilde{t}^*$ and $\tilde{t}$) turning on $\lambda''_{312}$, $\lambda''_{313}$ and $\lambda''_{323}$ one at a time.\footnote{To simulate collisions, we used the Mathematica package FeynRules 2.0 \cite{Alloul:2013bka} to produce our own MRSSM MadGraph models, one with the $U(1)_R$ symmetry preserved and another with the symmetry broken.} In this fashion, all constraints placed throughout this section assume only a single $\lambda''_{3ij}$ is non-zero. Our limits are then conservative compared to the case where multiple $\lambda''_{3ij}$ are non-zero.

\begin{figure}[t!]
  \centering
  \begin{subfigure}[b]{0.48\textwidth}
    \centering
    \includegraphics[width=\textwidth, bb = 0 0 584 453]{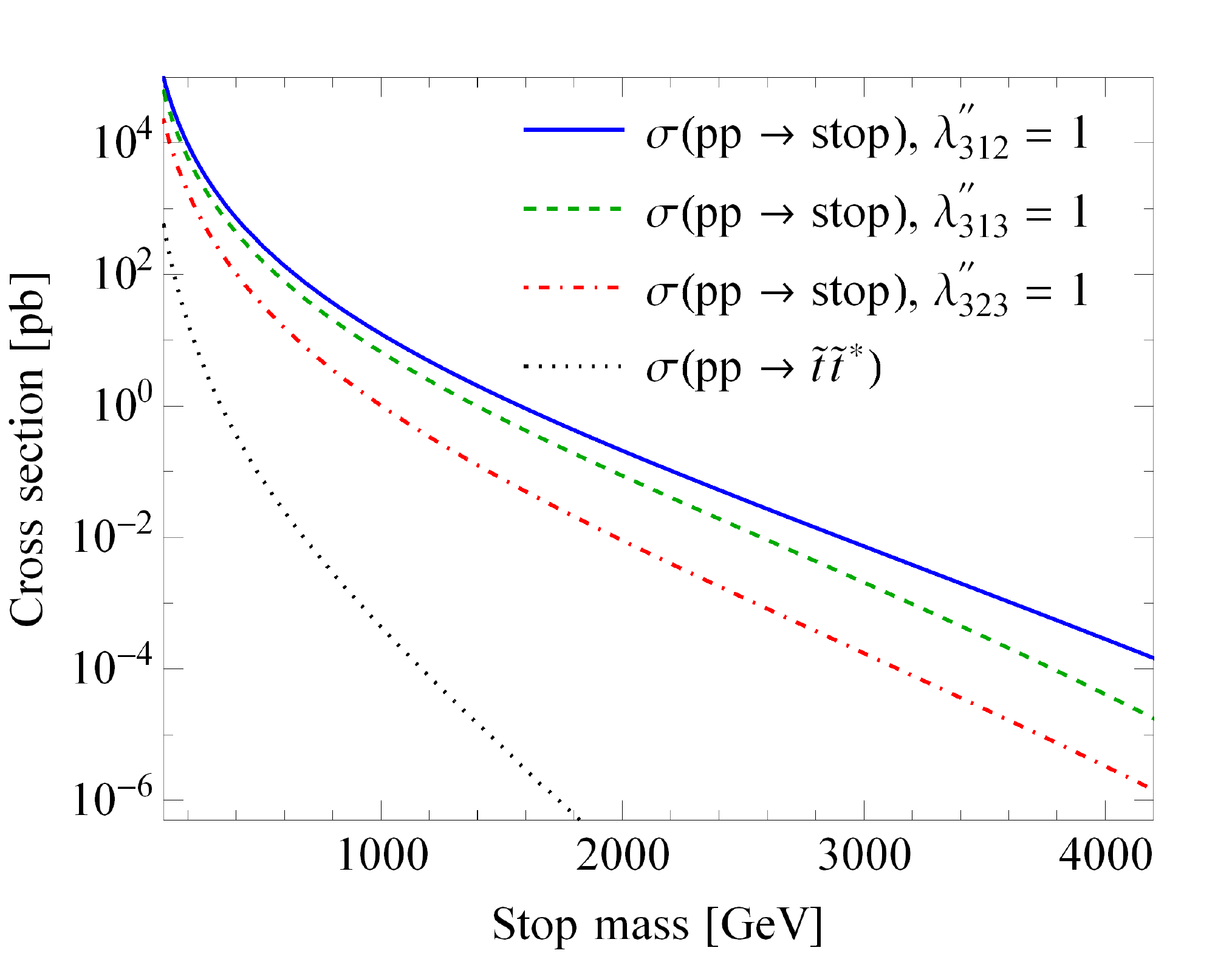}
    \caption{8 TeV}
    \label{fig:stop_production_8TeV}
  \end{subfigure}
  ~
  \begin{subfigure}[b]{0.48\textwidth}
    \centering
    \includegraphics[width=\textwidth, bb = 0 0 584 453]{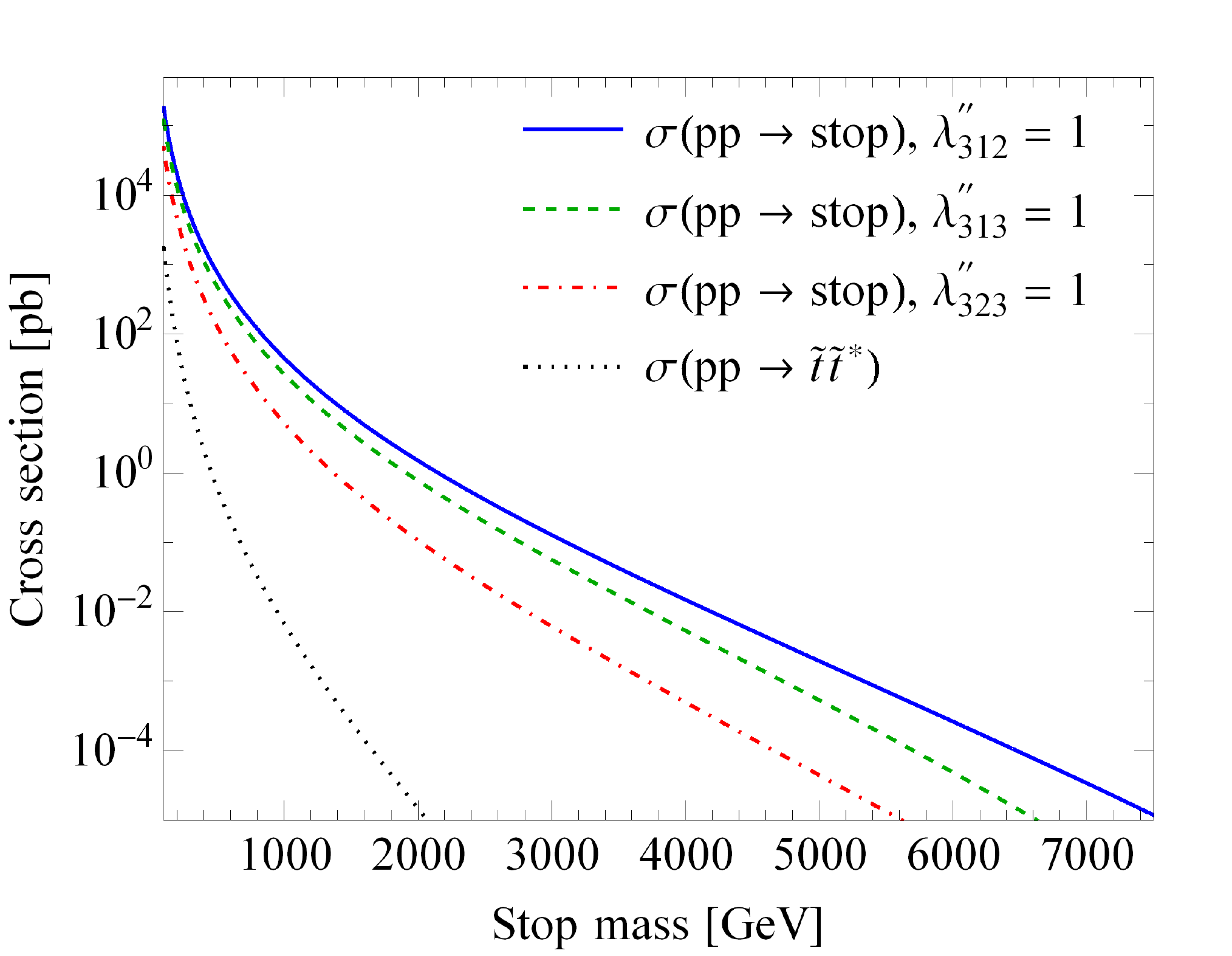}
    \caption{13 TeV}
    \label{fig:stop_production_13TeV}
  \end{subfigure}

  \caption{Stop production at centre of mass energy of 8 and 13 TeV. Here $\sigma(\text{pp} \rightarrow \text{stop})$ stands for $\sigma(\text{pp} \rightarrow \tilde{t}^*) + \sigma(\text{pp} \rightarrow \tilde{t})$. For resonant stop production, only one $\lambda''_{3ij}$ is non-zero at a time.}\label{fig:stop_production}
\end{figure}

Naturally, the LO cross section will be corrected by next-to-leading order (NLO) QCD effects. The NLO cross section for single stop production has been calculated in Ref$.$ \cite{Plehn:2000be}. There, K-factors for each of the $\lambda''_{3ij}$ are presented for stop masses between 200 and 800 GeV at a centre of mass energy of 14 TeV. It was found that the K-factors varied between approximately 1.2 and 1.4. To account for this, we simply multiply the LO cross sections computed with MadGraph by a constant K-factor of 1.3 for all stop masses. Figure \ref{fig:stop_production} shows the resulting cross sections where each $\lambda''_{3ij}$ has been set individually to one.

As in the MSSM, stops will also be produced in pairs. However, due to the $\lambda''_{3ij}$ coupling, there are new diagrams that contribute. These diagrams consist of two $\lambda''_{3ij}$ vertices, two initial state quarks and a t-channel quark. Couplings of order one can give significant contributions to the cross sections. For example, using MadGraph to compute the LO pair production cross section for 200 GeV stops at 13 TeV, we find a 20\% increase when $\lambda''_{312}$ is set to one compared to when it is zero. However, as far as we know, NLO corrections have not been computed for these new diagrams. Moreover, for $\lambda''_{3ij}$ of order one, single top production dominates the exclusion in most of the parameter space. For these reasons, we choose not to include this new contribution to our stop pair production cross section. Instead, we compute the cross section for this process using NLL-fast \cite{Beenakker:1997ut, Beenakker:2010nq, Beenakker:2011fu} and NNLL-fast \cite{Beenakker:2016lwe, Beenakker:1997ut, Beenakker:2010nq, Beenakker:2016gmf} for centre of mass energies of 8 and 13 TeV, respectively. We verify the results using Prospino \cite{Beenakker:1996ed}. Our limits from stop pair production are then conservative, particularly when the $\lambda''_{3ij}$ are of order one. The cross section is also shown in figure \ref{fig:stop_production}. As can be seen, resonant stop production is quite a bit larger than stop pair production for any of the $\lambda''_{3ij}$ set to unity.

\subsubsection{Stop LSP}\label{ssSec:Stop_LSP}
If the stop is the LSP, then it will decay directly into two quarks through the $\lambda''_{3ij}$ coupling with a branching ratio of one. In this situation, there are two processes of interest:
\begin{align*}
(1) \ 
\begin{array}{l}
p p \rightarrow \tilde{t}^* \rightarrow d_i d_j
\end{array}
\quad \quad
(2) \
\begin{array}{l}
p p \rightarrow \tilde{t}^* \tilde{t} \rightarrow d_i d_j \bar{d}_i \bar{d}_j
\end{array}
\end{align*}
where the final state quarks depend on which one $\lambda''_{3ij}$ is non-zero. We now constrain the parameter space using these two processes.

Let us focus on the first process. This case is sensitive to dijet searches performed at the LHC. We examined many of these searches and selected the following ones to recast: \cite{Aad:2014aqa, ATLAS-CONF-2016-030, ATLAS-CONF-2016-069} from ATLAS and \cite{Khachatryan:2016ecr, CMS-PAS-EXO-16-032} from CMS. The procedures used to recast these searches are described below. Notably, each one of these searches is independent of the flavour of the final state quarks as they do not utilize b-tagging. We also considered the ATLAS dijet search \cite{ATLAS-CONF-2016-060} which does utilize b-tagging but found that the exclusion limits did not improve. Particularly, search \cite{ATLAS-CONF-2016-069} provides stronger limits than \cite{ATLAS-CONF-2016-060}.  The reason for this is that even though \cite{ATLAS-CONF-2016-060} requires a b-tagged jet, the limits on the cross section times branching ratio times acceptance between \cite{ATLAS-CONF-2016-069} and \cite{ATLAS-CONF-2016-060} are comparable. However, requiring a b-tagged jet results in the acceptance for \cite{ATLAS-CONF-2016-060} being about half that of \cite{ATLAS-CONF-2016-069}, thus making it less constraining. 

Both the ATLAS and CMS experiments have developed special techniques to place limits on low mass resonances decaying to dijets. The ATLAS technique is known as Trigger-object Level Analysis (TLA) and was implemented in \cite{ATLAS-CONF-2016-030} to constrain masses below 1.1 TeV. The CMS technique is known as data scouting and was implemented in searches \cite{Khachatryan:2016ecr, CMS-PAS-EXO-16-032} to constrain masses below 1.6 TeV. The low mass region is experimentally difficult due to a combination of the limited bandwidth available to record events to disk and the large Standard Model multijet rate. Either a large fraction of events must be discarded or stringent triggers must be used in order to keep the amount of recorded data to an acceptable level. However, both options limit the statistical power of the search. The TLA and data scouting approach is to record only the portion of the event data, such as jet four-momenta, needed to perform the dijet search. By doing so, event sizes can be reduced to 5\% (2\%) of what they would normally be for ATLAS \cite{ATLAS-CONF-2016-030} (CMS \cite{Khachatryan:2016ecr}). This allows for more statistics and hence stronger limits.

\begin{figure}[t!]
  \centering
  \begin{subfigure}[b]{0.3\textwidth}
    \centering
    \includegraphics[width=\textwidth, bb = 0 0 584 463]{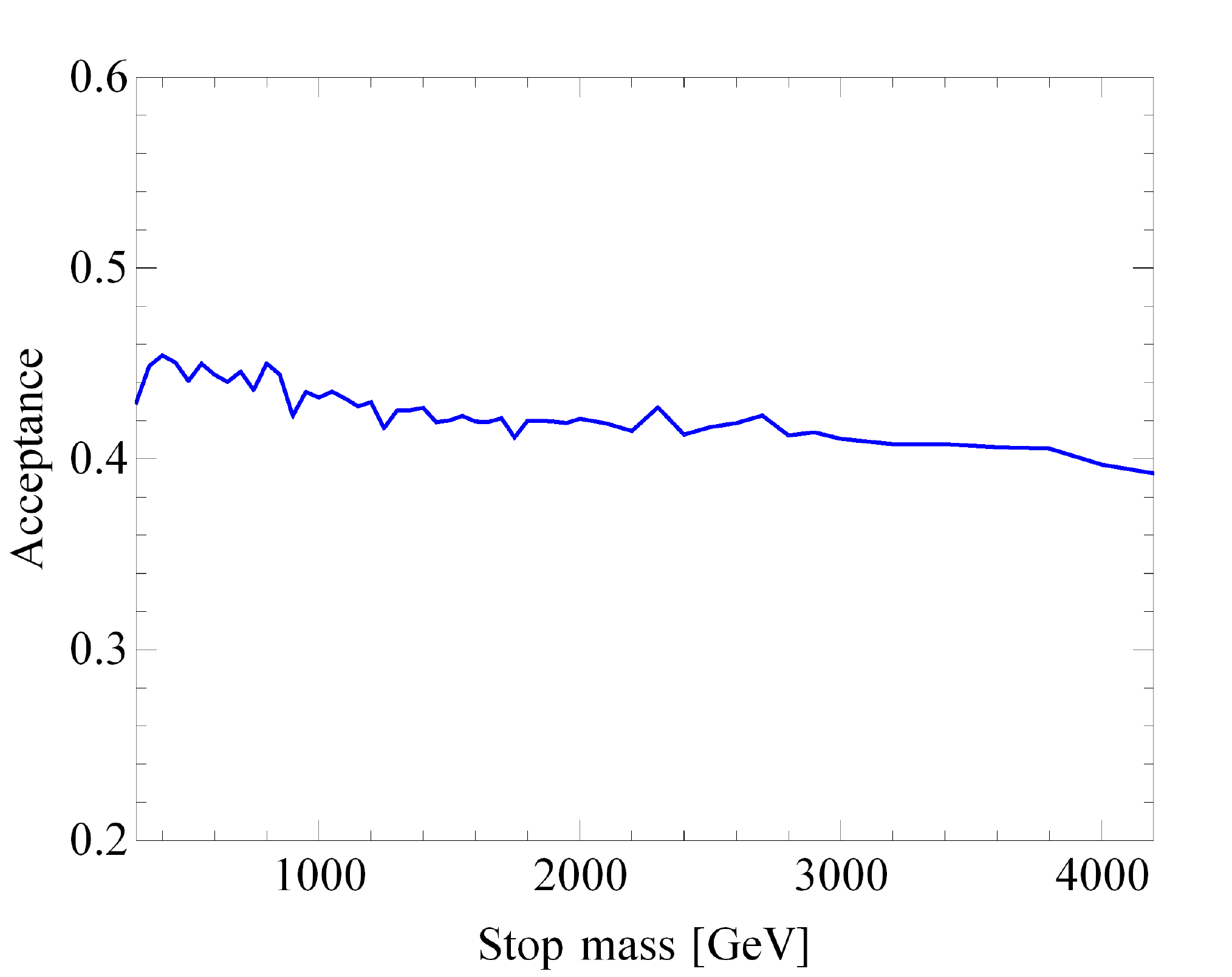}
    \caption{\cite{Aad:2014aqa}}
    \label{Fig:Acceptance:ATLAS_1407_1376}
  \end{subfigure}
  ~
  \begin{subfigure}[b]{0.3\textwidth}
    \centering
    \includegraphics[width=\textwidth, bb = 0 0 584 463]{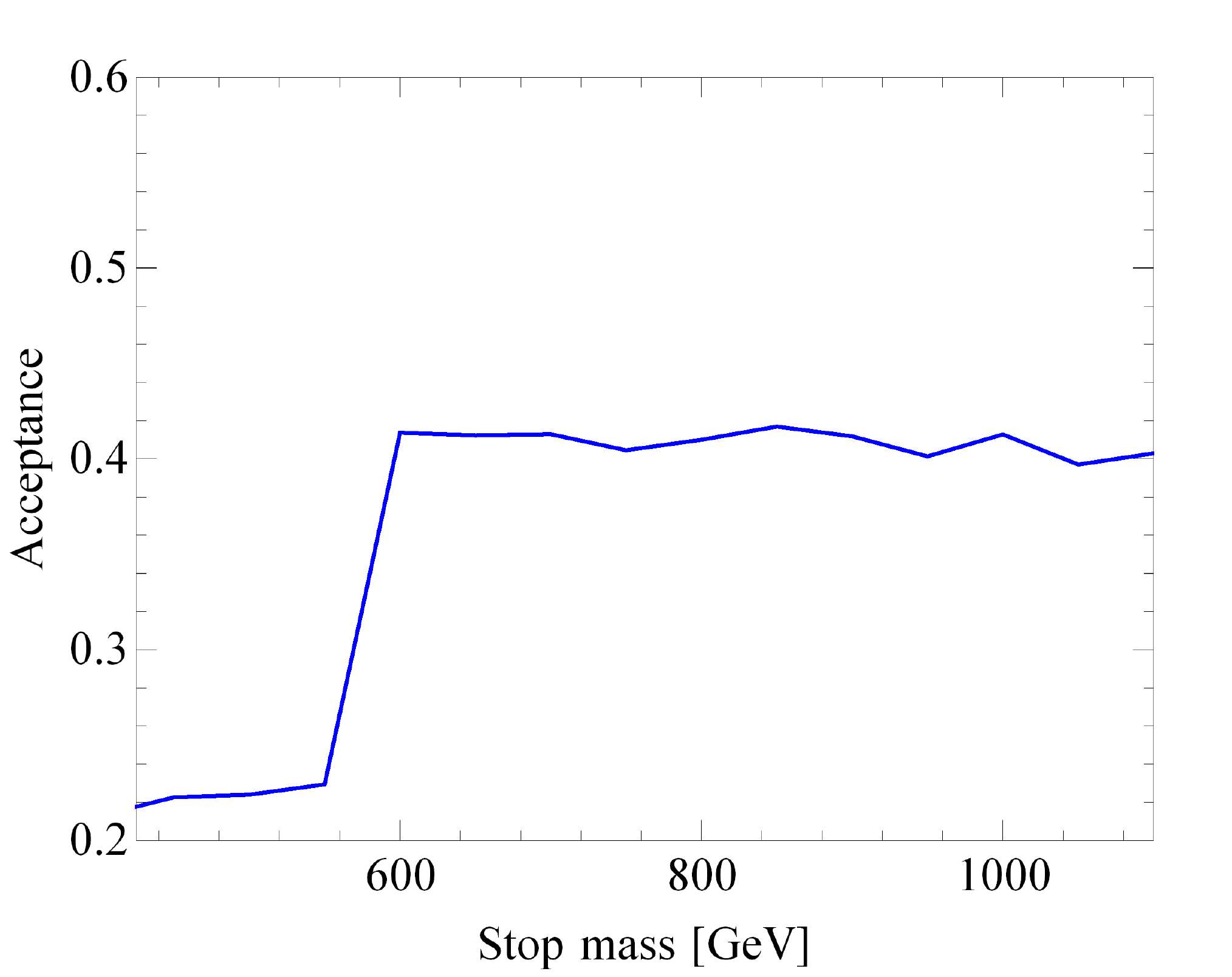}
    \caption{\cite{ATLAS-CONF-2016-030}}
    \label{Fig:Acceptance:ATLAS_CONF_2016_030}
  \end{subfigure}
  ~
  \begin{subfigure}[b]{0.3\textwidth}
    \centering
    \includegraphics[width=\textwidth, bb = 0 0 584 463]{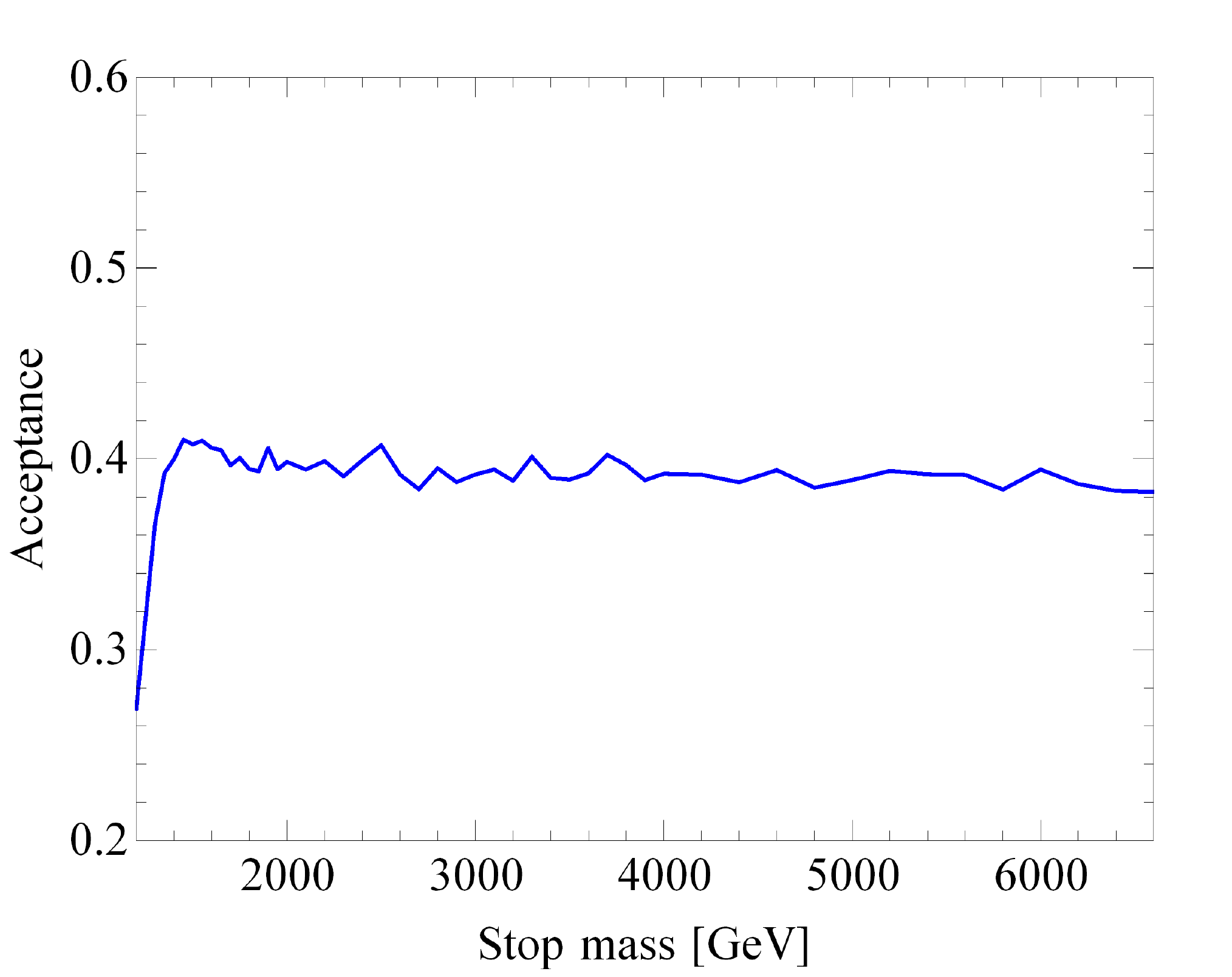}
    \caption{\cite{ATLAS-CONF-2016-069}}
    \label{Fig:Acceptance:ATLAS_CONF_2016_069}
  \end{subfigure}
  ~
  \begin{subfigure}[b]{0.45\textwidth}
    \centering
    \includegraphics[width=0.66\textwidth, bb = 0 0 584 463]{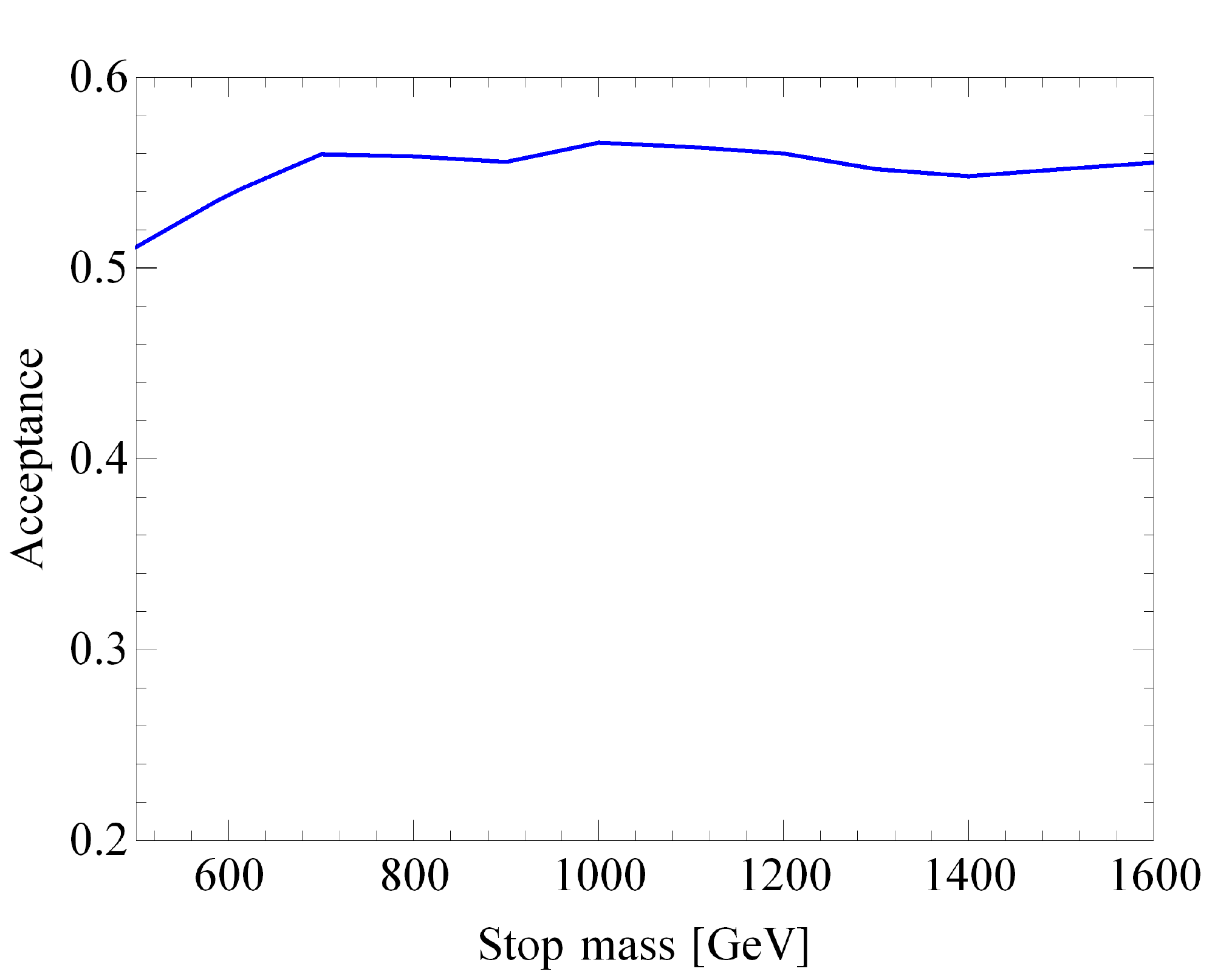}
    \caption{\cite{Khachatryan:2016ecr}}
    \label{Fig:Acceptance:CMS_1604_08907}
  \end{subfigure}
  ~
  \begin{subfigure}[b]{0.45\textwidth}
    \centering
    \includegraphics[width=0.66\textwidth, bb = 0 0 584 463]{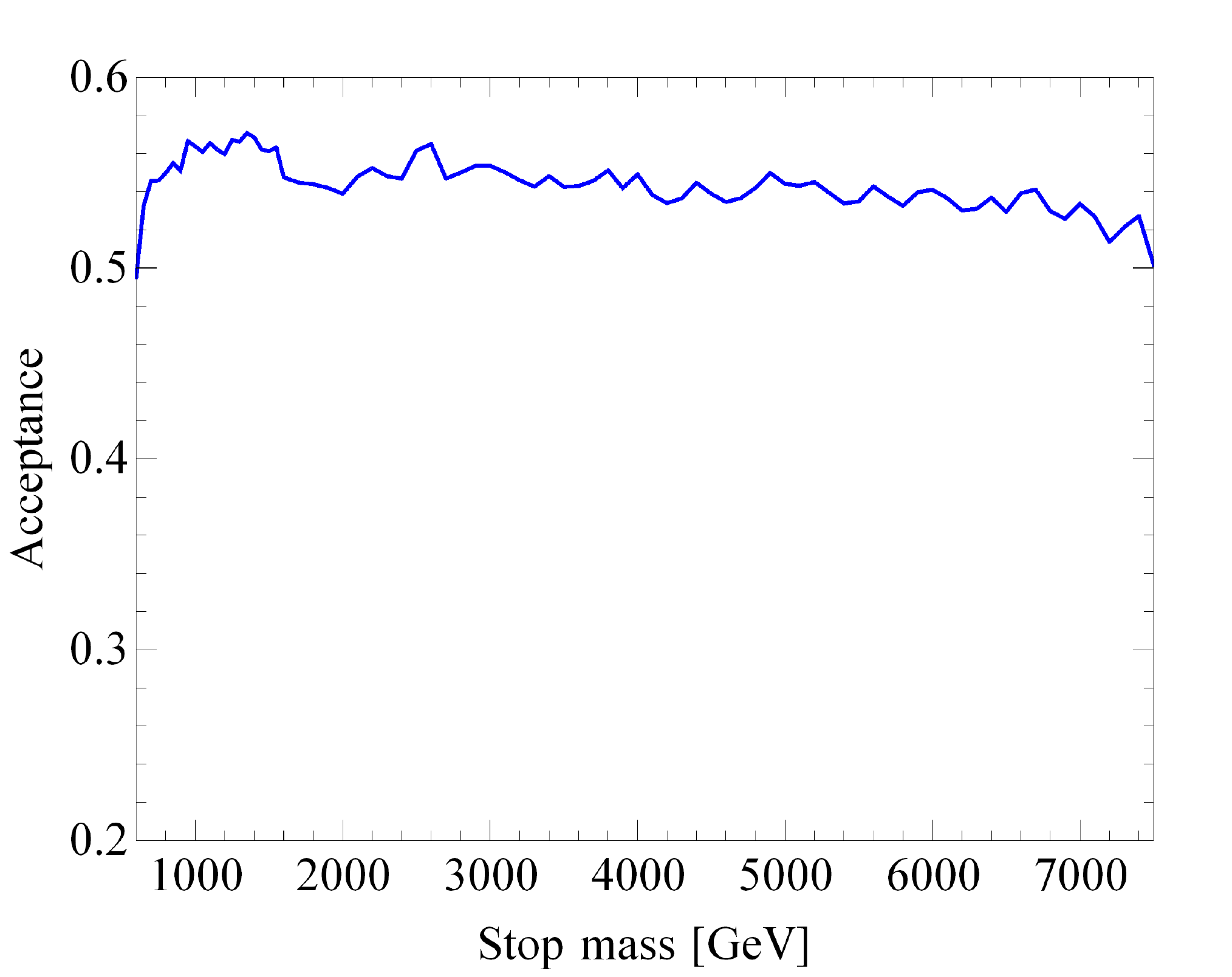}
    \caption{\cite{CMS-PAS-EXO-16-032}}
    \label{Fig:Acceptance:CMS_PAS_EXO_16_032}
  \end{subfigure}

  \caption{Acceptances for the dijet searches recasted in this analysis. Top row: ATLAS searches. Bottom row: CMS searches.}\label{fig:acceptance}
\end{figure}

To recast ATLAS dijet searches, we followed the procedure within Appendix A of \cite{Aad:2014aqa} to set limits on models of new physics with Gaussian resonances. First, for each search we chose a selection of stop masses $M$ to sample. Then, for each $M$, we used MadGraph to generate 10000 events of resonant stop production with the stop subsequently decaying into quarks. The events were given to PYTHIA 8.2 \cite{Sjostrand:2014zea} to simulate non-perturbative effects and then fed into Delphes 3 \cite{deFavereau:2013fsa} for detector simulation. The package HepMC2 \cite{Dobbs:2001ck} was used to interface between PYTHIA and Delphes. Next, code was written to implement the kinematic cuts. The cuts for each search were: 
\begin{alignat*}{2}
&\text{\cite{Aad:2014aqa}}: \ &&|y_{j_1}| < 2.8, \ |y_{j_2}| < 2.8, \ {p_T}_{j_1} > 50 \ \text{GeV}, \ {p_T}_{j_2} > 50 \ \text{GeV}, \\
& &&|\Delta y_{j_1 j_2}| < 1.2, \ m_{j_1 j_2} > 250 \ \text{GeV}, \ 0.8M < m_{j_1 j_2} < 1.2M, \\
&\text{\cite{ATLAS-CONF-2016-030}}: \ &&|\eta_{j_1}| < 2.8, \ |\eta_{j_2}| < 2.8, \ {p_T}_{j_1} > 185 \ \text{GeV}, \ {p_T}_{j_2} > 85 \ \text{GeV},\\
& &&|\Delta y_{j_1 j_2}| < \begin{cases} 0.6 \ \text{if} \ 425 \ \text{GeV} < m_G < 550 \ \text{GeV}, \\ 1.2 \ \text{if} \ 550 \ \text{GeV} < m_G < 1100 \ \text{GeV}, \end{cases} \\
& &&0.8M < m_{j_1 j_2} < 1.2M, \\
&\text{\cite{ATLAS-CONF-2016-069}}: \ &&{p_T}_{j_1} > 440 \ \text{GeV}, \ {p_T}_{j_2} > 60 \ \text{GeV}, \\
& &&|\Delta y_{j_1 j_2}| < 1.2, \ m_{j_1 j_2} > 1100 \ \text{GeV}, \ 0.8M < m_{j_1 j_2} < 1.2M,
\end{alignat*}
where, for the two leading jets $j_1$ and $j_2$: $y_{j_1}$ and $y_{j_2}$ are their rapidities, $\eta_{j_1}$ and $\eta_{j_2}$ are their pseudorapidities, ${p_T}_{j_1}$ and ${p_T}_{j_2}$ are their transverse momenta, $\Delta y_{j_1 j_2}$ is the difference between their rapidities and $m_{j_1 j_2}$ is their invariant mass. The cut $0.8M < m_{j_1 j_2} < 1.2M$ is designed to remove any long tails in the reconstructed $m_{j_1 j_2}$ distribution which has been assumed to be Gaussian. The acceptance for a search is then the fraction of the events to pass its cuts. The acceptances are shown in the top row of figure \ref{fig:acceptance}. Additionally, events that passed had their values of $m_{j_1 j_2}$ recorded in a histogram. A Gaussian distribution was then fit to the histogram and the standard deviation, $\sigma_G$, and mean, $m_G$, were determined. Finally, each search provided 95\% CL upper limits on the cross section times branching ratio times acceptance as a function of $m_G$ for different values of $\sigma_G/m_G$. We found that the vast majority of $\sigma_G/m_G$ values fell between 0.05 and 0.07.

Recasting CMS dijet searches followed a similar procedure up until applying the cuts. A major component  of the cuts centred around reconstructing two ``wide jets". The two leading jets served as the seeds for the two wide jets and the four-momentum of any other jet would be added to the closest leading jet if the two were separated by less than $\Delta R = 1.1$. Then, for a stop with mass $M$, the cuts for each search were:
\begin{alignat*}{2}
&\text{\cite{Khachatryan:2016ecr}}: \ &&H_T = \sum_j {p_T}_j > 250 \ \text{GeV}, \ \Delta\phi_{j_1 j_2} > \pi/3, \ |\Delta \eta_{J_1 J_2}| < 1.3, \ m_{J_1 J_2} > 390 \ \text{GeV}, \\
&\text{\cite{CMS-PAS-EXO-16-032}}: \ &&H_T = \sum_j {p_T}_j > \begin{cases} 250 \ \text{GeV} \ \text{if} \ 0.6 \ \text{TeV} < M < 1.6 \ \text{TeV}, \\ 800 \ \text{GeV} \ \text{if} \ 1.6 \ \text{TeV} < M < 7.5 \ \text{TeV}, \end{cases} \\
& &&|\Delta \eta_{J_1 J_2}| < 1.3, \\
& &&m_{J_1 J_2} > \begin{cases} 453 \ \text{GeV} \ \text{if} \ 0.6 \ \text{TeV} < M < 1.6 \ \text{TeV}, \\ 1058 \ \text{GeV} \ \text{if} \ 1.6 \ \text{TeV} < M < 7.5 \ \text{TeV}, \end{cases}
\end{alignat*}
where $H_T$ is the scalar sum of the transverse momenta of all the jets, $\Delta\phi_{j_1 j_2}$ is the azimuthal angle between the two leading jets, and, for the two wide jets $J_1$ and $J_2$, $\Delta \eta_{J_1 J_2}$ is the difference between their pseudorapidities and $m_{J_1 J_2}$ is their invariant mass. Once again, the acceptance for a search is the fraction of the events to pass its cuts. The acceptances are shown in the bottom row of figure \ref{fig:acceptance}. Both CMS searches provided 95\% CL upper limits on the cross section times branching ratio times acceptance for dijets originating from two quarks.

It is also possible to constrain the parameter space using the second process outlined at the beginning of this section, stop pair production with subsequent decay into four quarks. As a matter of fact, there have been several experimental searches looking for exactly this signature. These include searches \cite{Aad:2016kww, ATLAS-CONF-2016-022, ATLAS-CONF-2016-084} from ATLAS and \cite{Khachatryan:2014lpa} from CMS. We directly read off the limits on the cross section times branching ratio as a function of stop mass. For stops decaying only into quarks, the most powerful search is \cite{ATLAS-CONF-2016-084}, which is independent of flavour of the final state quarks. That is, it does not explicitly require b-tagged jets. This is in contrast to the other searches, \cite{Aad:2016kww, ATLAS-CONF-2016-022} both require b-jets while \cite{Khachatryan:2014lpa} provides different limits depending on whether or not b-jets are produced. As a result, the limits on the stop mass are the same for all three $\lambda''_{3ij}$, once again assuming a branching ratio of one. 

\begin{figure}[t!]
  \centering
  \includegraphics[width=0.66\textwidth, bb = 0 0 584 458]{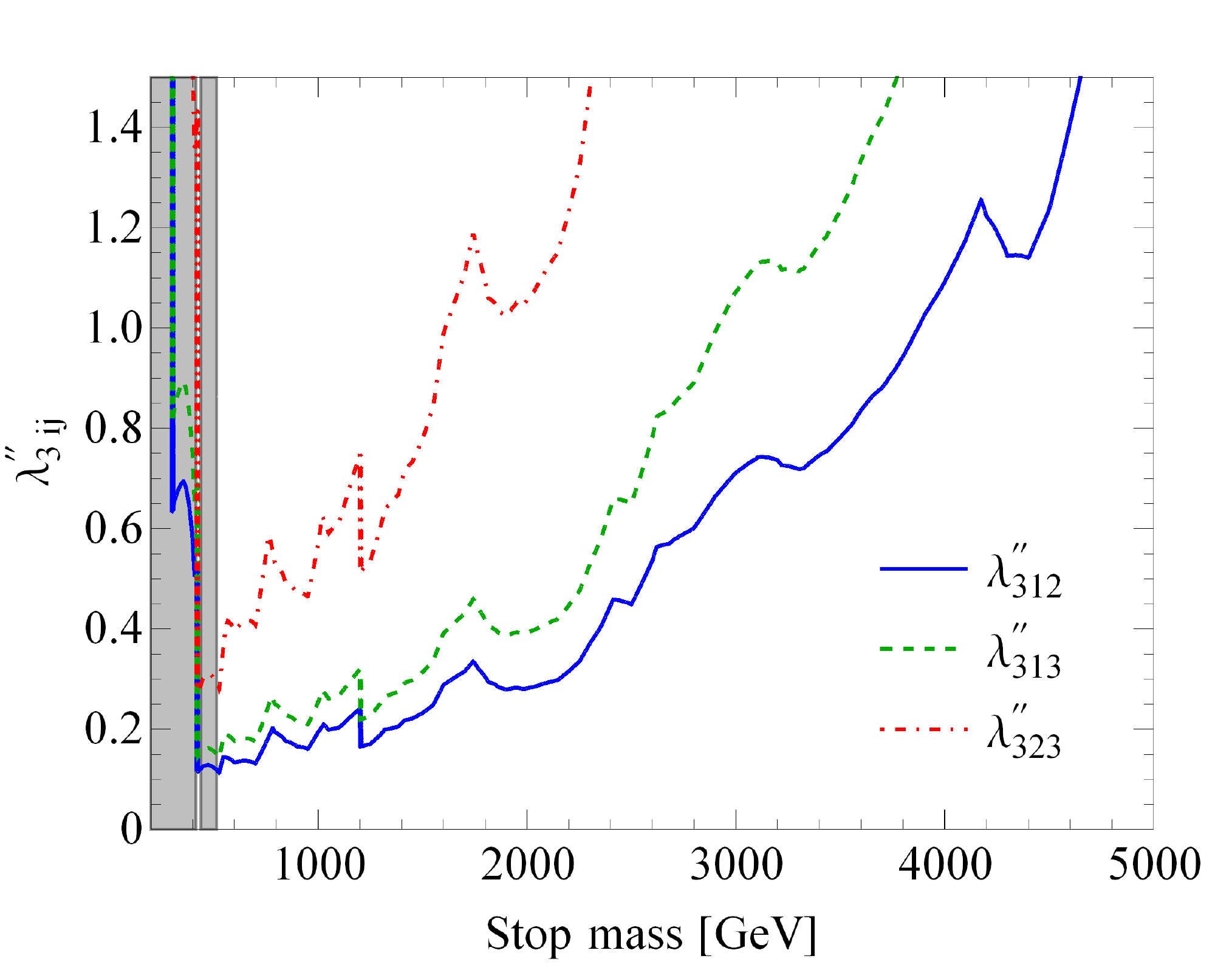}
  \caption{Exclusion plot for a LSP stop. The area above each curve is excluded by dijet searches. Additionally, the grey area is excluded by stop pair production with subsequent decay into four quarks. As explained in the text, it applies equally to all $\lambda''_{3ij}$.}
  \label{fig:stop_lSP}
\end{figure}

Combining the limits from the two types of searches considered in this section, we constrain the $\lambda''_{3ij}$ and stop mass parameter space. The result is shown in figure \ref{fig:stop_lSP}. The small white band of stop masses slightly above 400 GeV fails to be excluded due to an upward fluctuation of the signal in the search \cite{ATLAS-CONF-2016-084}. Interestingly, for each $\lambda''_{3ij}$ exclusion curve resulting from the dijet searches, at least a portion of its left edge happens to fall directly in this small unexcluded range. Future searches will likely close this gap. Disregarding this feature for a moment, we see that stop masses up to 3870, 2910 and 1610 GeV are excluded for $\lambda''_{312}$, $\lambda''_{313}$ and $\lambda''_{323}$ set to one, respectively. A similar plot is also presented within Ref$.$ \cite{Monteux:2016gag}. For comparison, Ref$.$ \cite{Monteux:2016gag} found that stop masses up to 3150, 2830 and 1500 (plus a small region between 1730 and 1870) GeV are excluded for $\lambda''_{312}$, $\lambda''_{313}$ and $\lambda''_{323}$ set to one, respectively. 

\subsubsection{Neutralino LSP}\label{ssSec:stop_neu_lsp}
If the LSP is a neutralino, additional phenomenological possibilities emerge. However, the stop is assumed to be right-handed, and as such couples only to the bino or the Higgsino-up. Therefore, we focus on two different possibilities: the LSP neutralino is essentially pure bino or essentially pure Higgsino-up. Naturally, for the Higgsino-up case, there is also an accompanying chargino with approximately the same mass. The next lightest neutralino or chargino is then taken to be heavier than the stop. This assures no cascade decays between neutralinos which would complicate the possible decay topologies. This type of spectrum, while not necessarily the most general, allows us to investigate the parameter space in a fairly straightforward and intuitive manner. We mention here that throughout this section and the section in which we constrain squarks, \ref{ssSec:squark_neu_lsp}, we set $\tan\beta = 10$.

\begin{figure}[t!]
  \centering
  \begin{subfigure}[b]{0.48\textwidth}
    \centering
    \includegraphics[width=\textwidth, bb = 0 0 584 458]{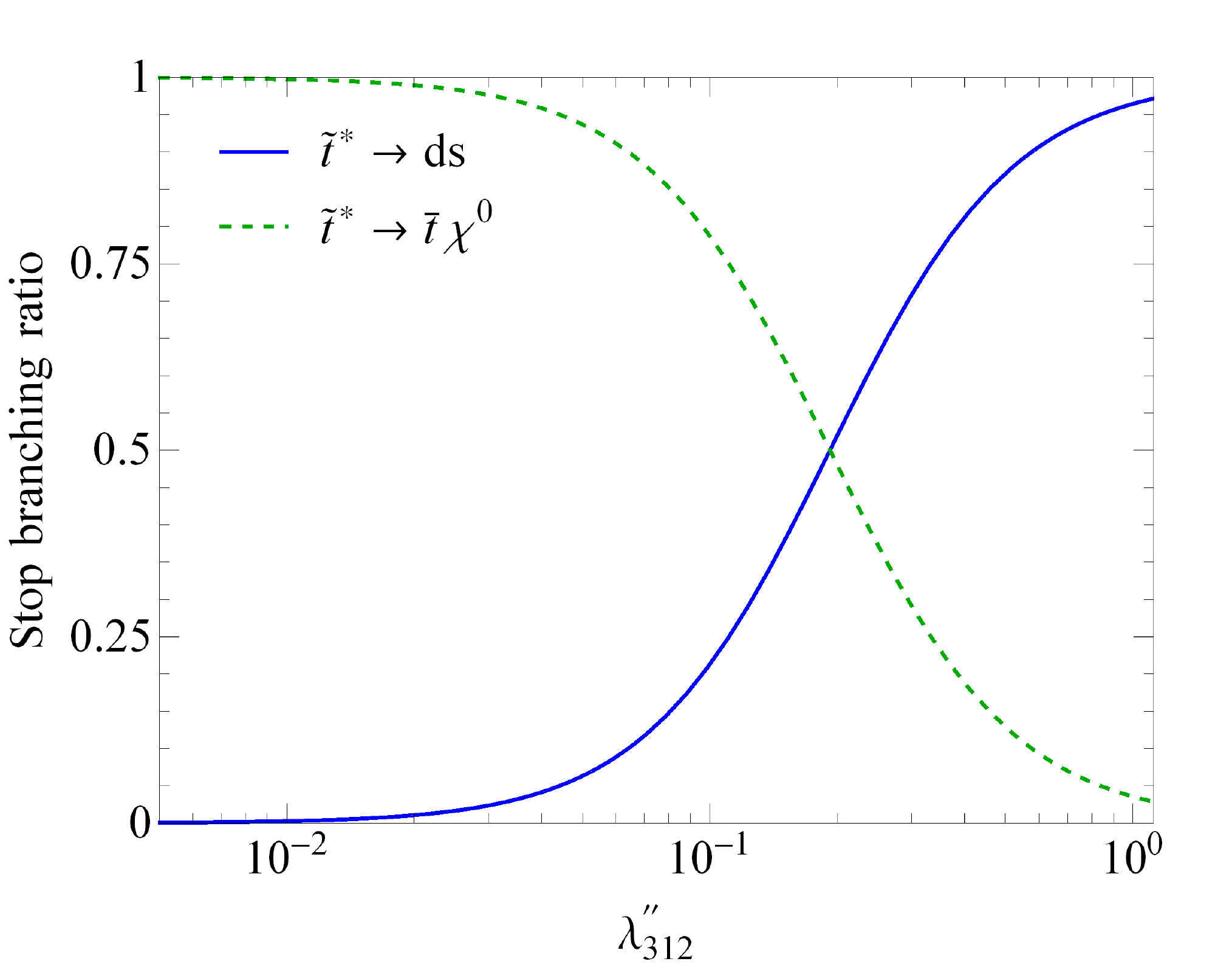}
    \caption{Bino neutralino}
    \label{fig:stop_BR_neuB_lambda}
  \end{subfigure}
  ~
  \begin{subfigure}[b]{0.48\textwidth}
    \centering
    \includegraphics[width=\textwidth, bb = 0 0 584 458]{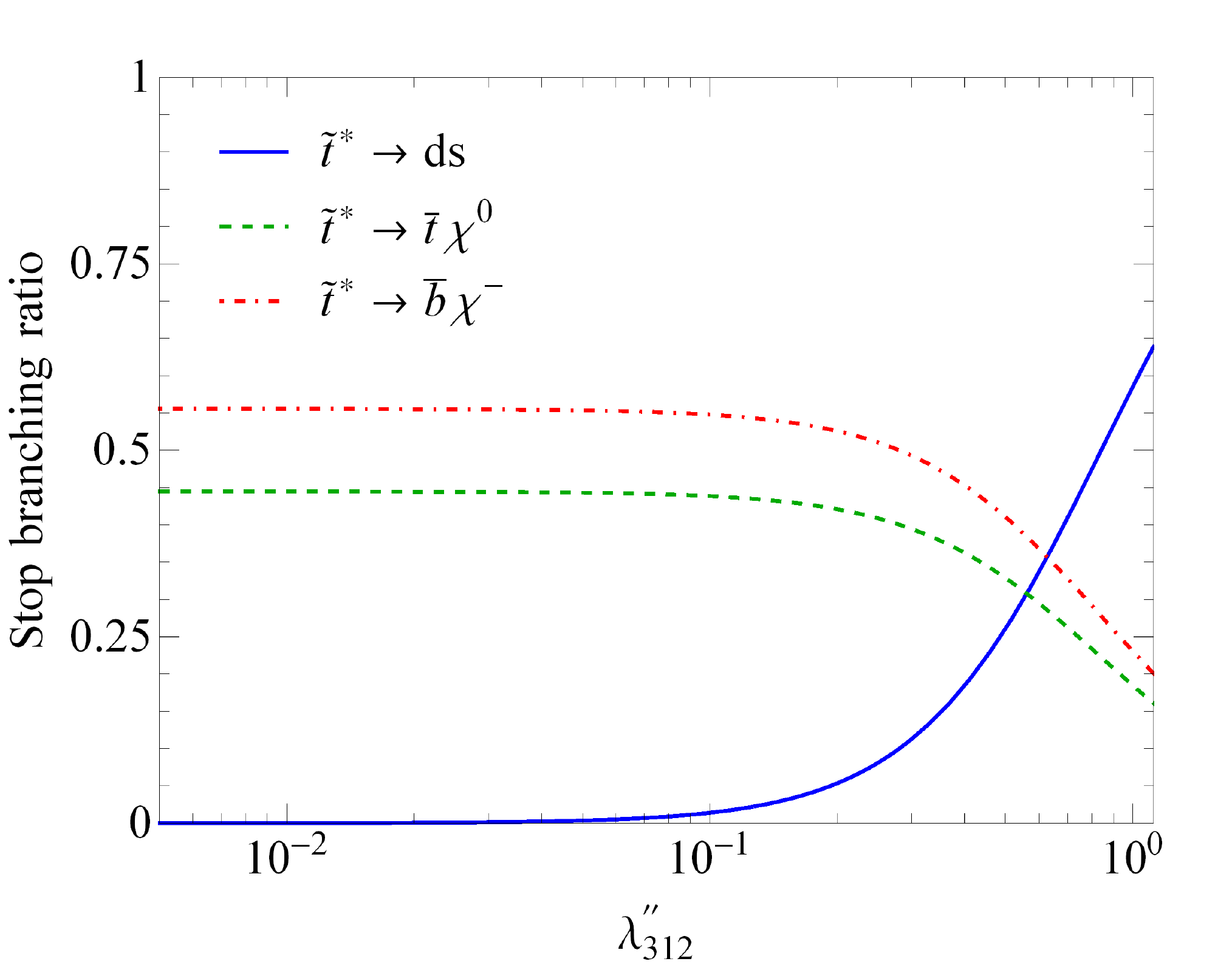}
    \caption{Higgsino-up neutralino}
    \label{fig:stop_BR_neuHu_lambda}
  \end{subfigure}
  \caption{Branching ratios for a $600$ GeV stop as a function of of $\lambda''_{312}$. For both plots, the neutralino mass has been set to 200 GeV.}\label{fig:stop_BR_neu_lambda}
\end{figure}

In general, there are now three different possibilities for how the stop can decay: $\tilde{t}^* \rightarrow d_i d_j$, $\tilde{t}^* \rightarrow \bar{t} \chi^0$ or $\tilde{t}^* \rightarrow \bar{b} \chi^-$. The first decay mode occurs, as before, through the $\lambda''_{3ij}$ coupling. For the last two decay modes, $\chi^0$ refers to the lightest neutralino and $\chi^-$ is the lightest chargino. Of course, if the LSP is a bino neutralino, only the first two decays will have non-zero branching ratios. On the other hand, if the Higgsino-up neutralino is the LSP then all three decay modes will occur. For both cases, we compute the branching ratios for the stop into each of the possible final states. For example, figure \ref{fig:stop_BR_neu_lambda} presents the branching ratios for a $600$ GeV stop as a function of $\lambda''_{312}$, with the neutralino mass set to $200$ GeV. As the figure shows, the stop decays mostly into dijets for $\lambda''_{312}$ of order one, while the other modes quickly begin to dominate for lesser values.

We now discuss the decay modes for the neutralinos and charginos, starting with the latter. Along with the normal decay of the chargino into the neutralino and a $W$ boson, the $\lambda''_{3ij}$ coupling allows for an additional decay into three quarks through an off-shell stop. Precisely, this new decay is $\chi^- \rightarrow b \tilde{t}^* \rightarrow b d_i d_j$. For the type of spectrum under consideration, the splitting between the chargino and neutralino is quite small. It then follows that the decay of the chargino into a neutralino and an off-shell $W$ boson is highly phase space suppressed. As a result, for essentially all values of $\lambda''_{3ij}$ and stop masses considered in this analysis, the RPV decay for the chargino dominates. We explicitly checked this by computing the branching ratios for the chargino and confirmed that this is indeed the case. Unless otherwise stated, we consider the chargino to decay into three quarks with a branching ratio of one.

Due to the $\lambda''_{3ij}$ coupling, the neutralino is also unstable and will decay into three quarks. This decay also occurs within the RPVMSSM but there is now an important difference. As explained in section \ref{subsec:spectrum}, our model has Dirac neutralinos which split into two pseudo-Dirac states once the small $U(1)_R$ breaking is taken into account. For Dirac neutralinos, there is only a single decay mode while for pseudo-Dirac neutralinos there are two. Specifically, the decay mode for Dirac neutralinos is $\chi^0 \rightarrow t \tilde{t}^* \rightarrow t d_i d_j$ (the antineutralino decay is $\bar{\chi}^0 \rightarrow \bar{t} \tilde{t} \rightarrow \bar{t} \bar{d}_i \bar{d}_j$). Pseudo-Dirac neutralinos can decay by $\chi^0 \rightarrow t \tilde{t}^* \rightarrow t d_i d_j$ or $\chi^0 \rightarrow \bar{t} \tilde{t} \rightarrow \bar{t} \bar{d}_i \bar{d}_j$. For a mass splitting, proportional to the scale of the $U(1)_R$ breaking, larger than the width, the two decay modes for the pseudo-Dirac neutralinos become equally relevant. The neutralinos will then behave similarly to the standard Majorana neutralinos of the RPVMSSM. Conversely, for mass splitting smaller than the width, the neutralino behaves as a purely Dirac state with a single decay mode.

\begin{figure}[t!]
  \centering
  \begin{subfigure}[b]{0.48\textwidth}
    \centering
    \includegraphics[width=\textwidth, bb = 0 0 584 479]{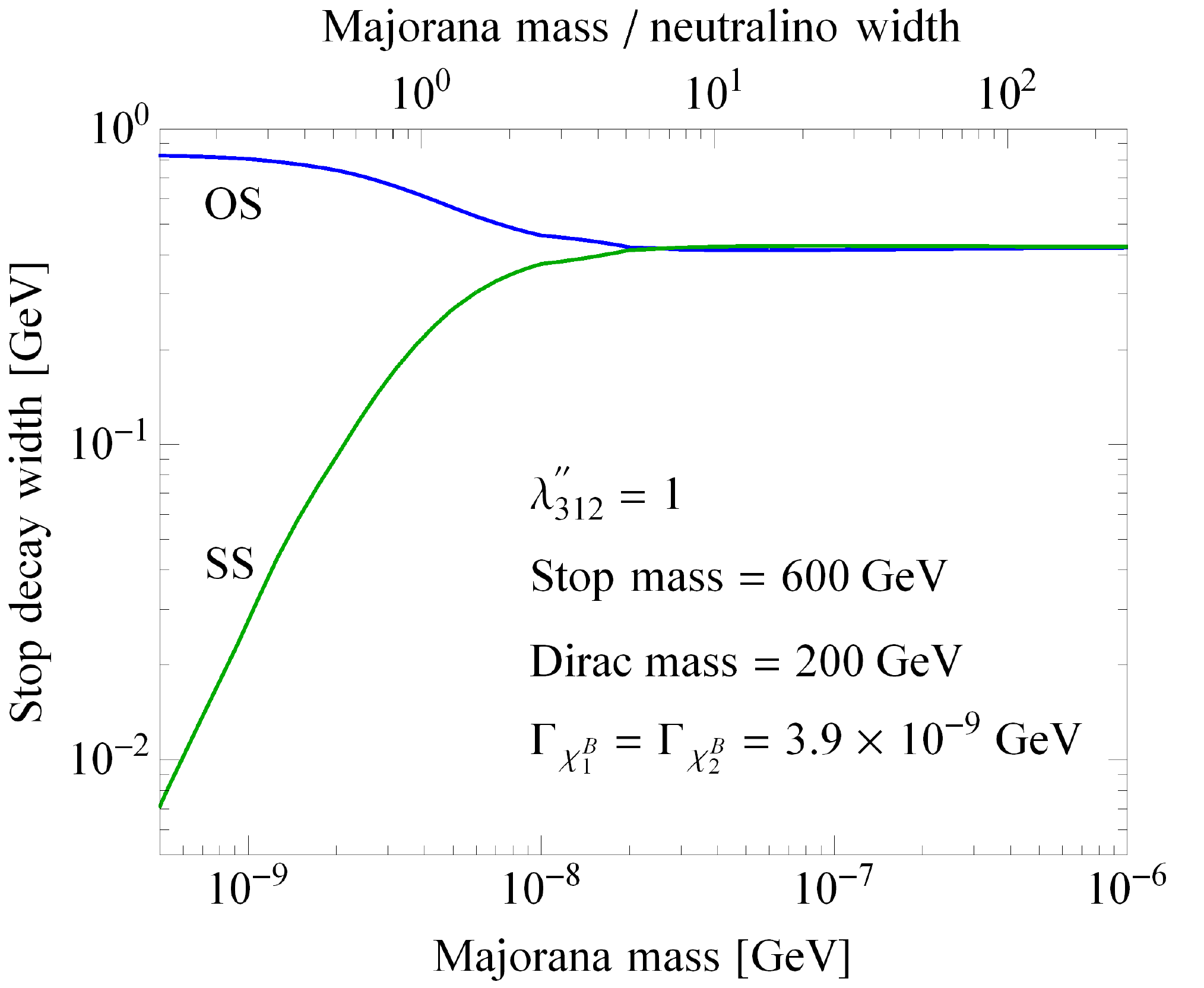}
    \caption{}
    \label{fig:stop_width_neuB_200}
  \end{subfigure}
  ~
  \begin{subfigure}[b]{0.48\textwidth}
    \centering
    \includegraphics[width=\textwidth, bb = 0 0 584 479]{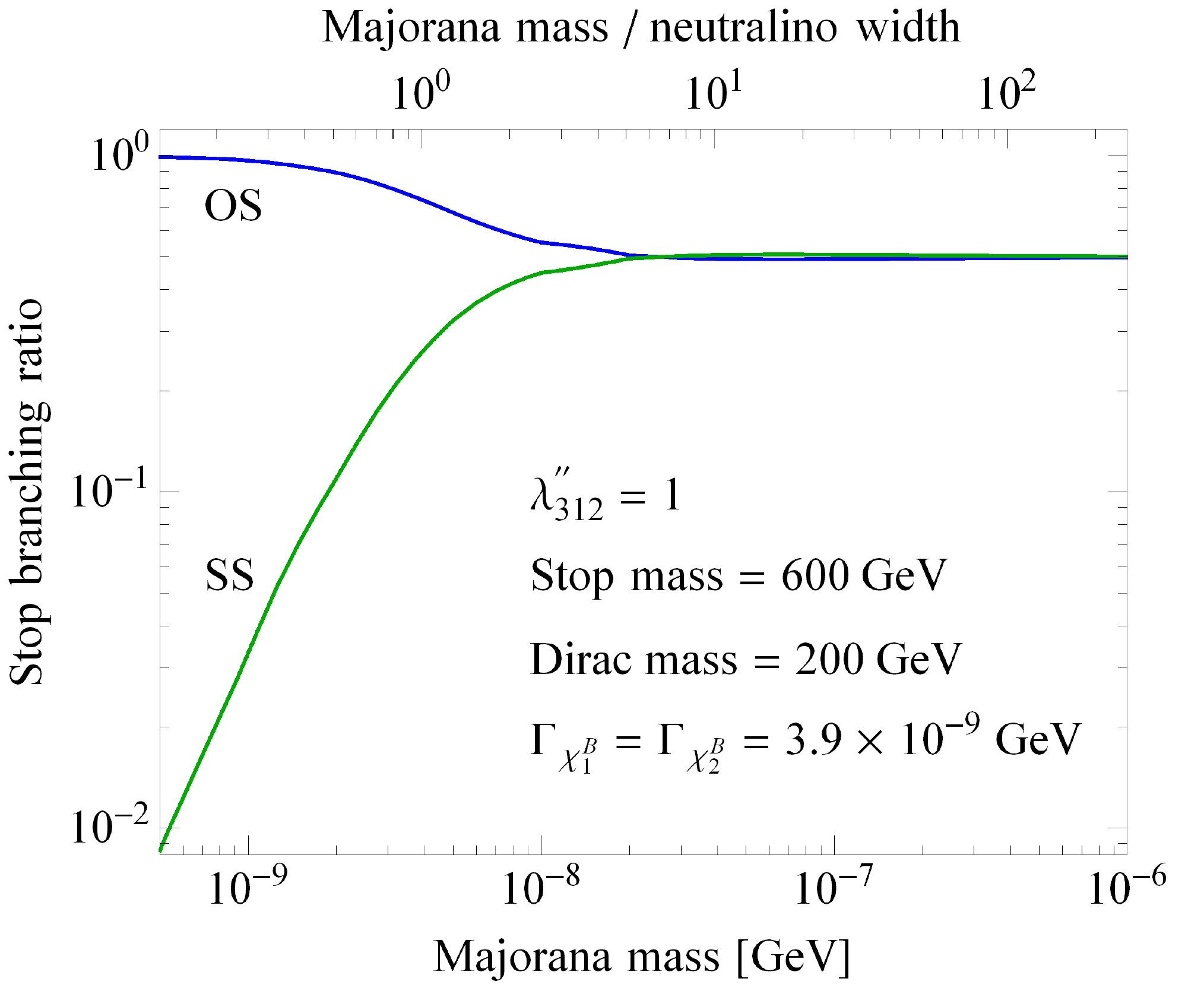}
    \caption{}
    \label{fig:stop_BR_neuB_200}
  \end{subfigure}
  \caption{Partial decay widths (a) and branching ratios (b) for $600$ GeV stops decaying into opposite sign (OS) and same sign (SS) tops through approximately $200$ GeV bino neutralinos with $\lambda''_{312} = 1$.}\label{fig:stop_BR_neuB}
\end{figure}

To demonstrate this feature, consider $600$ GeV stops decaying through approximately $200$ GeV binos with $\lambda''_{312} = 1$.
In figure \ref{fig:stop_BR_neuB} we show the partial decay widths and corresponding branching ratios for the stop as a function of the bino Majorana mass term. The branching ratios for the two different decays become equal when the Majorana mass is about five times the decay width of the neutralinos. We thoroughly explore the parameter space and find equivalent behaviour for the opposite sign and same sign decay widths. However, we note that this result crucially depends on the neutralinos being produced on-shell. If the stops decay through off-shell neutralinos, then the propagators of the neutralinos are not inversely proportional to their widths. In this case, the equality of branching ratios occurs when the mass splitting is comparable to the Dirac mass.\footnote{Although this discussion has been in terms of a bino LSP, similar results also hold for a Higgsino-up LSP. However, as a Majorana mass term for the Higgsino-up is not necessarily generated, the mass splitting between the two pseudo-Dirac Higgsino-up neutralinos results from a combination of Majorana gaugino masses for the bino and wino and mixing. For example, setting $\lambda''_{312} = 1$, $m_{\tilde{t}} = 600 \ \text{GeV}$, $\mu_u = 200 \ \text{GeV}$, $M_1^D = M_2^D = 10 \ \text{TeV}$ and $M_1$ and $M_2$ to their anomaly mediation masses, we find equal branching ratios for opposite sign and same sign tops resulting from stop decays for $m_{3/2} \gtrsim 7 \times 10^{-2} \ \text{GeV}$. If the Dirac masses for the bino and wino are lowered to $1 \ \text{TeV}$, then equal branching ratios occur for $m_{3/2} \gtrsim 3 \times 10^{-4} \ \text{GeV}$.}

To understand the phenomenological significance of this, suppose a $\tilde{t}^*$ is produced at the LHC and that it decays into a top and a neutralino. Then, for Dirac neutralinos, the final state quarks will always be $\bar{t} t d_i d_j$ while, for Majorana neutralinos, the final state quarks can either be $\bar{t} t d_i d_j$ or $\bar{t} \bar{t} \bar{d}_i \bar{d}_j$. Resonant stop production with Majorana neutralinos can lead to same sign tops whereas opposite sign tops are always produced for Dirac neutralinos. Same sign tops can potentially lead to same sign leptons, which is a powerful phenomenological signature for separating signal from background. In contrast, a final state of $\bar{t} t d_i d_j$ is difficult to distinguish from a background such as $t\bar{t}$ and jets. Similarly, stop pair production is also affected by whether or not the neutralino is Dirac or Majorana. If both stops decay into neutralinos, then a total of four tops will be produced. Dirac neutralinos will always result in two positively and two negatively charged tops. However, Majorana neutralinos will result in two positively and two negatively charged tops only half of the time. For the other half, three tops with the same sign will be produced, along with a single top with the opposite sign. Of note, the latter case has a larger probability of producing a same sign lepton pair. 

There is one more possible decay mode that we need to consider. If the Majorana mass term is large enough so that the mass splitting between the pseudo-Dirac neutralino states is non-negligible, then the decays $\chi_2^B \rightarrow \chi_1^B Z$ and $\chi_2^B \rightarrow \chi_1^B h$ potentially open up. In these decays the $Z$ and $h$ are off-shell for small mass splittings. However, for all neutralino masses and $\lambda''_{3ij}$ couplings considered in this analysis, the decay width for the neutralinos is relatively small (generally less than 1 GeV). As a result, only a modest Majorana mass term is needed to ensure that opposite sign and same sign tops are produced equally from stop decays. Thus, we make the following assumption. If the $U(1)_R$ symmetry is broken, then the Majorana mass term is large enough such that the stops decay into opposite sign and same sign tops with equal branching ratios, while, at the same time, is small enough so that the decays $\chi_2^B \rightarrow \chi_1^B Z$ and $\chi_2^B \rightarrow \chi_1^B h$ can be safely ignored. This also has the added benefit of making the analysis of the possible decay chains simpler. Finally, we also note that under this assumption, the phenomenology of the MRSSM with a broken $U(1)_R$ symmetry is essentially identical to the RPVMSSM.

Now that we have discussed the various decay modes for the stop, neutralino and chargino, we consider all processes involving stop production. First, consider the case were the $U(1)_R$ symmetry is strictly preserved. Then, enumerating all the possibilities, we get the following list:    
\begin{alignat*}{3}
&(1) \
\begin{array}{l}
p p \rightarrow \tilde{t}^* \rightarrow d_i d_j
\end{array}
\quad \quad
&&(2) \
\begin{array}{l}
p p \rightarrow \tilde{t}^* \rightarrow \bar{t} \chi^0 \rightarrow \bar{t} t d_i d_j
\end{array}
\quad \quad
&&(3) \
\begin{array}{l}
p p \rightarrow \tilde{t}^* \rightarrow \bar{b} \chi^- \rightarrow \bar{b} b d_i d_j 
\end{array} \\[1.0ex]
&(4) \
\begin{array}{l}
p p \rightarrow \tilde{t}^* \tilde{t} \\
\phantom{p p} \rightarrow d_i d_j \bar{d}_i \bar{d}_j
\end{array}
\quad \quad
&&(5) \
\begin{array}{l}
p p \rightarrow \tilde{t}^* \tilde{t} \rightarrow \bar{t} \chi^0 t \bar{\chi}^0 \\
\phantom{p p \rightarrow \tilde{t}^* \tilde{t}} \rightarrow \bar{t} t d_i d_j t \bar{t} \bar{d}_i \bar{d}_j
\end{array}
\quad \quad
&&(6) \
\begin{array}{l}
p p \rightarrow \tilde{t}^* \tilde{t} \rightarrow \bar{b} \chi^- b \chi^+ \\
\phantom{p p \rightarrow \tilde{t}^* \tilde{t}} \rightarrow \bar{b} b d_i d_j b \bar{b} \bar{d}_i \bar{d}_j
\end{array} \\[1.0ex]
&(7) \
\begin{array}{l}
p p \rightarrow \tilde{t}^* \tilde{t} \rightarrow d_i d_j t \bar{\chi}^0  \\
\phantom{p p \rightarrow \tilde{t}^* \tilde{t}} \rightarrow d_i d_j t \bar{t} \bar{d}_i \bar{d}_j 
\end{array}
\quad \quad
&&(8) \
\begin{array}{l}
p p \rightarrow \tilde{t}^* \tilde{t} \rightarrow d_i d_j b \chi^+  \\
\phantom{p p \rightarrow \tilde{t}^* \tilde{t}} \rightarrow d_i d_j b \bar{b} \bar{d}_i \bar{d}_j
\end{array}
\quad \quad
&&(9) \ 
\begin{array}{l}
p p \rightarrow \tilde{t}^* \tilde{t} \rightarrow \bar{t} \chi^0 b \chi^+ \\
\phantom{p p \rightarrow \tilde{t}^* \tilde{t}} \rightarrow \bar{t} t d_i d_j b \bar{b} \bar{d}_i \bar{d}_j.
\end{array}
\end{alignat*} 
If, instead, the $U(1)_R$ symmetry is broken, then processes 2, 5, 7 and 9 need to be modified:
\begin{alignat*}{3}
&(2) \
\begin{array}{l}
p p \rightarrow \tilde{t}^* \rightarrow \bar{t} \chi^0 \rightarrow \begin{cases} \bar{t} t d_i d_j \\ \bar{t} \bar{t} \bar{d}_i \bar{d}_j \end{cases}
\end{array}
\quad \quad
&&(5) \
\begin{array}{l}
p p \rightarrow \tilde{t}^* \tilde{t} \rightarrow \bar{t} \chi^0 t \chi^0 \rightarrow \begin{cases} \bar{t} t d_i d_j t \bar{t} \bar{d}_i \bar{d}_j \\ \bar{t} t d_i d_j t t d_i d_j \\ \bar{t} \bar{t} \bar{d}_i \bar{d}_j t \bar{t} \bar{d}_i \bar{d}_j \\ \bar{t} \bar{t} \bar{d}_i \bar{d}_j t t d_i d_j \end{cases}
\end{array} \\[1.0ex]
&(7) \
\begin{array}{l}
p p \rightarrow \tilde{t}^* \tilde{t} \rightarrow d_i d_j t \chi^0 \rightarrow \begin{cases} d_i d_j t \bar{t} \bar{d}_i \bar{d}_j \\ d_i d_j t t d_i d_j \end{cases}
\end{array} 
\quad \quad
&&(9) \
\begin{array}{l}
p p \rightarrow \tilde{t}^* \tilde{t} \rightarrow \bar{t} \chi^0 b \chi^+ \rightarrow \begin{cases} \bar{t} t d_i d_j b \bar{b} \bar{d}_i \bar{d}_j \\ \bar{t} \bar{t} \bar{d}_i \bar{d}_j b \bar{b} \bar{d}_i \bar{d}_j. \end{cases}
\end{array}
\end{alignat*}
Where appropriate, each process also includes its charge conjugated version. 

Processes 1 and 4 can be constrained by using the results for stop LSP (section \ref{ssSec:Stop_LSP}) with appropriate modifications to the branching ratios. When presenting plots of the parameter space, the region ruled out by process 1 is referred to as the region excluded by dijets searches. Likewise, the region ruled out by process 4 is referred to as the region excluded by paired dijet searches. Further exclusion is possible if other types of experimental searches are considered. Our methodology for choosing which searches to recast is as follows. First, there have been several experimental searches featuring supersymmetric particles decaying through the $\lambda''_{3ij}$ couplings. We select three of the most recent searches of this variety. These are \cite{ATLAS-CONF-2016-037, ATLAS-CONF-2016-057, ATLAS-CONF-2016-094}, of which all are from ATLAS. Next, notice that many of the different possible final states contain either four tops or two same sign tops. We therefore examine searches that constrain these types of final states. This led us to choosing searches \cite{ATLAS-CONF-2016-013, ATLAS-CONF-2016-032} from ATLAS and \cite{CMS-PAS-SUS-16-020} from CMS. A brief outline of the strategy for each search is summarized in table \ref{table:searches}. The region of parameter space ruled out by these searches is referred to as the region excluded by neutralino LSP searches.

\begin{table}[t!]
\begin{center}
\begin{tabular}{ |C{0.18\textwidth}|C{0.18\textwidth}|C{0.54\textwidth}| }
\hline
Collaboration & Search & Strategy \\ \hline
ATLAS & \cite{ATLAS-CONF-2016-037} & \vspace{-9pt} \parbox{0.54\textwidth}{\centering 2 (potentially negative) same sign leptons,\\total number of leptons,\\jets with $p_T > 25$, $40$ or $50 $ GeV, b-jets,\\MET, $m_{\text{eff}} = \sum\limits_ {\substack{\text{jets}\\\text{leptons}}} {p_T} + \text{MET}$} \vspace{-3pt} \\ \hline

ATLAS & \cite{ATLAS-CONF-2016-057} & \vspace{-9pt} \parbox{0.54\textwidth}{\centering large ($R = 1.0$) jets $J_i$,\\${{p_T}_{J_1}} > 440$ GeV, $|\Delta \eta_{J_1 J_2}| < 1.4$,\\$M_J^\Sigma = \sum\limits_{i=1}^4 m_{J_i}$, small ($R = 0.4$) b-jets} \vspace{-3pt} \\ \hline

ATLAS & \cite{ATLAS-CONF-2016-094} & \vspace{-9pt} \parbox{0.54\textwidth}{\centering at least 1 lepton,\\jets with $p_T > 40$ or $60 $ GeV,\\b-jets with $p_T > 40$ or $60 $ GeV} \vspace{-3pt} \\ \hline

ATLAS & \cite{ATLAS-CONF-2016-013} & \vspace{-9pt} \parbox{0.54\textwidth}{\centering exactly 1 lepton, jets, b-jets,\\mass-tagged jets = large ($R = 1.0$) jets with cuts, $m_{bb}^{\min\Delta R} = \text{invariant mass of closest b-jets}$,\\MET, MET + $M_T(\ell,\text{MET})$\\where $M_T = \text{transverse mass}$} \vspace{-3pt} \\ \hline

ATLAS & \cite{ATLAS-CONF-2016-032} & \vspace{-9pt} \parbox{0.54\textwidth}{\centering 2 same sign leptons, jets, b-jets,\\MET, $H_T = \sum\limits_ {\substack{\text{jets}\\\text{leptons}}} {p_T}$} \vspace{-3pt} \\ \hline

CMS & \cite{CMS-PAS-SUS-16-020} & \vspace{-9pt} \parbox{0.54\textwidth}{\centering 2 same sign leptons, jets, b-jets,\\$M_T^{\text{min}} = \min(M_T(\ell_1,\text{MET}), M_T(\ell_2,\text{MET}))$\\where $M_T = \text{transverse mass}$,\\MET, $H_T = \sum\limits_ {\text{jets}} {p_T}$} \vspace{-3pt} \\ \hline
\end{tabular}%
\caption{Neutralino LSP searches. For searches \cite{ATLAS-CONF-2016-057, ATLAS-CONF-2016-013}, we use FastJet \cite{Cacciari:2011ma, Cacciari:2005hq} for manipulating large ($R = 1$) jets. This mainly involves jet reclustering and jet trimming. Additionally, searches that feature missing transverse energy (MET) either have very lenient cuts on this quantity or also contain signal regions probing $R$-parity conserving (RPC) supersymmetry signatures.}\label{table:searches}
\end{center}
\end{table}

The procedure used to recast these searches is similar to the procedures used to recast dijets searches described above. First, the neutralino mass is set to 200 GeV and the stop mass is scanned between 200 and 1000 GeV. For all combinations, we use MadGraph, PYTHIA and Delphes to simulate 10000 events for each of the nine possible decay chains. This was done twice for processes 2, 5, 7 and 9, once with the $U(1)_R$ symmetry preserved and a second time with it broken. Code was implemented to simulate the cuts for each of the six searches. We verified our code by reproducing each search with good accuracy. Using the simulated events, our code produced acceptances for every signal region described within each search. Then, within the stop mass and $\lambda''_{3ij}$ parameter space, the acceptances are combined with production cross sections and appropriate branching ratios to determine the number of expected signals for each signal region. The 95\% CL upper limit for each signal region were then determined. Searches \cite{ATLAS-CONF-2016-037, ATLAS-CONF-2016-057, ATLAS-CONF-2016-094} explicitly provided these upper limits. Conversely, searches \cite{ATLAS-CONF-2016-013, ATLAS-CONF-2016-032, CMS-PAS-SUS-16-020} did not, and so we calculate the upper limits using the $CL_S$ technique \cite{Read:2002hq, Junk:1999kv}. A point in parameter space is then excluded if the expected number of signals in any of the signal regions exceed its upper limit.

\begin{figure}[t!]
  \centering
  \begin{subfigure}[b]{0.48\textwidth}
    \centering
    \includegraphics[width=\textwidth, bb = 0 0 584 455]{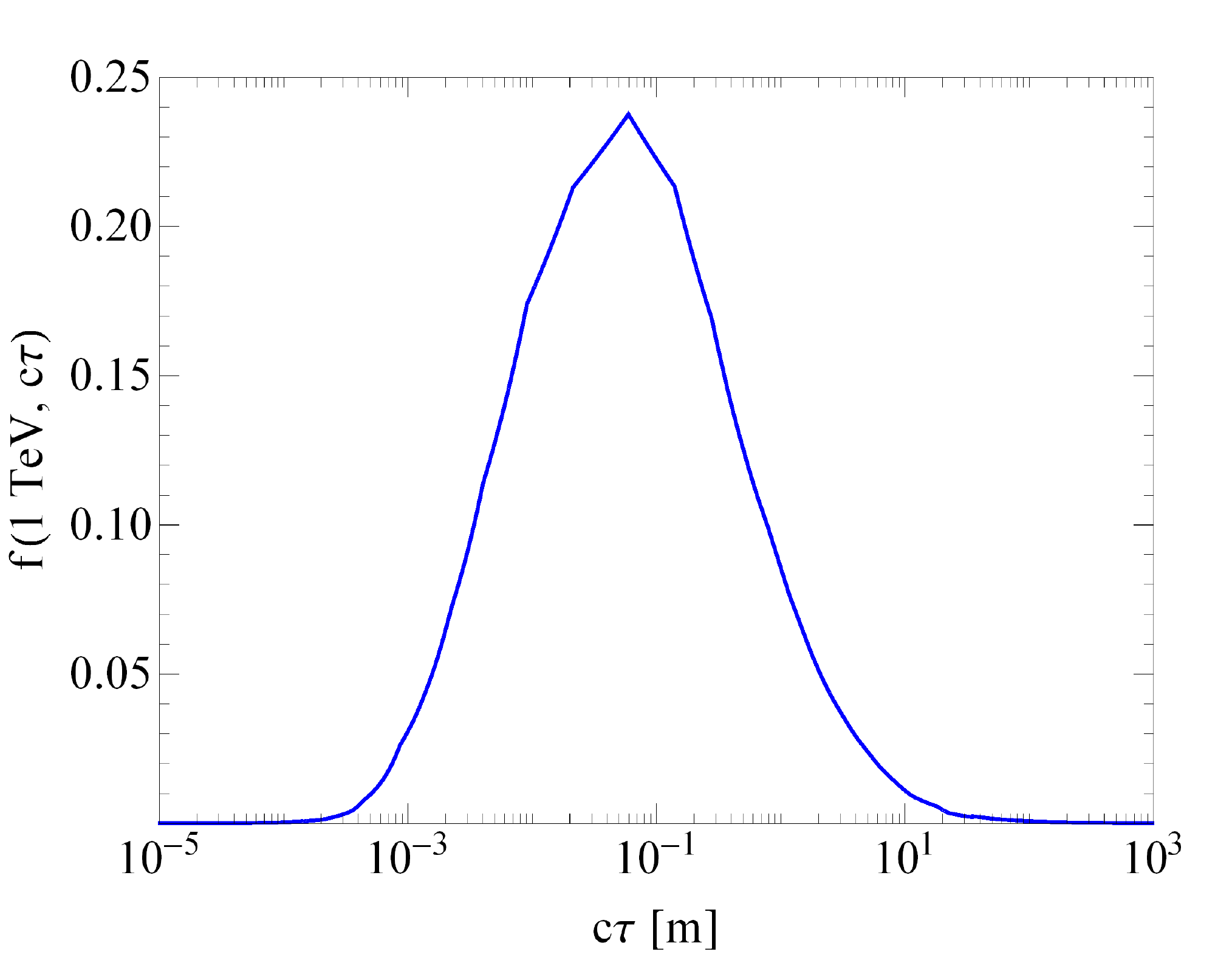}
    \caption{}
    \label{fig:f1}
  \end{subfigure}
  ~
  \begin{subfigure}[b]{0.48\textwidth}
    \centering
    \includegraphics[width=\textwidth, bb = 0 0 584 455]{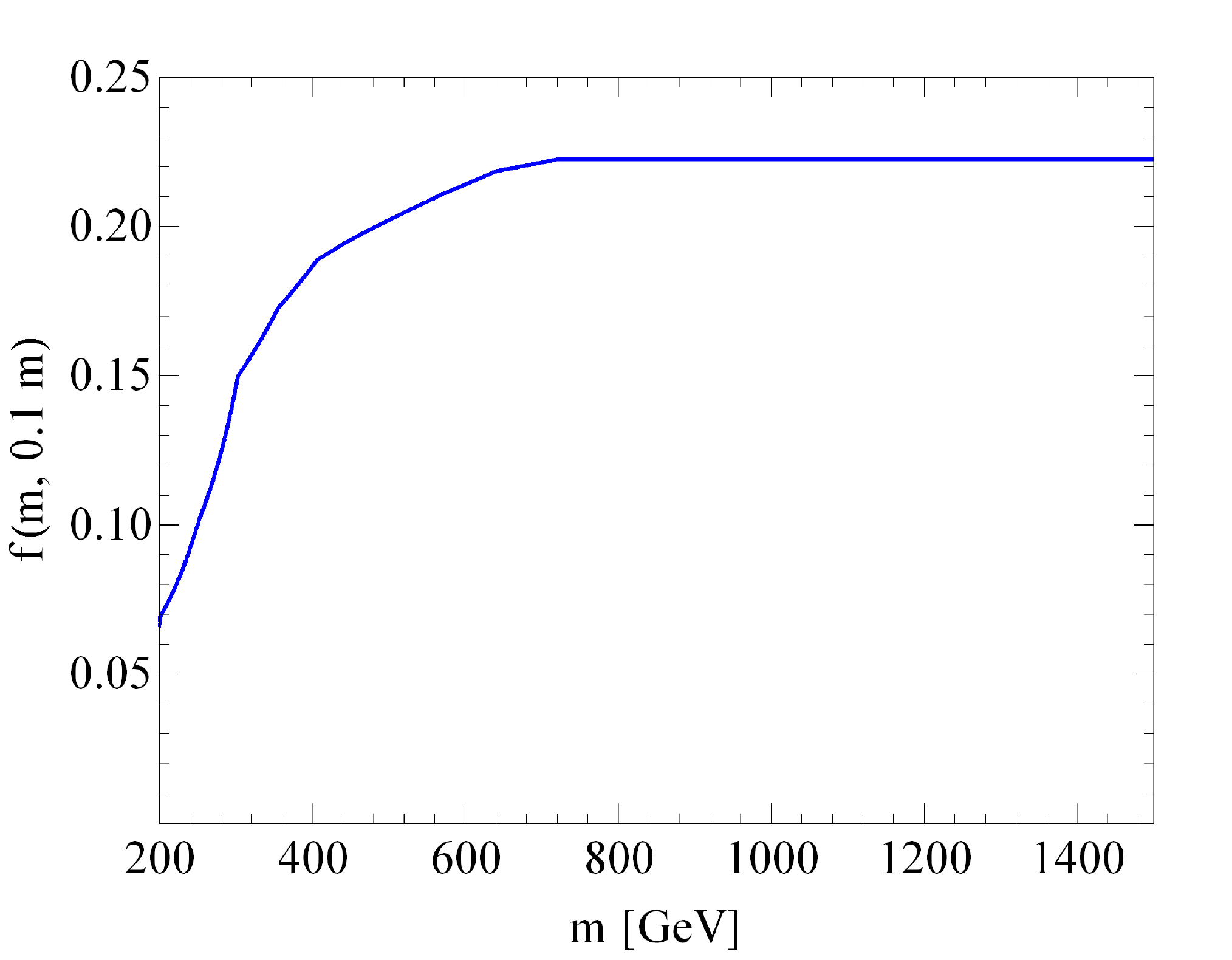}
    \caption{}
    \label{fig:f2}
  \end{subfigure}
  \caption{The efficiency for reconstructing a single vertex $f(m,c\tau)$ for (a) fixed $m = 1$ TeV and (b) fixed $c\tau = 0.1$ m for the CMS search \cite{CMS:2014wda}.}\label{DV:f1f2}
\end{figure}

The region of small $\lambda''_{3ij}$ can additionally be constrained by searches for displaced vertices. The efficiency for reconstructing a single displaced vertex is to good approximation only a function of the mass $m$ and decay length $c\tau$ of the particle involved. We make the further approximation that this function, which we call $f(m,c\tau)$, can be factorized as $f_1(m) f_2(c\tau)$. This is justified by the results of Ref$.$ \cite{Cui:2014twa}, which presents upper limits on the cross section for pair production of hadronically decaying neutralinos as a function of their mass for a fixed decay length. These results are based on the CMS search \cite{CMS:2014wda}. The function $f_1(m)$ can be read from Ref$.$ \cite{Cui:2014twa} up to a multiplicative factor that can be absorbed in $f_2(c\tau)$. The latter function can be extracted from Ref$.$ \cite{Liu:2015bma}, which presents exclusion limits on displaced vertices for Higgsino LSP in the parameter space of Higgsino mass and decay length. These results are based on the same search and assume charged Higgsinos decay promptly to the almost degenerate neutral one. Higgsinos are again assumed to be pair produced and for the lightest one to decay hadronically. Knowing the cross section involved and upper limit on the signal, $f_2(c\tau)$ can be reconstructed everywhere except for very short and very long decay lengths. In these regions, the efficiency decreases exponentially as expected and we extrapolate this behaviour. This allows for a complete reconstruction of $f(m,c\tau)$, which is shown in figure \ref{DV:f1f2}. 

Next, note that the displaced vertices in our model result from neutralino decays. We consider both neutralino pair production and neutralinos produced from stop decays. Furthermore, in this part of the parameter space, we assume the Higgsino-up charginos decay dominantly into neutralinos. This is in contrast to our previous benchmark points where the RPV decay for charginos was assumed. For small values of $\lambda''_{3ij}$ the charginos will decay into neutralinos provided that the spectrum is not too degenerate. A large enough splitting can easily be generated provided that the wino is not exceptionally heavy. To compute the cross sections for Higgsino-up and bino pair production we use Prospino. The bino cross section depends on the masses for the first and second generations of squarks. We consider two different cases. For the first case, we decouple the squarks by setting their masses to 10 TeV. For the second case, we set their masses to 1 TeV. Combining these cross sections with the known $f(m,c\tau)$, limits on displaced vertices can easily be applied to our parameter spaces using once again Ref$.$ \cite{CMS:2014wda}.

\begin{figure}[t!]
  \centering
  \begin{subfigure}[b]{0.48\textwidth}
    \centering
    \includegraphics[width=\textwidth, bb = 0 0 584 451]{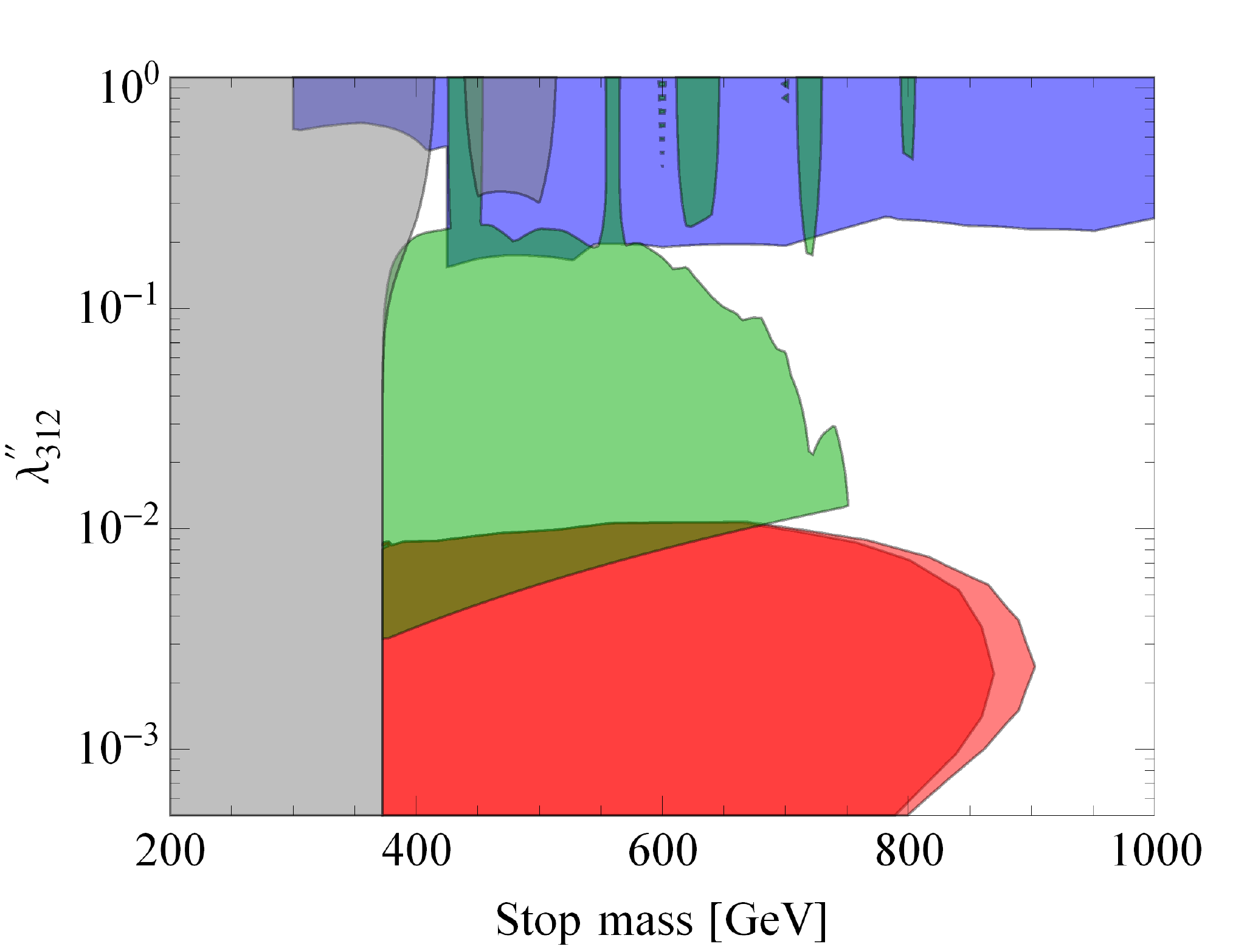}
    \label{fig:stop_exclusion_U1R_bino_312}
  \end{subfigure}
  ~
  \begin{subfigure}[b]{0.48\textwidth}
    \centering
    \includegraphics[width=\textwidth, bb = 0 0 584 451]{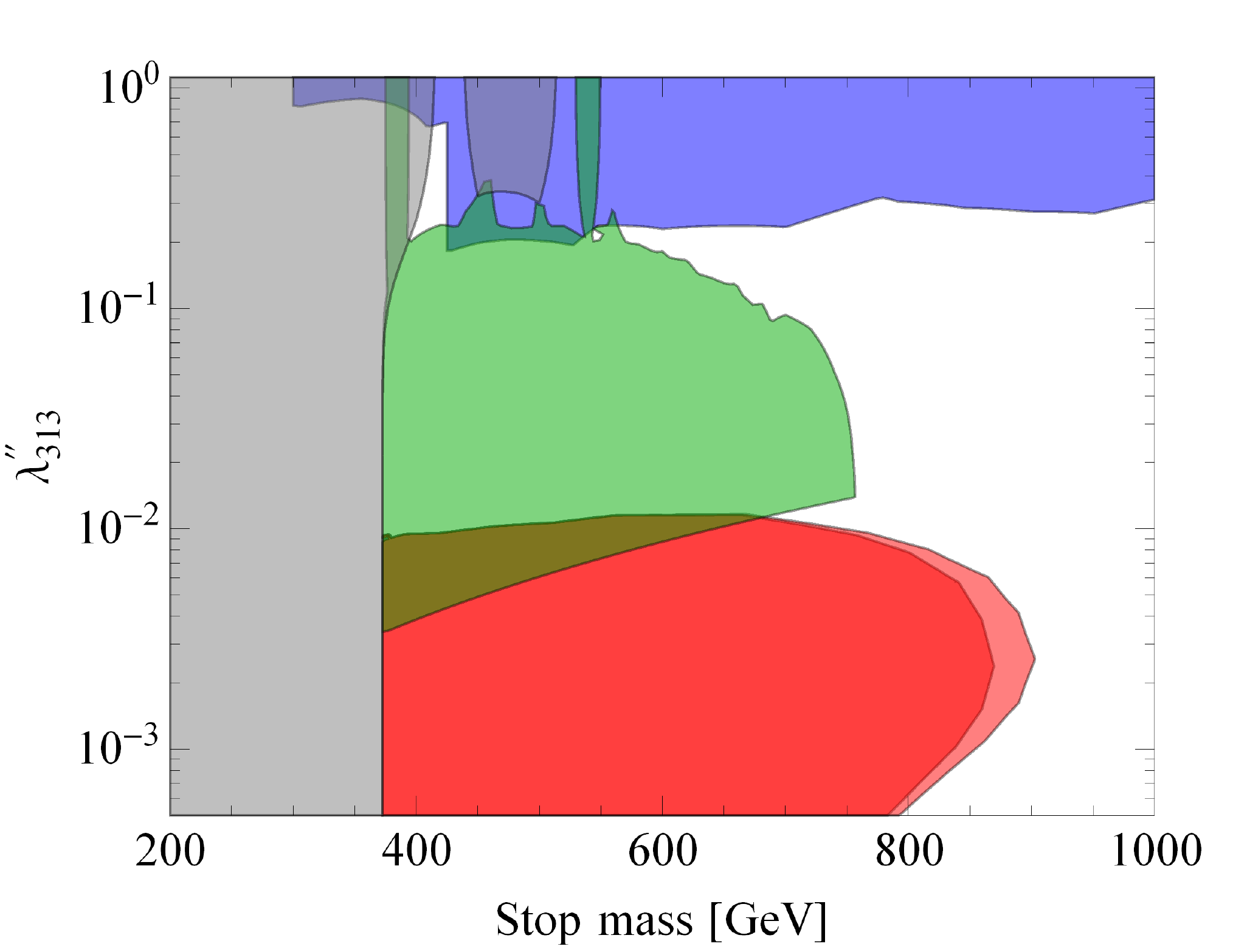}
    \label{fig:stop_exclusion_U1R_bino_313}
  \end{subfigure}
  ~
   \begin{subfigure}[b]{0.48\textwidth}
    \centering
    \includegraphics[width=\textwidth, bb = 0 0 584 451]{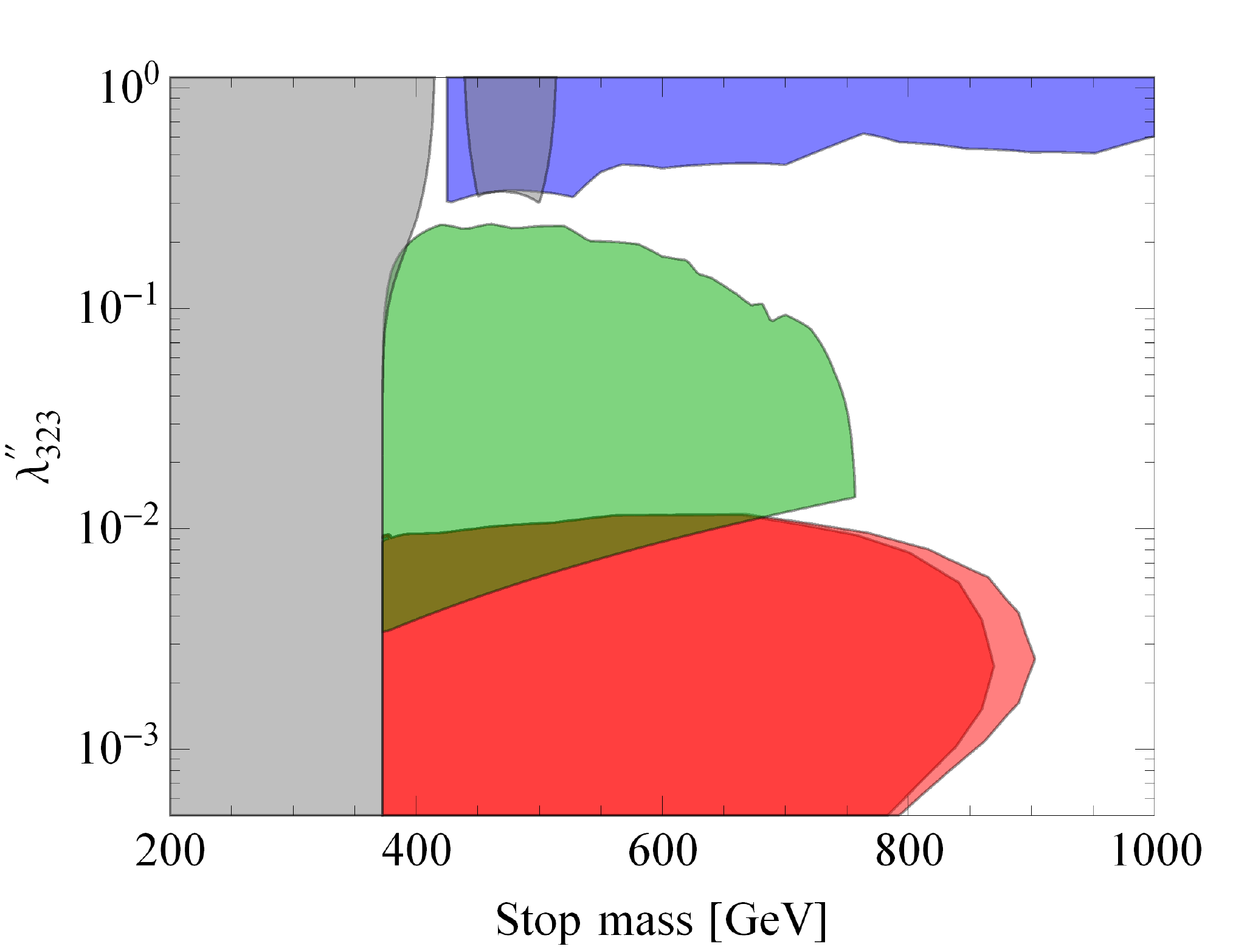}
    \label{fig:stop_exclusion_U1R_bino_323}
  \end{subfigure}
\caption{Exclusion plots for a 200 GeV bino neutralino LSP with the $U(1)_R$ symmetry strictly preserved. The grey region on the left side of the plots ($m_{\tilde{t}} \lesssim 375$ GeV) is excluded by paired dijet searches. Next, consider the middle region of the plots. Starting from large $\lambda''_{3ij}$ couplings and working downwards, the blue region is excluded by dijet searches, the green region is excluded by neutralino LSP searches and the red regions are excluded by displaced vertices searches. Bino pair production, which contributes to the displaced vertices limits, depends on the masses of the first and second generations of squarks. Setting the masses of these squarks to 10 TeV results in the darker red region being excluded. Instead, setting the masses of these squarks to 1 TeV excludes both the darker red and lighter red regions.}\label{fig:stop_exclusion_Bino_U1R}
\end{figure}

\begin{figure}[t!]
  \centering
  \begin{subfigure}[b]{0.48\textwidth}
    \centering
    \includegraphics[width=\textwidth, bb = 0 0 584 451]{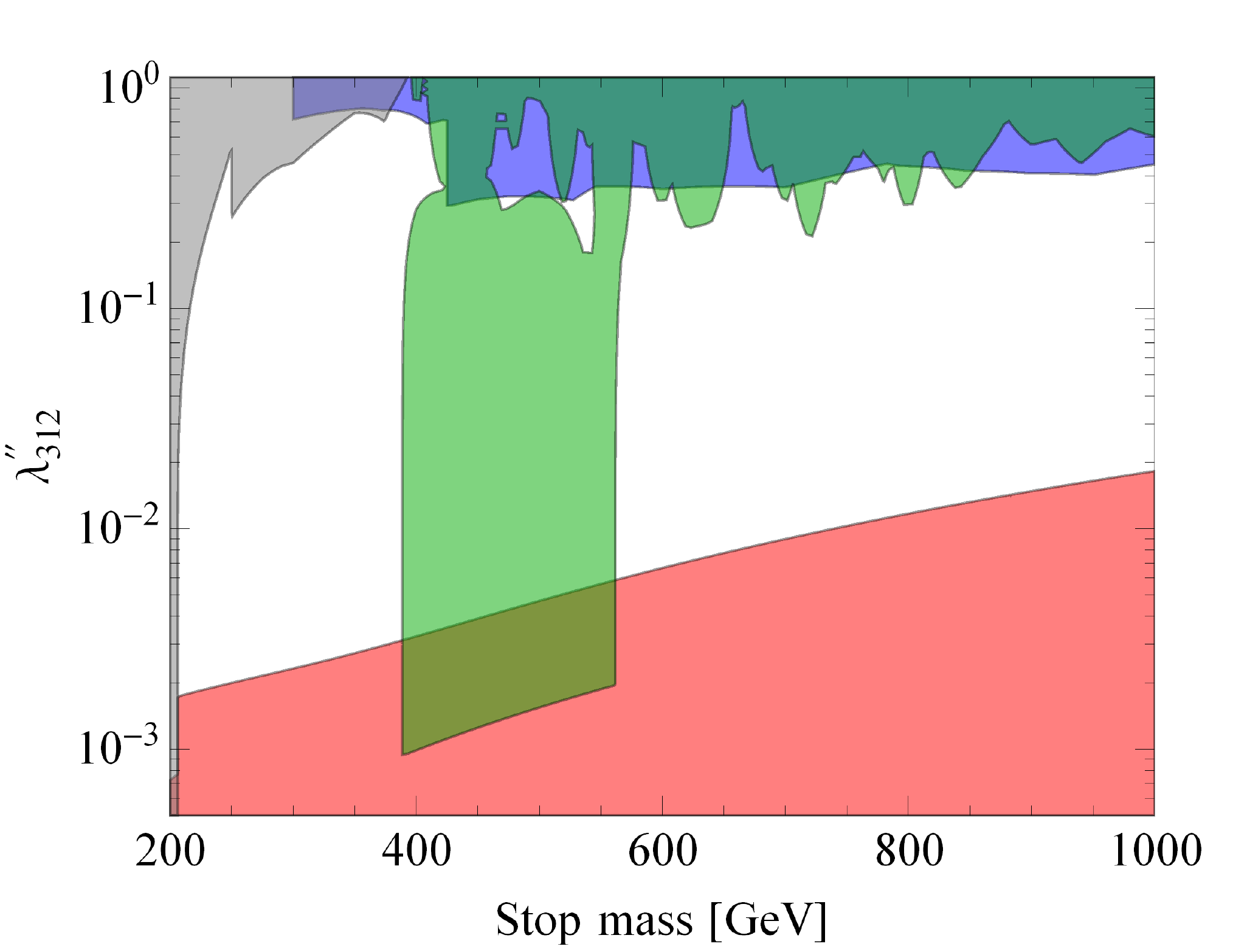}
    \label{fig:stop_exclusion_U1R_Higgsino_up_312}
  \end{subfigure}
  ~
  \begin{subfigure}[b]{0.48\textwidth}
    \centering
    \includegraphics[width=\textwidth, bb = 0 0 584 451]{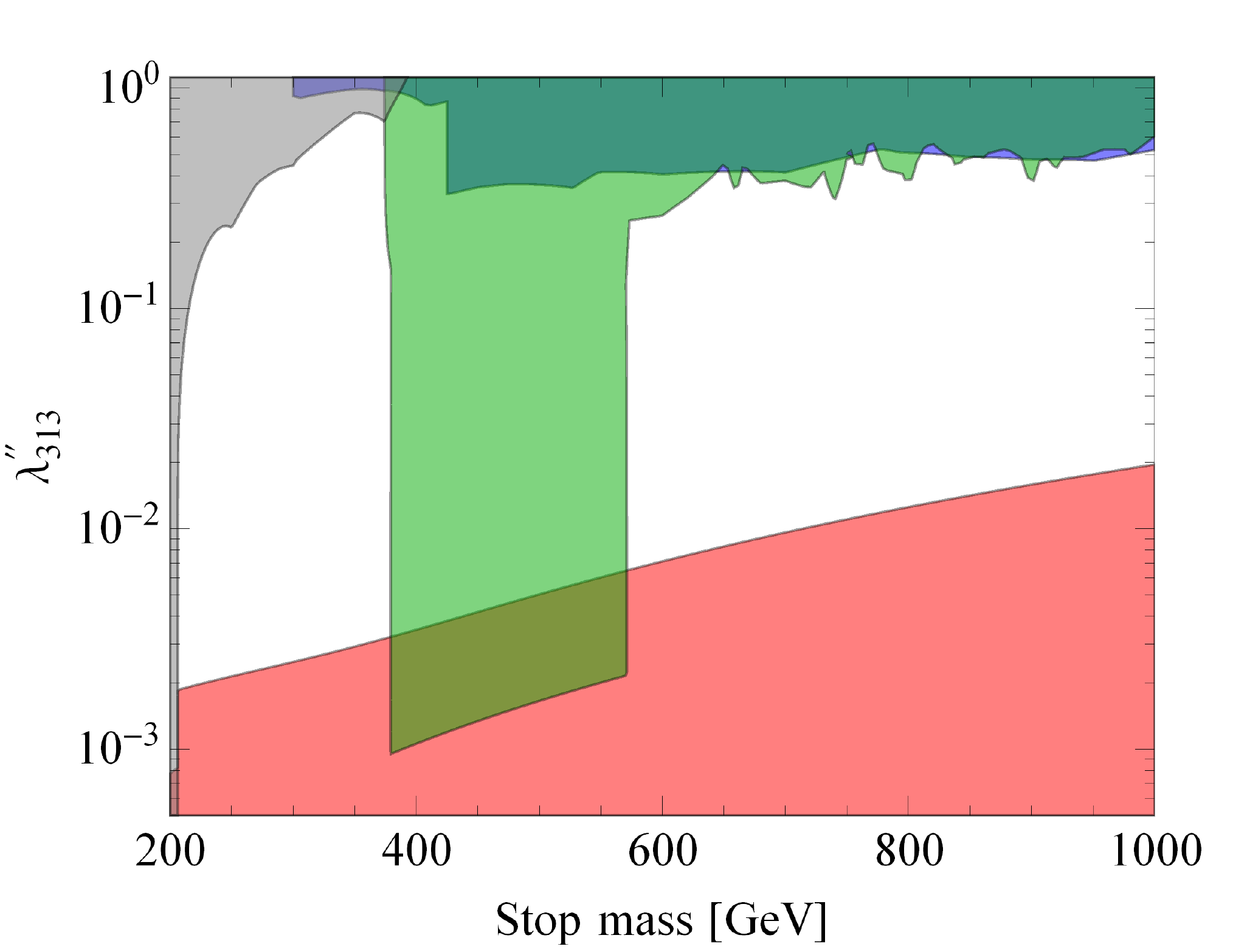}
    \label{fig:stop_exclusion_U1R_Higgsino_up_313}
  \end{subfigure}
  ~
  \begin{subfigure}[b]{0.48\textwidth}
    \centering
    \includegraphics[width=\textwidth, bb = 0 0 584 451]{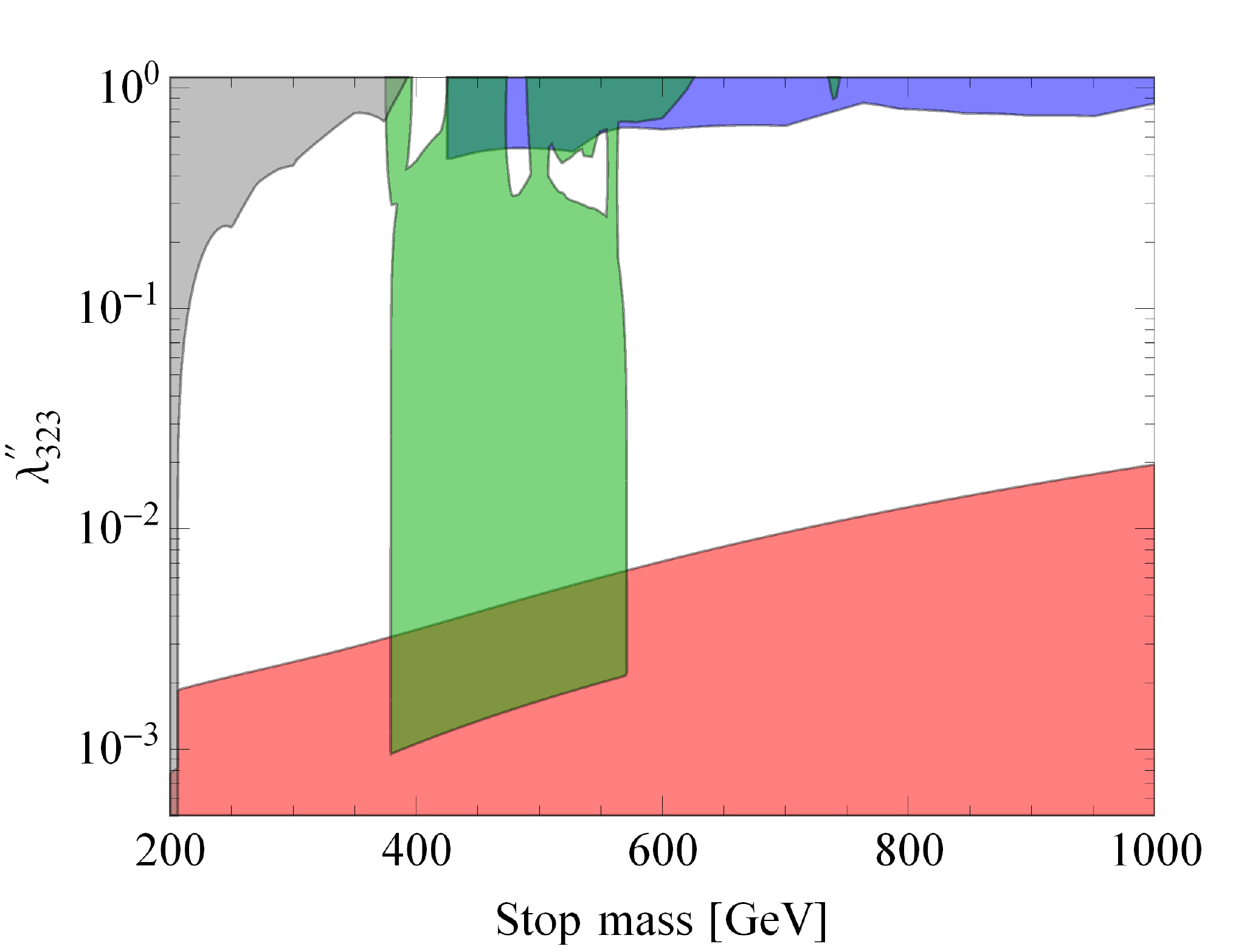}
    \label{fig:stop_exclusion_U1R_Higgsino_up_323}
  \end{subfigure}
  \caption{Exclusion plots for a 200 GeV Higgsino-up neutralino LSP with the $U(1)_R$ symmetry strictly preserved. The grey region on the left side of the plots ($m_{\tilde{t}} \lesssim 205$ GeV) is excluded by paired dijet searches. Next, consider the middle region of the plots. Starting from large $\lambda''_{3ij}$ couplings and working downwards, the blue region is excluded by dijet searches, the green region is excluded by neutralino LSP searches and the red region is excluded by displaced vertices searches.}\label{fig:stop_exclusion_Hu_U1R}
\end{figure}

\begin{figure}[t!]
  \centering
  \begin{subfigure}[b]{0.48\textwidth}
    \centering
    \includegraphics[width=\textwidth, bb = 0 0 584 451]{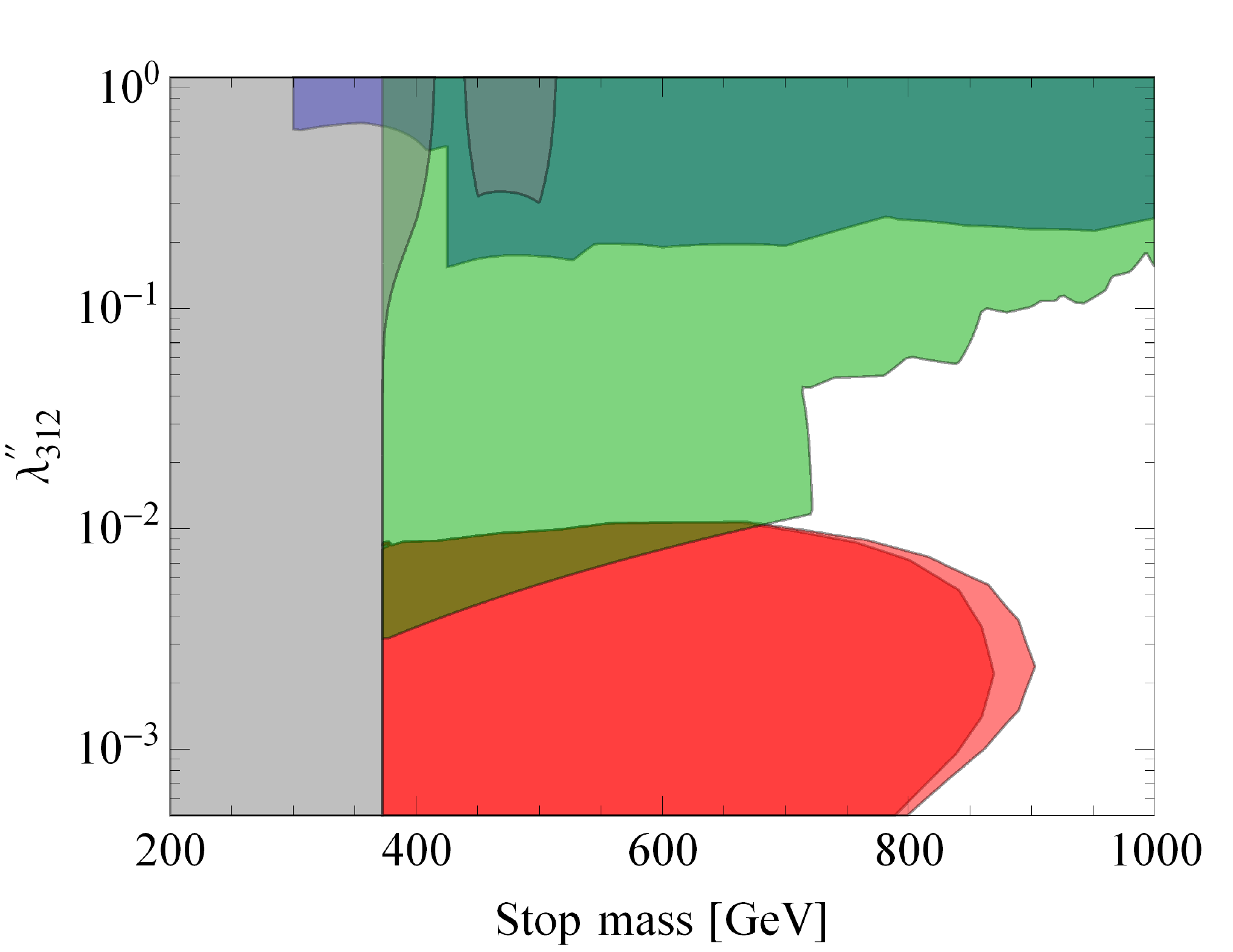}
    \label{fig:stop_exclusion_U1R_B_bino_312}
  \end{subfigure}
  ~
  \begin{subfigure}[b]{0.48\textwidth}
    \centering
    \includegraphics[width=\textwidth, bb = 0 0 584 451]{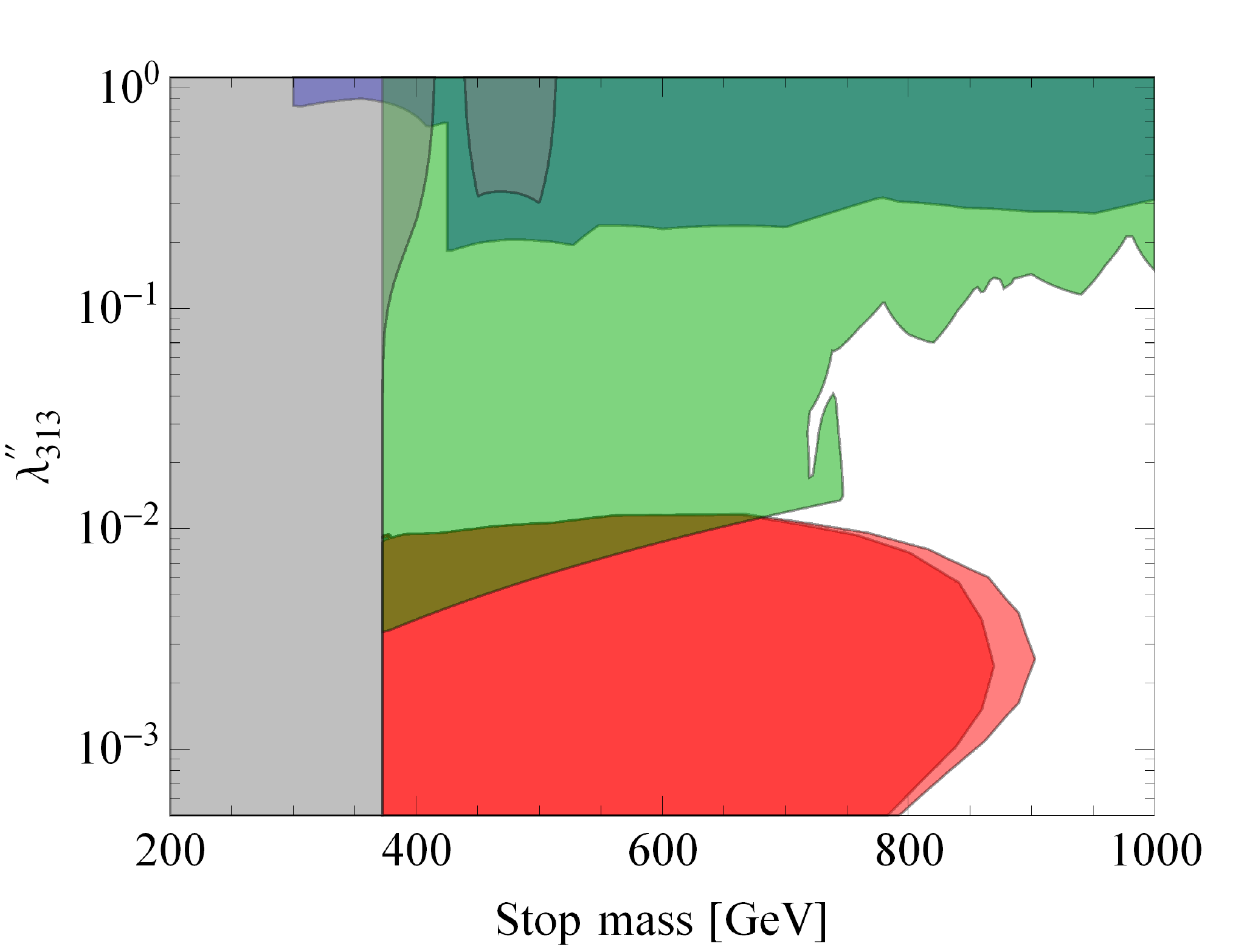}
    \label{fig:stop_exclusion_U1R_B_bino_313}
  \end{subfigure}
  ~
  \begin{subfigure}[b]{0.48\textwidth}
    \centering
    \includegraphics[width=\textwidth, bb = 0 0 584 451]{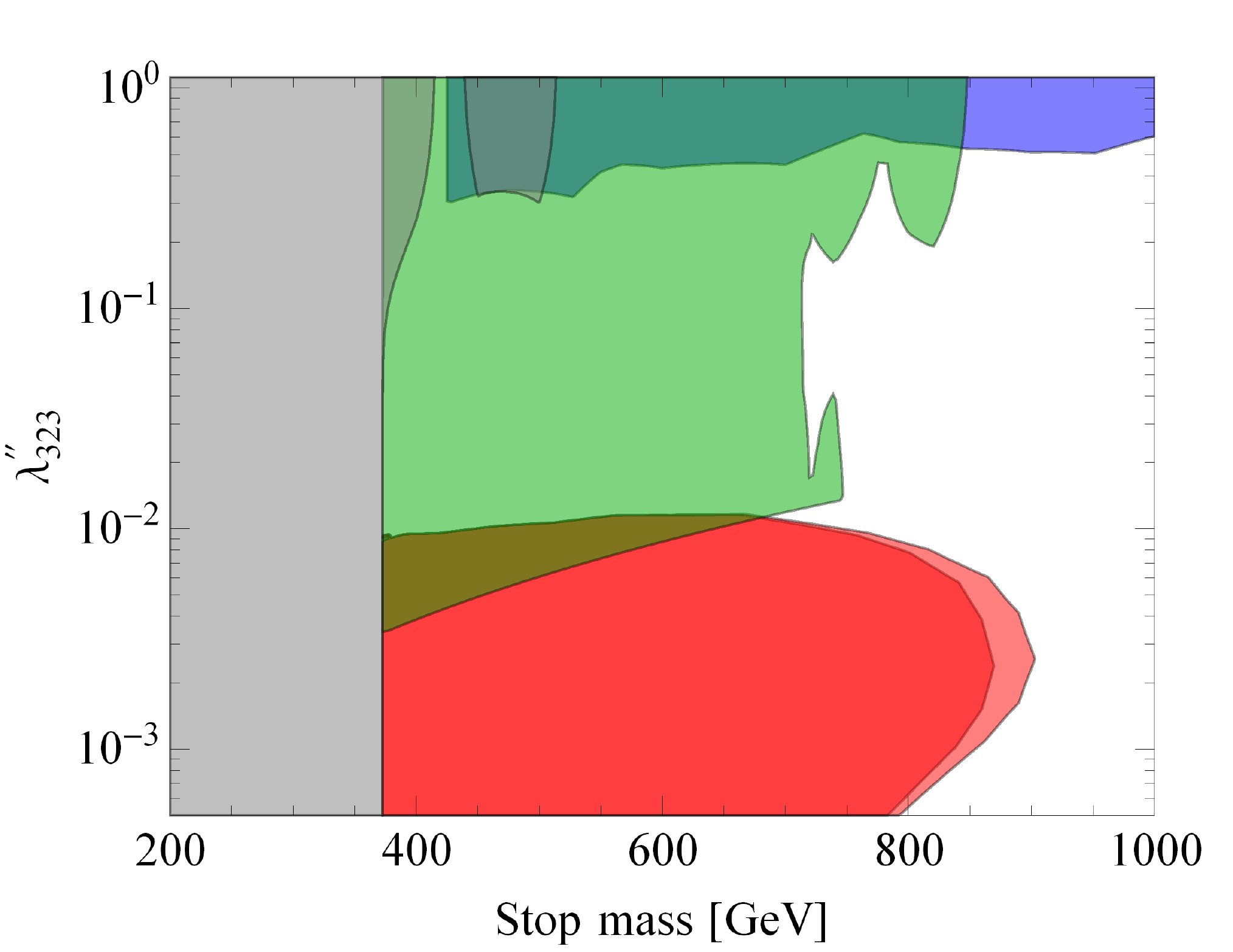}
    \label{fig:stop_exclusion_U1R_B_bino_323}
  \end{subfigure}
  \caption{Exclusion plots for a 200 GeV bino neutralino LSP with the $U(1)_R$ symmetry broken. The grey region on the left side of the plots ($m_{\tilde{t}} \lesssim 375$ GeV) is excluded by paired dijet searches. Next, consider the middle region of the plots. Starting from large $\lambda''_{3ij}$ couplings and working downwards, the blue region is excluded by dijet searches, the green region is excluded by neutralino LSP searches and the red regions are excluded by displaced vertices searches. Bino pair production, which contributes to the displaced vertices limits, depends on the masses of the first and second generations of squarks. Setting the masses of these squarks to 10 TeV results in the darker red region being excluded. Instead, setting the masses of these squarks to 1 TeV excludes both the darker red and lighter red regions.}\label{fig:stop_exclusion_Bino_U1R_B}
\end{figure}

\begin{figure}[t!]
  \centering
  \begin{subfigure}[b]{0.48\textwidth}
    \centering
    \includegraphics[width=\textwidth, bb = 0 0 584 451]{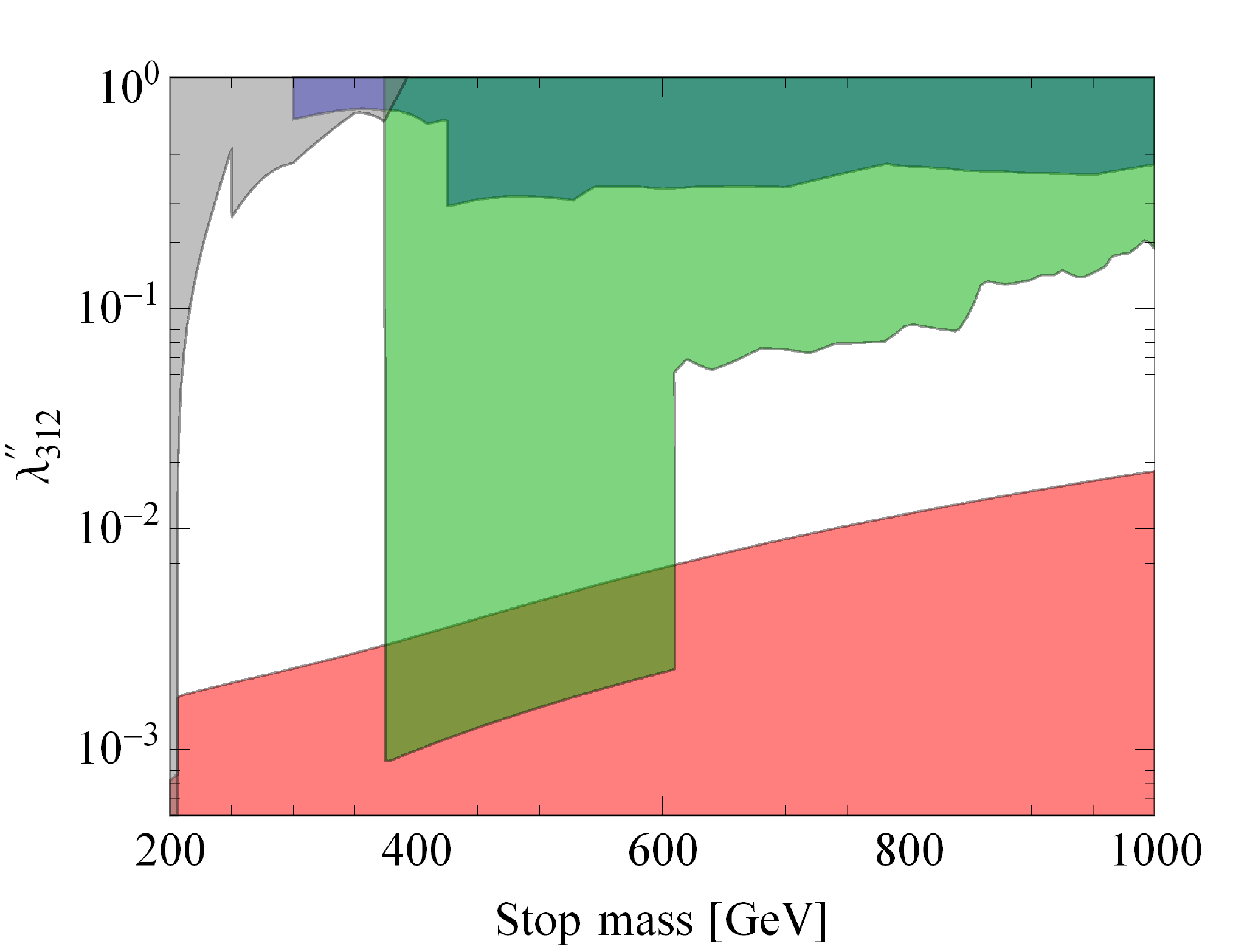}
    \label{fig:stop_exclusion_U1R_B_Higgsino_up_312}
  \end{subfigure}
  ~
  \begin{subfigure}[b]{0.48\textwidth}
    \centering
    \includegraphics[width=\textwidth, bb = 0 0 584 451]{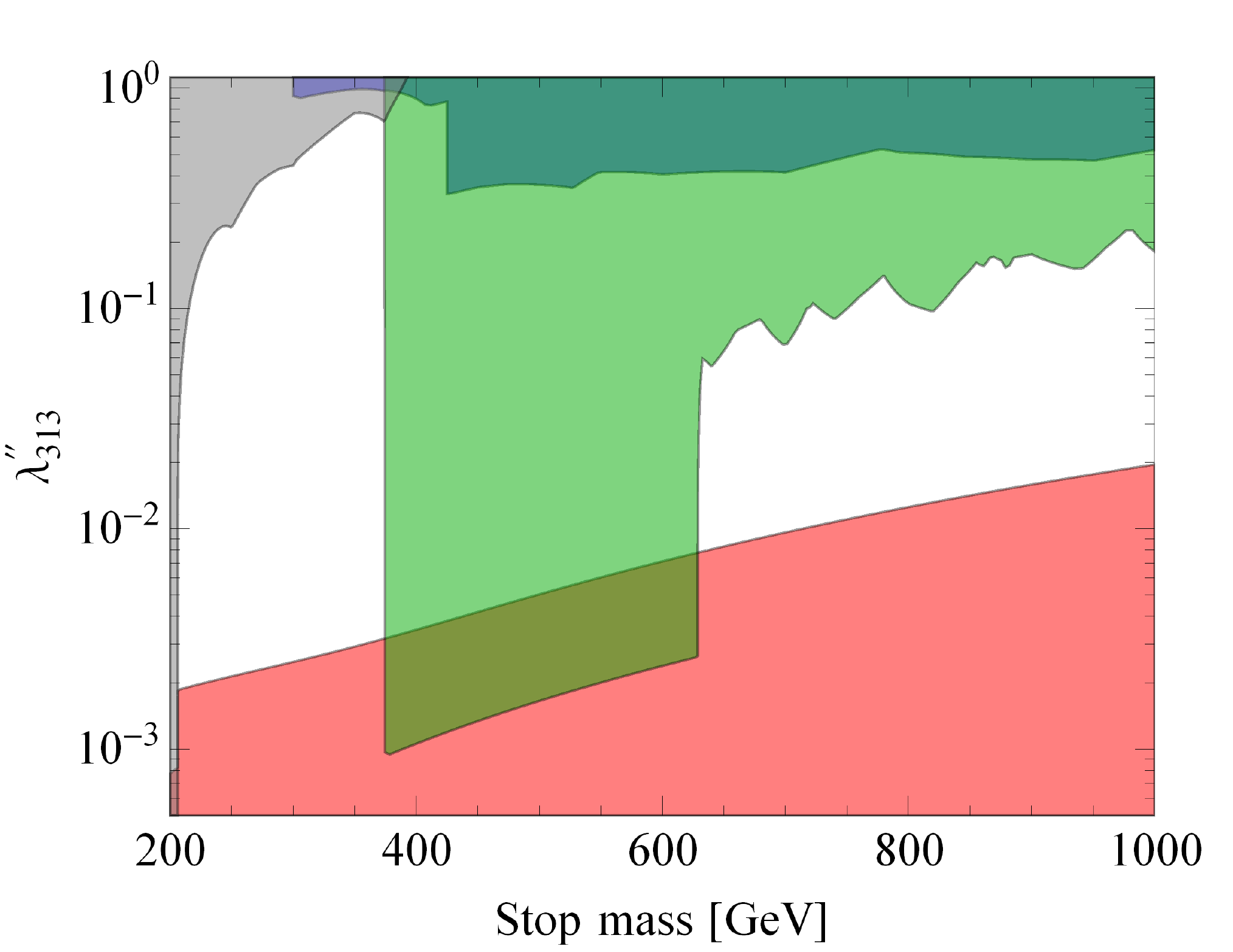}
    \label{fig:stop_exclusion_U1R_B_Higgsino_up_313}
  \end{subfigure}
  ~
  \begin{subfigure}[b]{0.48\textwidth}
    \centering
    \includegraphics[width=\textwidth, bb = 0 0 584 451]{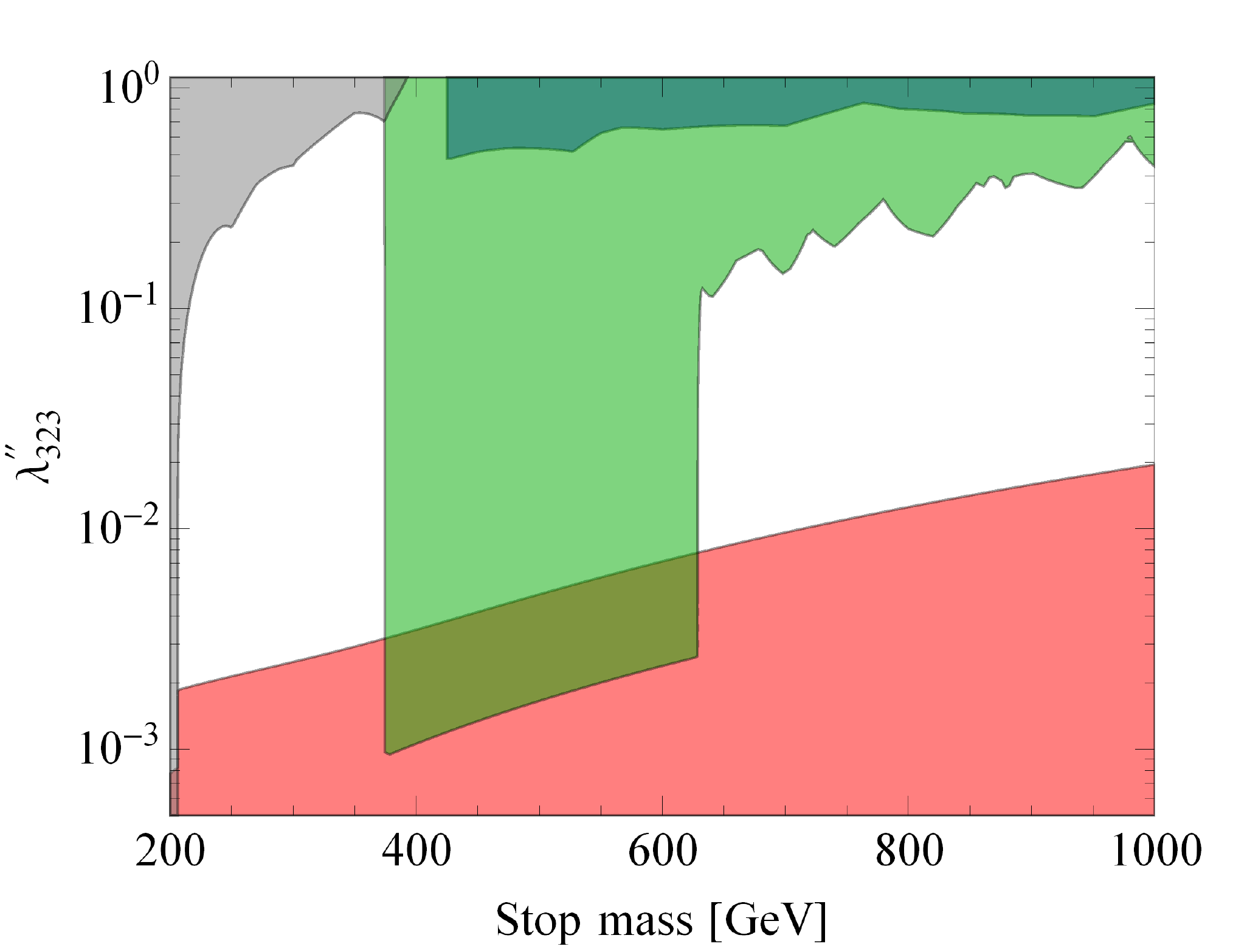}
    \label{fig:stop_exclusion_U1R_B_Higgsino_up_323}
  \end{subfigure}
  \caption{Exclusion plots for a 200 GeV Higgsino-up neutralino LSP with the $U(1)_R$ symmetry broken. The grey region on the left side of the plots ($m_{\tilde{t}} \lesssim 205$ GeV) is excluded by paired dijet searches. Next, consider the middle region of the plots. Starting from large $\lambda''_{3ij}$ couplings and working downwards, the blue region is excluded by dijet searches, the green region is excluded by neutralino LSP searches and the red region is excluded by displaced vertices searches.}\label{fig:stop_exclusion_Hu_U1R_B}
\end{figure}

Combining all types of constraints discussed above, we present exclusion plots within the stop mass and $\lambda''_{3ij}$ parameter space. Figures \ref{fig:stop_exclusion_Bino_U1R} and \ref{fig:stop_exclusion_Hu_U1R} show the regions excluded provided that the $U(1)_R$ symmetry is strictly preserved for bino LSP and Higgsino-up LSP, respectively. Similarly, figures \ref{fig:stop_exclusion_Bino_U1R_B} and \ref{fig:stop_exclusion_Hu_U1R_B} show the regions excluded when the $U(1)_R$ symmetry is broken, again for bino LSP and Higgsino-up LSP, respectively. Notice that the limits coming from the neutralino LSP searches (green area) do not extend into the smallest values of $\lambda''_{3ij}$ shown in the plots. These searches rely on promptly decaying particles and so we conservatively cutoff their exclusion capabilities when the neutralino's decay length becomes longer than 1 mm \cite{Liu:2015bma}. The green area excluded for $\lambda''_{3ij} \lesssim 0.1$  mostly results from stop pair production with subsequent decay into neutralinos. Larger stop masses are excluded for the bino than the Higgsino-up for this range of $\lambda''_{3ij}$ coupling because there is no competing chargino decay. The green area for $\lambda''_{3ij} \gtrsim 0.1$ is mostly excluded by resonant stop production with subsequent decay through neutralinos. Note that quite a bit more of this parameter space is excluded when the $U(1)_R$ symmetry is broken. This is largely due to the production of same sign tops which is absent when the $U(1)_R$ symmetry is preserved. Another interesting feature for this part of the parameter space is that approximately equal areas are excluded for $\lambda''_{312}$ and $\lambda''_{313}$. The reason for this is that while the cross section for resonant stop production is smaller for $\lambda''_{313}$, the efficiencies are generally larger than for $\lambda''_{312}$ due to the production of extra bottom quarks. These two effects are seen to approximately compensate one another.

Figures \ref{fig:stop_exclusion_Bino_U1R_B} and \ref{fig:stop_exclusion_Hu_U1R_B}, with the $U(1)_R$ symmetry broken, are similar to figures shown in Ref$.$ \cite{Monteux:2016gag}. (The figures within Ref$.$ \cite{Monteux:2016gag} are presented within a RPVMSSM context. However, as previously noted, the phenomenology of the MRSSM with the $U(1)_R$ symmetry broken is nearly identical to the RPVMSSM.) Our neutralino LSP searches exclude larger amounts of the parameter space than the corresponding searches considered by Ref$.$ \cite{Monteux:2016gag}. This is simply because we use more recent, and, in particular, 13 TeV experimental searches. Our exclusion regions for displaced vertices searches are, on the other hand, significantly smaller than Ref$.$ \cite{Monteux:2016gag}.

\subsection{Placing limits on first and second generation squarks}\label{sSec:squarks}

\subsubsection{Squark production}\label{ssSec:squark_production}
As previously mentioned, the label squarks refers to only the first and second generations. Explicitly, we consider the states $\tilde{d}_L$, $\tilde{u}_L$, $\tilde{s}_L$, $\tilde{c}_L$, $\tilde{d}_R$, $\tilde{u}_R$, $\tilde{s}_R$ and $\tilde{c}_R$ and their charge conjugates. The squarks are taken to be mass degenerate.

As we are considering the $\lambda''_{3ij}$ couplings, squarks can only be produced in pairs. In general, squark pairs are produced either from initial state gluons or initial state quarks with a t-channel gluino propagator. (We again ignore potential squark production involving two $\lambda''_{3ij}$ couplings.) Although, due to the Dirac nature of gluino, some of the production mechanisms present in the MSSM are forbidden within the MRSSM. In particular, diagrams which require a gluino Majorana mass insertion are forbidden \cite{Heikinheimo:2011fk,Kribs:2012gx}. This prevents the production of $\tilde{q}_L \tilde{q}_L$, $\tilde{q}_R \tilde{q}_R$ and $\tilde{q}_L \tilde{q}_R^*$ and their charge conjugates. Additionally, breaking the $U(1)_R$ symmetry with a small gluino mass will only reintroduce the forbidden diagrams by negligible amounts. There is no enhancement comparable to stops decaying into same sign tops. As noted above, this enhancement requires the neutralino from the stop decay to be produced on-shell. Here, the four-momentum of the t-channel gluino is spacelike and thus the gluino is never on-shell. 

As a result, we are interested in the production of $\tilde{q}_L \tilde{q}_L^*$, $\tilde{q}_R \tilde{q}_R^*$, $\tilde{q}_L \tilde{q}_R$ and $\tilde{q}_L^* \tilde{q}_R^*$. We use MadGraph to calculate the LO cross sections for these final states. To estimate higher order effects, we use NNLL-fast \cite{Beenakker:2016lwe, Beenakker:1996ch, Kulesza:2008jb, Kulesza:2009kq, Beenakker:2009ha, Beenakker:2011sf, Beenakker:2013mva, Beenakker:2014sma} to compute MSSM K-factors for squark-antisquark and squark-squark production as a function of the mass of the squarks. The gluino mass is set to 2 TeV for both steps. Below, we present plots using both the LO and the MSSM K-factor improved cross sections. 

\subsubsection{Neutralino LSP}\label{ssSec:squark_neu_lsp}
To avoid long lived squarks, we require a neutralino LSP. We again only consider a bino neutralino or a Higgsino-up neutralino. Furthermore, we consider the stop heavier than the squarks but light enough so that the neutralinos and charginos decay promptly. Then, the possible decay chains are:
\begin{alignat*}{3}
&(1) \
\begin{array}{l}
p p \rightarrow \tilde{q}^* \tilde{q} \\
\phantom{p p} \rightarrow d_{i/j} t \bar{d}_{i/j} \bar{t}
\end{array}
\quad 
&&(2) \
\begin{array}{l}
p p \rightarrow \tilde{q}^* \tilde{q} \rightarrow \bar{q} \chi^0 q \bar{\chi}^0 \\
\phantom{p p \rightarrow \tilde{q}^* \tilde{q}} \rightarrow \bar{q} t d_i d_j q \bar{t} \bar{d}_i \bar{d}_j
\end{array}
\quad 
&&(3) \
\begin{array}{l}
p p \rightarrow \tilde{q}^* \tilde{q} \rightarrow \bar{q}' \chi^+ q' \chi^- \\
\phantom{p p \rightarrow \tilde{q}^* \tilde{q}} \rightarrow \bar{q}' \bar{b} \bar{d}_i \bar{d}_j q' b d_i d_j
\end{array} \\[1.0ex]
&(4) \
\begin{array}{l}
p p \rightarrow \tilde{q}^* \tilde{q} \rightarrow d_{i/j} t q \bar{\chi}^0  \\
\phantom{p p \rightarrow \tilde{q}^* \tilde{q}} \rightarrow d_{i/j} t q \bar{t} \bar{d}_i \bar{d}_j 
\end{array}
\quad 
&&(5) \
\begin{array}{l}
p p \rightarrow \tilde{q}^* \tilde{q} \rightarrow d_{i/j} t q' \chi^-  \\
\phantom{p p \rightarrow \tilde{q}^* \tilde{q}} \rightarrow d_{i/j} t q' b d_i d_j
\end{array}
\quad 
&&(6) \ 
\begin{array}{l}
p p \rightarrow \tilde{q}^* \tilde{q} \rightarrow \bar{q} \chi^0 q' \chi^- \\
\phantom{p p \rightarrow \tilde{q}^* \tilde{q}} \rightarrow \bar{q} t d_i d_j q' b d_i d_j.
\end{array}
\end{alignat*} 
The decays involving the $\lambda''_{3ij}$ coupling can only occur for squarks $\tilde{d}_R$ and $\tilde{s}_R$ and their charge conjugates. The notation $d_{i/j}$ stands for $d_i$ or $d_j$. Whether the final state is $d_i$ or $d_j$ depends on which squark is decaying and which one of the $\lambda''_{3ij}$ is non-zero. The other squarks are required to decay through either a neutralino or chargino. If, instead, the $U(1)_R$ symmetry is broken, then processes 2, 4 and 6 are modified:
\begin{alignat*}{2}
&(2) \
\begin{array}{l}
p p \rightarrow \tilde{q}^* \tilde{q} \rightarrow \bar{q} \chi^0 q \chi^0 \rightarrow \begin{cases} \bar{q} t d_i d_j q \bar{t} \bar{d}_i \bar{d}_j \\ \bar{q} t d_i d_j q t d_i d_j \\ \bar{q} \bar{t} \bar{d}_i \bar{d}_j q \bar{t} \bar{d}_i \bar{d}_j \\ \bar{q} \bar{t} \bar{d}_i \bar{d}_j q t d_i d_j \end{cases}
\end{array}
\quad \quad
&&(4) \
\begin{array}{l}
p p \rightarrow \tilde{q}^* \tilde{q} \rightarrow d_{i/j} t q \chi^0 \rightarrow \begin{cases} d_{i/j} t q \bar{t} \bar{d}_i \bar{d}_j \\ d_{i/j} t q t d_i d_j \end{cases}
\end{array} \\[1.0ex]
&(6) \
\begin{array}{l}
p p \rightarrow \tilde{q}^* \tilde{q} \rightarrow \bar{q} \chi^0 q' \chi^- \rightarrow \begin{cases} \bar{q} t d_i d_j q' b d_i d_j \\ \bar{q} \bar{t} \bar{d}_i \bar{d}_j q' b d_i d_j. \end{cases}
\end{array}
\end{alignat*}
Note that these are very similar to the decay chains for stop pair production. Although, a major difference is that at most two tops are produced whereas stop pair production resulted in four tops. If the $U(1)_R$ symmetry is broken, then production of two same sign leptons can still potentially take place, but now with a much lower probability than in the stop scenario. 

A nearly identical procedure to the one described above is used to constrain the parameter space. Here, we scan the stop mass and neutralino mass parameter space simulating each of the decay chains above. The acceptances are once again determined for the neutralino LSP searches of table \ref{table:searches}. We then compute the branching ratios for each of the squarks. Combining the cross sections, branching ratios and acceptances, we produce exclusion curves within the stop mass and neutralino mass parameter space. 

\begin{figure}[t!]
  \centering
  \begin{subfigure}[b]{0.48\textwidth}
    \centering
    \includegraphics[width=\textwidth, bb = 0 0 584 459]{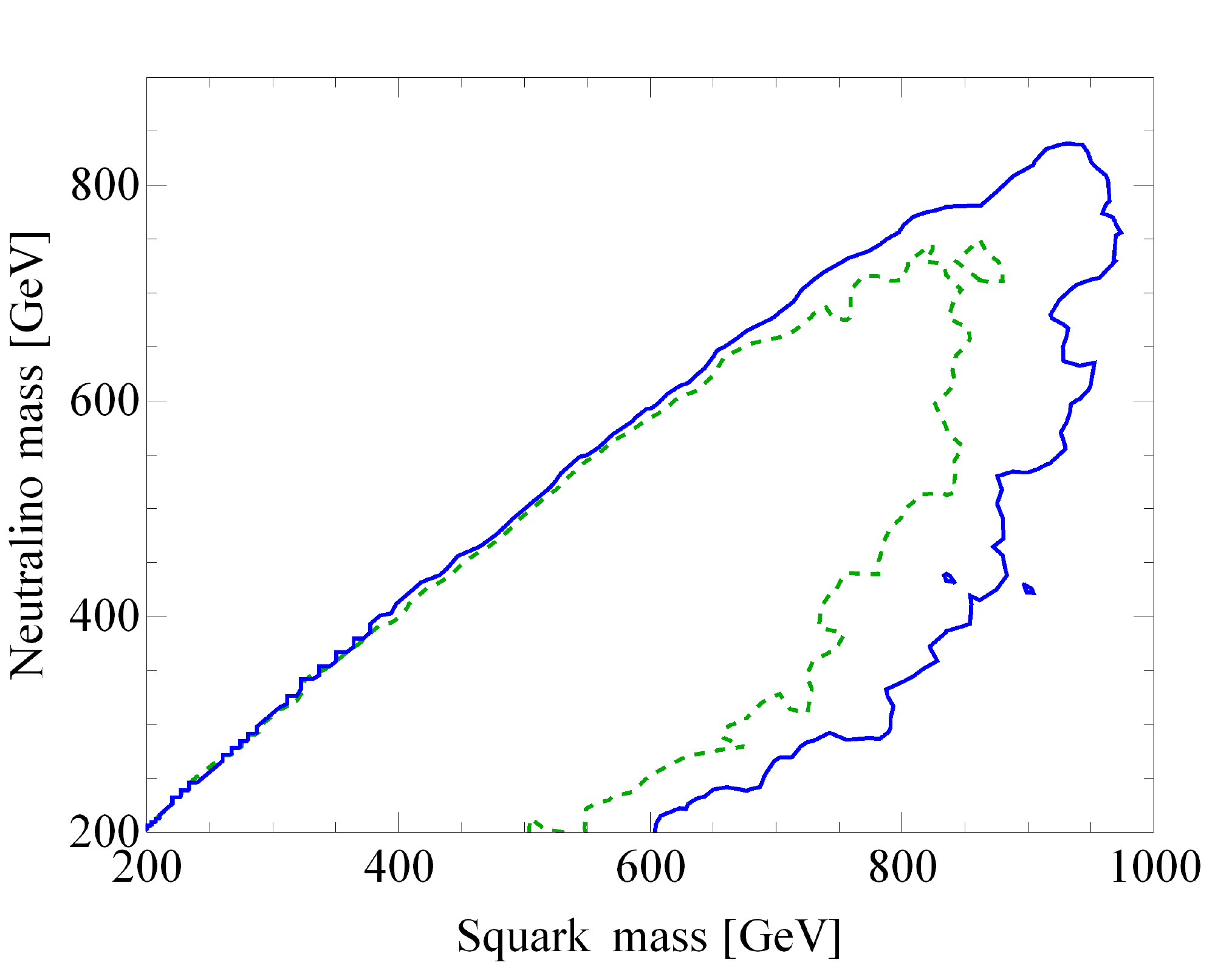}
    \caption{$\lambda''_{312}$}
    \label{fig:squark_exclusion_neuB_312}
  \end{subfigure}
  ~
  \begin{subfigure}[b]{0.48\textwidth}
    \centering
    \includegraphics[width=\textwidth, bb = 0 0 584 451]{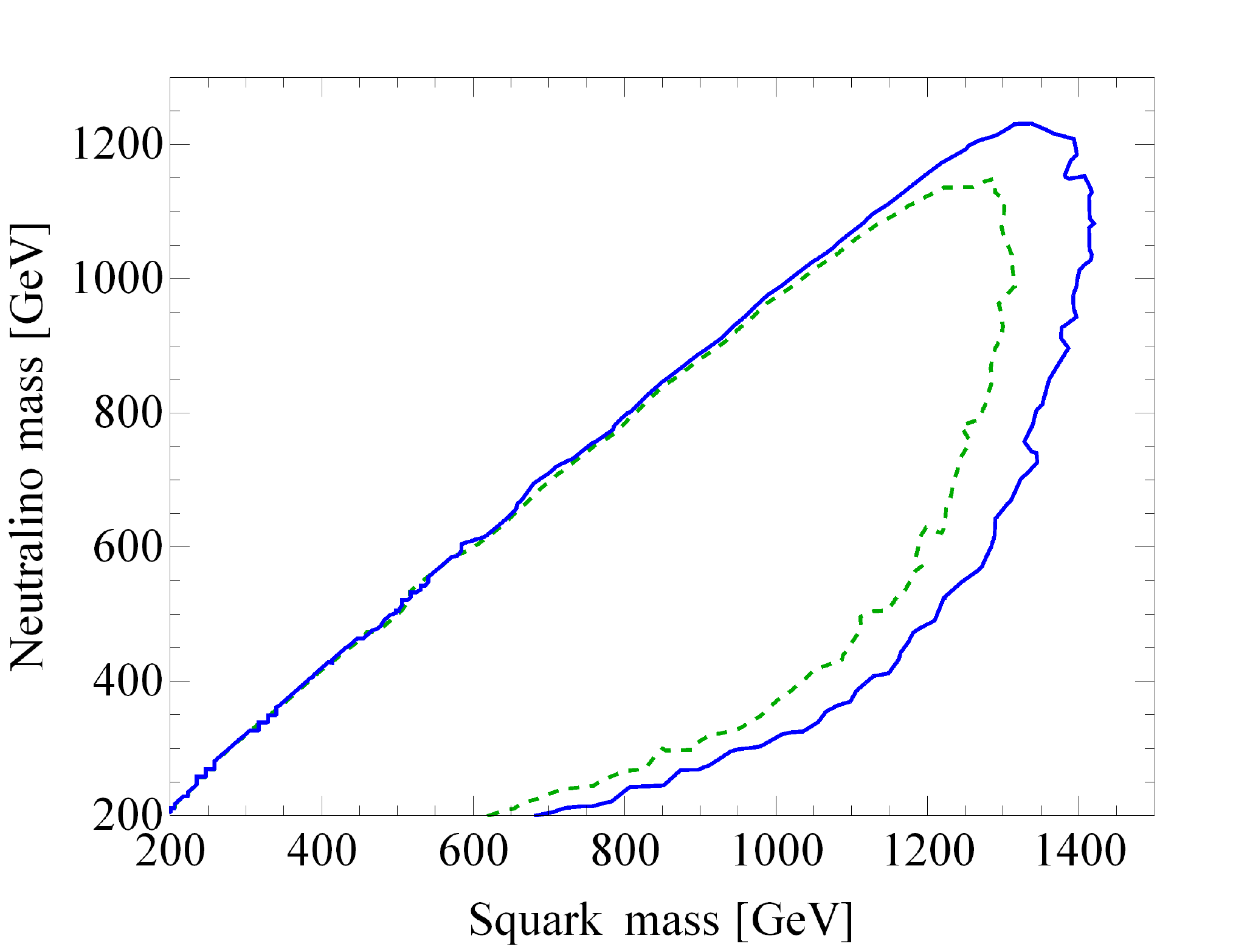}
    \caption{$\lambda''_{313} \ / \ \lambda''_{323}$}
    \label{fig:squark_exclusion_neuB_323}
  \end{subfigure}
  \caption{Exclusion curves within the neutralino-squark mass parameter space for a bino neutralino. Figure (a) presents $\lambda''_{312} = 1$ while (b) presents either $\lambda''_{313} = 1$ or $\lambda''_{323} = 1$. The dashed green curves assume leading order squark production whereas the solid blue curves include MSSM K-factors.}\label{fig:squark_exclusion_neuB}
\end{figure}

\begin{figure}[t!]
  \centering
  \begin{subfigure}[b]{0.48\textwidth}
    \centering
    \includegraphics[width=\textwidth, bb = 0 0 584 459]{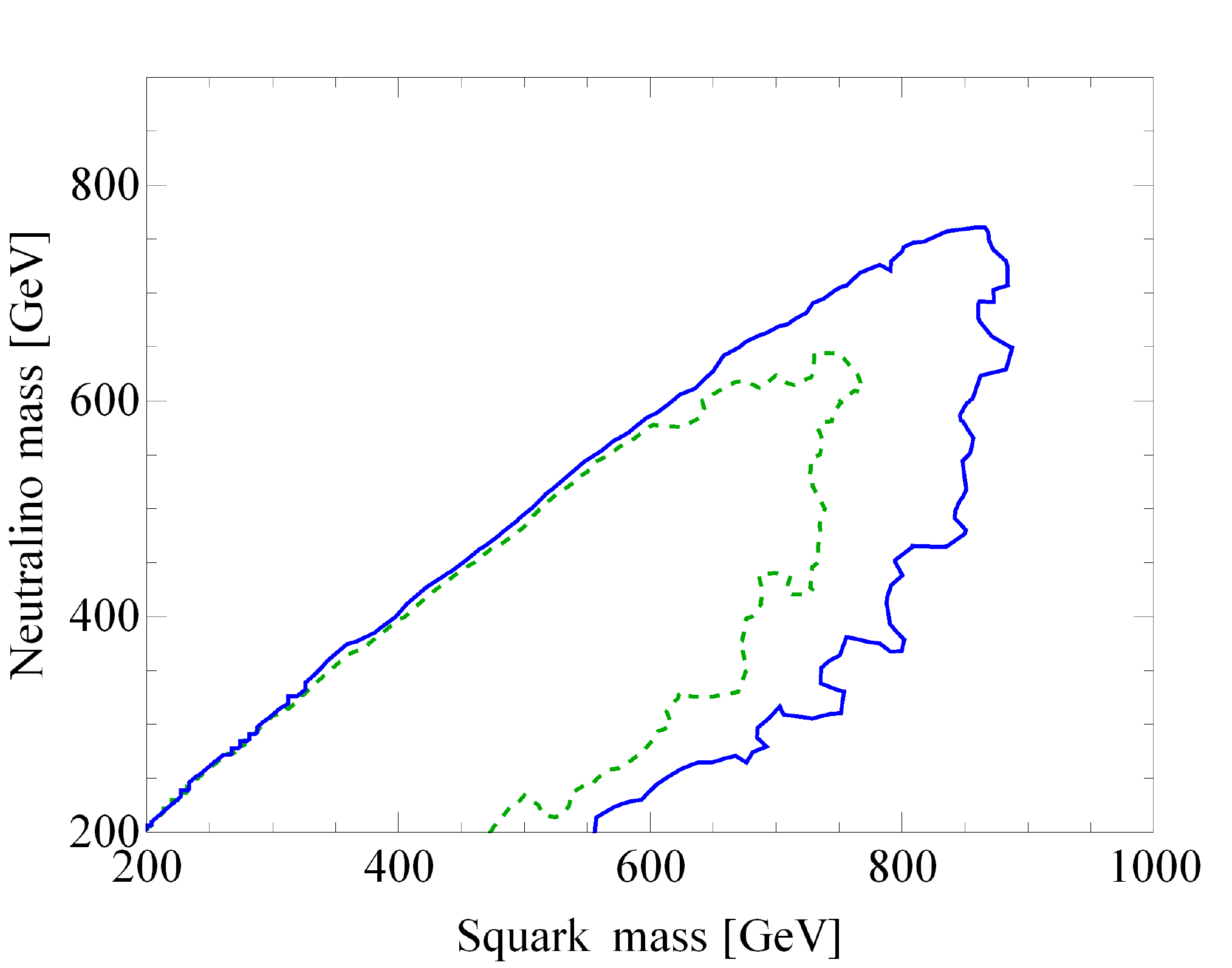}
    \caption{$\lambda''_{312}$}
    \label{fig:squark_exclusion_neuHu_312}
  \end{subfigure}
  ~
  \begin{subfigure}[b]{0.48\textwidth}
    \centering
    \includegraphics[width=\textwidth, bb = 0 0 584 451]{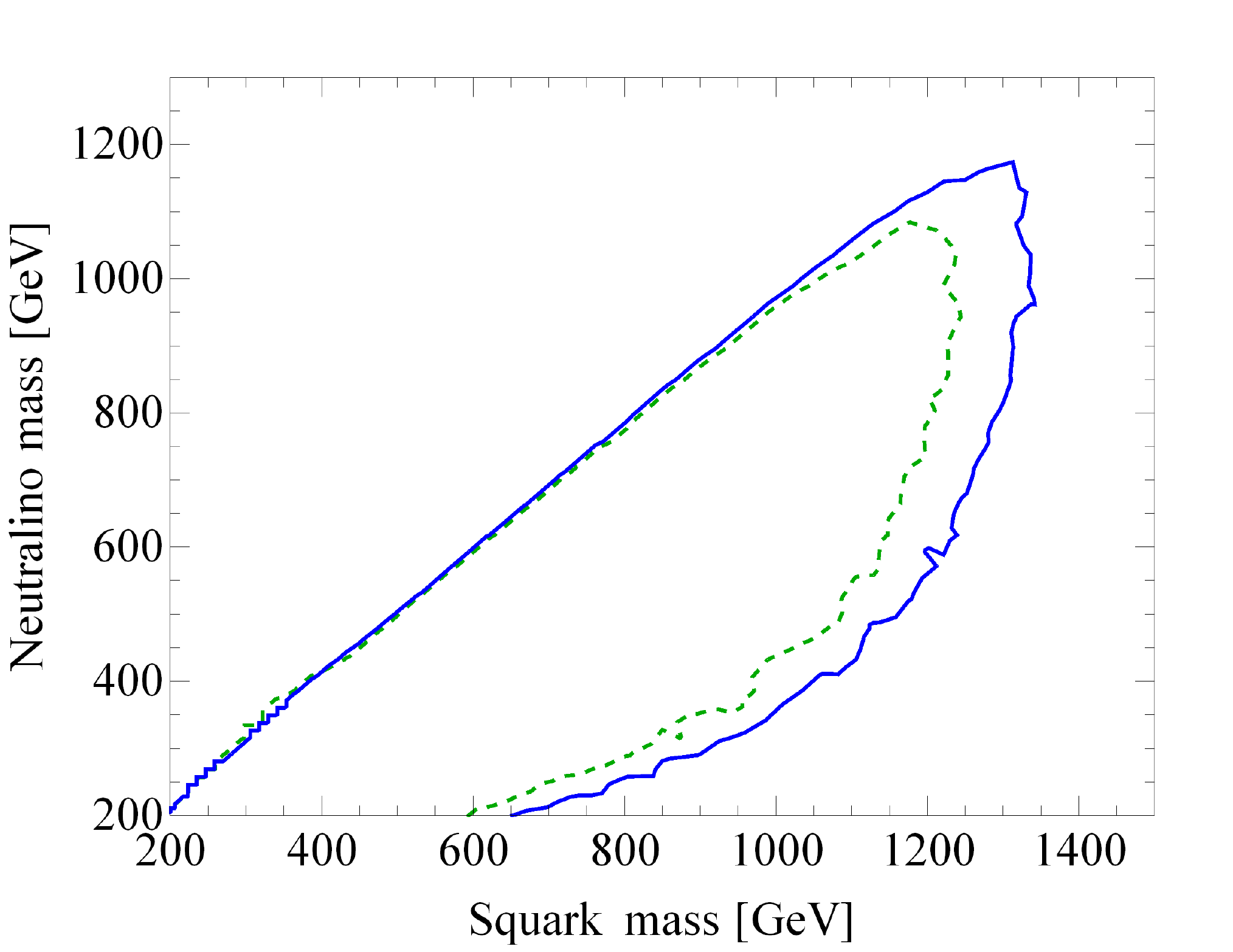}
    \caption{$\lambda''_{313} \ / \ \lambda''_{323}$}
    \label{fig:squark_exclusion_neuHu_323}
  \end{subfigure}
  \caption{Exclusion curves within the neutralino-squark mass parameter space for a Higgsino-up neutralino. Figure (a) presents $\lambda''_{312} = 1$ while (b) presents either $\lambda''_{313} = 1$ or $\lambda''_{323} = 1$. The dashed green curves assume leading order squark production whereas the solid blue curves include MSSM K-factors.}\label{fig:squark_exclusion_neuHu}
\end{figure}

Figures \ref{fig:squark_exclusion_neuB} and \ref{fig:squark_exclusion_neuHu} present the exclusions curves for bino and Higgsino-up LSP, respectively. For each plot, the $\lambda''_{3ij}$ have been set to one. Note that curves are shown only for the case in which the $U(1)_R$ symmetry is preserved. As previously noted, breaking the symmetry only introduces a small probability of producing a same sign lepton pair. Consequently, the searches that require same sign leptons are less constraining then the searches that require multiple jets. The ATLAS search \cite{ATLAS-CONF-2016-094}, which does not rely on a same sign lepton pair, dominates for the entirety of the exclusion curve for both the $U(1)_R$ symmetry preserved and broken. As a result, the exclusion curves are for the two cases are the same. Additionally, the cases $\lambda''_{313}$ and $\lambda''_{323}$ are presented together. The only difference between these two cases is the production of down versus strange quarks, which is irrelevant to the searches involved. An interesting feature is that the excluded region prefers large neutralino masses. This follows from the decay chains as most of the final state quarks come from decaying neutralinos or charginos. This is in contrast to RPC MSSM searches with decaying squarks, which exclude light neutralino masses preferentially (see figure 11(a) of the ATLAS search \cite{ATLAS-CONF-2016-078} for example). Finally, these limits on squark production are presented in a MRSSM framework. However, the exclusion curves can also be seen as lower limits for the RPVMSSM as the major difference is the exclusion of some of the possible production cross sections.  

\section{Breaking of \texorpdfstring{$U(1)_R$}{U(1)R} baryon number as an explanation of baryogenesis}\label{Sec:Baryogenesis}
Supersymmetry with an $U(1)_R$ baryon number presents a unique possibility to address baryogenesis. Indeed, the $U(1)_R$ symmetry is broken by the gravitino mass and the breaking could be transmitted to the SM sector through anomaly mediation or through Planck scale suppressed operators. This signifies a breaking of baryon number conservation, which can potentially lead to matter-antimatter asymmetry.

In this section, we discuss whether this can lead to successful baryogenesis. Essentially, Dirac binos are split into two Majorana gauginos by the introduction of Majorana masses. The resulting binos decouple early and decay out of equilibrium. The heaviest bino presents an asymmetry in its decay to baryons and antibaryons, while the lightest one can exhibit similar behaviour in the presence of light gluinos. This leads to a net baryon density. For the mechanism to be successful, a Mini-Split \cite{Arvanitaki:2012ps} spectrum is however required. The mechanism is therefore similar to previous works on baryogenesis within a Mini-Split scenario \cite{Cui:2013bta, Arcadi:2013jza, Arcadi:2015ffa}. See also \cite{Cui:2016rqt} for the LHC phenomenology of such models.

This section goes as follows. We first discuss the impact of $U(1)_R$ breaking on the gaugino sector and the assumptions on the parameter space. We then calculate the decay widths of the binos and their scattering cross sections. We finally discuss our calculation of the baryon relic density and present some results.

\subsection{\texorpdfstring{$U(1)_R$}{U(1)R} breaking}
The $U(1)_R$ breaking manifests itself in the gaugino sector mainly via the introduction of Majorana masses which modify the mass eigenstates structure of the gauginos. We consider the effect of this for both binos and gluinos. For binos, the mass Lagrangian becomes:
\begin{equation}\label{BG:Eq:Masses1}
	\mathcal{L}_{\text{masses}}=-\frac{1}{2}\left(\begin{matrix} \tilde{B} & \tilde{S} \end{matrix}\right)\left(\begin{matrix} M_1 & M_1^D \\ M_1^D & \rho_1 \end{matrix}\right)\left(\begin{matrix} \tilde{B} \\ \tilde{S} \end{matrix}\right)+\text{h.c.},
\end{equation}
where $\rho_1$ is the singlino Majorana mass which we add for generality's sake. The Majorana masses cause the Dirac bino to split into two Majorana particles of different masses.  We label the lightest one by $\chi_1^B$ and the heaviest by $\chi_2^B$. We refer to their masses as $m_1^B$ and $m_2^B$ respectively. The bino and the singlino will then be a linear combination of mass eigenstates of the form:
\begin{equation}\label{BG:Eq:Masses3}
\begin{aligned}
  \tilde{B} &=a_1 \chi^B_{1}+a_2 \chi^B_{2},\\
  \tilde{S} &=b_1 \chi^B_{1}+b_2 \chi^B_{2}.
\end{aligned}
\end{equation}
For convenience, we refer to $\chi_1^B$ and $\chi_2^B$ as binos when the context is clear.

Similar results hold for gluinos, where masses are instead labeled as $M_3^D$, $M_3$ and $\rho_3$. The result is two eigenstates $\chi_1^g$ and $\chi_2^g$ of mass $m_1^g$ and $m_2^g$ respectively. They are related to gauge eigenstates by:
\begin{equation}\label{BG:Eq:Masses4}
\begin{aligned}
  \tilde{g} &=c_1 \chi^g_{1}+c_2 \chi^g_{2},\\
  \tilde{O} &=d_1 \chi^g_{1}+d_2 \chi^g_{2}.
\end{aligned}
\end{equation}

Other $U(1)_{R}$ breaking terms could also potentially affect our results. First, $A$-terms could be introduced, but their effects are typically suppressed by the scalar masses which are assumed large in Mini-Split. Even if they were important, they would not spoil any mechanism and could in fact be used for generating additional baryon asymmetry. Second, the $\mu$-term of the MSSM could reappear in the superpotential. As will be further discussed in section \ref{sSec:BinoDecay}, this would spoil a mechanism that allows for the Higgsinos to be lighter than in simple Mini-Split versions of the MSSM. The $\mu$-term can however be naturally small as it is a coefficient in the superpotential.  Finally, the most dangerous possibility is soft-terms that mix Higgses such as $H_u R_d+\text{h.c.}$ and $H_u^\dagger R_u+\text{h.c.}$. These terms can lead to the lightest Higgs containing parts of $R_u$ and $R_d$, which would also reintroduce the need for heavy Higgsinos. This effect can however be suppressed by $R_u$ and $R_d$ having large soft masses, which we assume from now on to be the case. The only $U(1)_R$ breaking that we consider is then the Majorana masses. One property of anomaly mediation worth mentioning is that the problematic terms are either not generated or are suppressed.

\subsection{Assumptions on the parameter space}
In the mechanism that we consider, the baryon asymmetry originates from the decay of the binos to three quarks through the $\lambda''_{ijk}$ coupling. However, the Nanopoulos-Weinberg theorem \cite{Nanopoulos:1979gx} states that, if a particle is to exhibit an asymmetry in its decay to baryons and antibaryons, it must also be able to decay via a baryon number conserving channel. In our case, this corresponds to decays to quarks and lighter gauginos. Nonetheless, it is necessary for the baryon number breaking decays to dominate. Else, the baryon asymmetry will be suppressed by a small branching ratio. This will require some of the $\lambda''_{ijk}$ to be of $\mathcal{O}(0.1)$ or more. As explained in section \ref{sSec:BoundsOnLambdapp}, this is only possible for a few of them, though the fact that the optimal region of parameter space that we will obtain corresponds to a Mini-Split spectrum relaxes the constraints on several $\lambda''_{ijk}$. As such, we assume that a single $\lambda''_{ijk}$ is non-zero and refer to it as $\lambda''$. The generalization to several non-zero $\lambda''_{ijk}$'s is trivial. We refer to the associated up quark as $u_1$ and the associated down quarks as $d_1$ and $d_2$.  We choose to concentrate on the case of a single relevant right-handed sdown-type squark and take it to be $\tilde{d}_2$. Analytical results can easily be converted to multiple sdown-type squarks or right-handed sup-type squarks.

We also assume that $\tilde{d}_2$ is considerably heavier than the binos. This is necessary for two reasons. First, binos are required to decay after they decouple to avoid washout effects. This is simply stating that the decay must be out of equilibrium to satisfy the Sakharov conditions \cite{Sakharov:1967dj}. We will use this fact to calculate would-be relic number densities of binos which corresponds to what the relic densities would be if the binos were stable. Second, the binos are also required to decouple early. This is simply a question that the baryon number density coming from bino decay is several orders of magnitude smaller than the bino would-be number relic density. We define $x=m_2^B/T$ and label the value of $x$ around which the bino decouples by $x_f$. A viable baryogenesis will typically require $x_f<5$ \cite{Cui:2013bta}.

Additionally, we assume that winos are heavy. We also consider Higgsinos to be heavy but not so much as to be irrelevant. This opens a decay channel to Higgses. Finally, we assume for simplicity's sake that $\chi_2^B$ is considerably heavier than the electroweak scale.

\subsection{Binos decays}\label{sSec:BinoDecay}
We now proceed to list the widths associated to all $\chi_1^B$ and $\chi_2^B$ decay channels. We neglect the mass of all quarks.

\subsubsection{Baryon breaking decay at tree-level}
The binos can decay to three quarks with a tree-level decay width of:
\begin{equation}\label{BG:Eq:DecayWidthRPV}
  \Gamma_{\chi^B_{i}\to u_1 d_1 d_2}=\frac{g'^2Y_d^2|a_i \lambda''|^2}{512 \pi^3}\frac{(m_i^B)^5}{m_{\tilde{d}_2}^4},
\end{equation}
where $Y_d=1/3$ is the weak hypercharge of $d_2$. The tree-level decay width to antiquarks is the same.

\subsubsection{Decay of \texorpdfstring{$\chi_2^B$}{chi2B} to \texorpdfstring{$\chi_1^B$}{chi1B} and quarks}
The baryon conserving decay of $\chi_2^B$ to $d_2$, $\overline{d}_2$ and $\chi_1^B$ corresponds to a decay width of:
\begin{equation}\label{BG:Eq:DecayWidthB2B1bbar}
  \Gamma_{\chi^B_{2}\to \chi^B_{1} d_2 \overline{d}_2}=\frac{g'^4 Y_d^4}{256\pi^3}\left[|a_1 a_2|^2 f\left(\frac{m^B_1}{m^B_2}\right)+2 \text{Re}\{a_1^{2} a_2^{*2}\}\frac{m^B_1}{m^B_2}g\left(\frac{m^B_1}{m^B_2}\right)\right]\frac{(m^B_2)^5}{m_{\tilde{d}_2}^4},
\end{equation}
where:
\begin{equation}\label{BG:Eq:KinematicFunctions}
  \begin{aligned}
    f(x) &= (1-8x^2-12x^4\ln x^2+8x^6-x^8)\theta(1-x),\\
    g(x) &= (1+9x^2+6x^2(1+x^2)\ln x^2-9x^4-x^6)\theta(1-x).
  \end{aligned}
\end{equation}

\subsubsection{Decay of \texorpdfstring{$\chi_i^B$}{chiiB} to \texorpdfstring{$\chi_j^g$}{chijg} and quarks}
If a bino $\chi_i^B$ is heavier than a gluino $\chi_j^g$, it can also decay to this gluino and quarks with a decay width of:
\begin{equation}\label{BG:Eq:DecayWidthBiGjbbar}
	\Gamma_{\chi^B_{i}\to \chi^g_{j} d_2 \overline{d}_2}=\frac{g'^2 g_s^2 Y_d^2}{192\pi^3}\left[|a_i c_j|^2 f\left(\frac{m_j^g}{m_i^B}\right)+2 \text{Re}\{a_i^{2} c_j^{*2}\}\frac{m_j^g}{m_i^B}g\left(\frac{m_j^g}{m_i^B}\right)\right]\frac{(m_i^B)^5}{m_{\tilde{d}_2}^4}.
\end{equation}

\subsubsection{Decay of \texorpdfstring{$\chi_2^B$}{chi2B} to \texorpdfstring{$\chi_1^B$}{chi1B} and Higgses}
\begin{figure}[t!] 
  \centering 
  \includegraphics[width=0.33\textwidth, bb = 0 0 209 178]{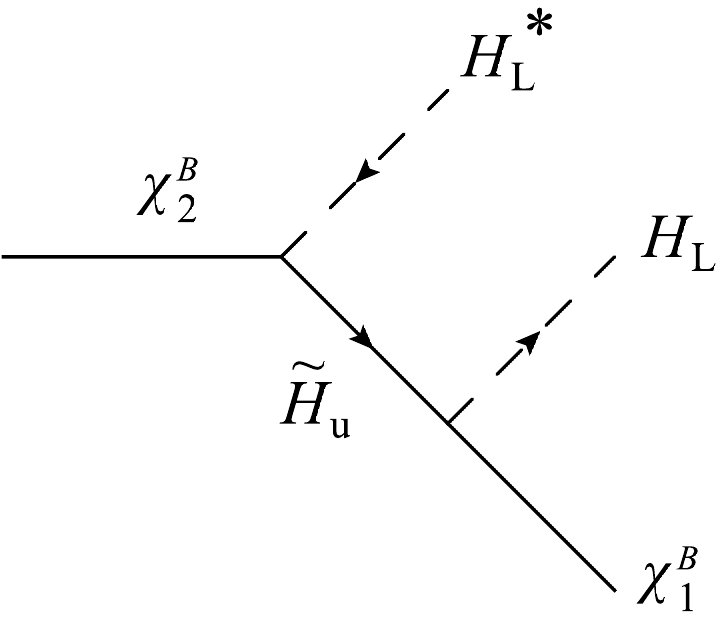} 
  \caption{Example of diagram contributing to the baryon conserving decays of $\chi_2^B$ to $\chi_1^B$ and Higgses.}
  \label{BG:Fig:BconservingDecayHH}
\end{figure}
An example of a diagram that leads to decay of $\chi_2^B$ to $\chi_1^B$ and two Higgses is shown in figure \ref{BG:Fig:BconservingDecayHH}. This decay leads to the only width that is only suppressed by two powers of a superpartner mass. Other decay processes are instead suppressed by four. As such, the Higgsinos are in general required to be considerably heavier than the scalars. The masses of the Higgs doublets are then approximatively given by:
\begin{equation}\label{BG:Eq:HiggsesMasses}
	\mathcal{L}_{\text{masses}}=-\left(\begin{matrix} H_u^\dagger & \tilde{H}_d^\dagger \end{matrix}\right)\left(\begin{matrix} \mu_u^2 & B_\mu \\ B_\mu & \mu_d^2 \end{matrix}\right)\left(\begin{matrix} H_u \\ \tilde{H}_d \end{matrix}\right),
\end{equation}
where $\tilde{H}_d = i\sigma^2H_d^*$ and where we assumed that $\mu_u$, $\mu_d$ and $B_\mu$ are real. Requiring one of the Higgses to be light necessitates $B_\mu^2=\mu_u^2 \mu_d^2$. The resulting light Higgs $H_L$ is then given by:
\begin{equation}\label{BG:Eq:LightHiggs}
	H_L = \frac{1}{\sqrt{\mu_u^2+\mu_d^2}}\left(\mu_d H_u + \mu_u \tilde{H}_d \right).
\end{equation}
In this limit, the corresponding decay width is given by:
\begin{equation}\label{BG:Eq:DecayToHiggs}
  \Gamma_{\chi^B_{2}\to \chi^B_{1} H_L H_L^*}=\frac{1}{768\pi^3}\left[|C_{12}|^2 u\left(\frac{m_1^B}{m_2^B}\right)+3\text{Re}\{C_{12}^2\}\frac{m_1^B}{m_2^B}v\left(\frac{m_1^B}{m_2^B}\right)\right](m_2^B)^3,
\end{equation}
where:
\begin{equation}\label{BG:Eq:u_and_v}
  u(x)=(1-x^2)^3\theta(1-x), \hspace{1cm} v(x)=(1+2x^2\ln x^2-x^4)\theta(1-x),
\end{equation}
and:
\begin{equation}\label{BG:Eq:cH}
  C_{ij}=\frac{g'}{\mu_u^2+\mu_d^2}\left(\lambda^s_u\frac{\mu_d^2}{\mu_u}-\lambda^s_d\frac{\mu_u^2}{\mu_d}\right)(a_i b_j + a_j b_i).
\end{equation}
A similar result exists for Mini-Split leptogenesis. In this case, the wino is required to be lighter than the bino for leptogenesis to occur. The bino can then decay to the wino and Higgses with a decay width of \cite{Cui:2013bta}:
\begin{equation}\label{BG:Eq:SMleptogenesis}
  \Gamma_{\tilde{B}\to \tilde{W} H_L H_L^*}^{\text{MSSM}}=\frac{(Y_H g_1 g_2)^2}{384\pi^3}\frac{M_1^3}{\mu^2},
\end{equation}
where $Y_H=1/2$ is the weak hypercharge of the Higgs doublet and the wino mass was neglected. The main difference of eq$.$ (\ref{BG:Eq:DecayToHiggs}) is the presence of $\lambda^s_u$ and $\lambda^s_d$, which is an effect of the extended Higgs sector. Being coefficients in the superpotential, $\lambda^s_u$ and $\lambda^s_d$ can naturally be small and the Higgsinos are not required to be as heavy as in the MSSM. As was alluded to earlier, this mechanism can however be spoiled by the presence of either a $\mu$-term or soft-terms like $H_u  R_d+\text{h.c.}$. These terms would allow for diagrams that do not require $\lambda^s_u$ and $\lambda^s_d$. That is why we needed to make assumptions to limit their effects.

We note that the width of this decay channel does not go to zero when $\lambda^s_u$ and $\lambda^s_d$ are zero, but that it would instead be suppressed by higher powers of $\mu_u$ and $\mu_d$. We also note that this result is not exact as the decay will typically take place after electroweak phase transition (EWPT). The degrees of freedom involved will not be the same and the exact expression depends on the precise details of the scalar sector. As we are more interested in a proof of principle, we will be satisfied with this result.

\subsubsection{Decay of \texorpdfstring{$\chi_2^B$}{chi2B} to \texorpdfstring{$\chi_1^B$}{chi1B} and a photon}
\begin{figure}[t!] 
  \centering 
  \includegraphics[width=0.4\textwidth, bb = 0 0 258 180]{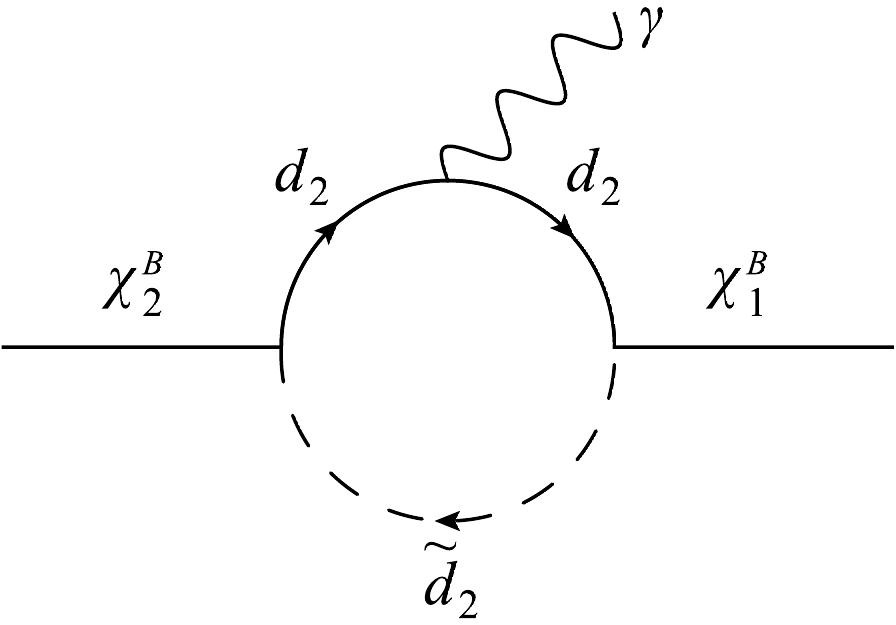} 
  \caption{Example of diagram contributing to the decay of $\chi^B_{2}$ to $\chi^B_{1}$ and a photon.}
  \label{BG:Fig:PhotonN2}
\end{figure}

Finally, $\chi_2^B$ can also decay to $\chi_1^B$ and a photon as shown in figure \ref{BG:Fig:PhotonN2}. The decay width is:
\begin{equation}\label{BG:Eq:DecayWidthB2B1Bboson}
  \begin{aligned}
  \Gamma_{\chi^B_{2}\to \chi^B_{1} \gamma} &=\frac{e^2 g'^4Y_d^4}{8192\pi^5} \left[|a_1 a_2|^2 +2 \text{Re}\{a_1^{2} a_2^{*2}\} \frac{m_1^B m_2^B}{(m_1^B)^2+(m_2^B)^2} \right]\\
  &\hspace{1.9cm}\left(1+\left(\frac{m_1^B}{m_2^B}\right)^2\right)\left(1-\left(\frac{m_1^B}{m_2^B}\right)^2\right)^3\frac{(m_2^B)^5}{m_{\tilde{d}_2}^4},
  \end{aligned}
\end{equation}
which is negligible. Note that it was assumed that the decay takes place after EWPT, which will turn out to be case in most of the successful region of parameter space. If the decay were to take place before phase transition, it would instead involve a $B$ boson. The answer would change slightly but would still remain negligible. The decay to a $Z$ boson is also negligible.

\subsubsection{Net decay width to baryons}\label{BG:Sec:NetDecay}
\begin{figure}[t!] 
  \centering
    \begin{subfigure}[b]{0.33\textwidth}
       \centering
       \includegraphics[width=\textwidth, bb = 0 0 230 148]{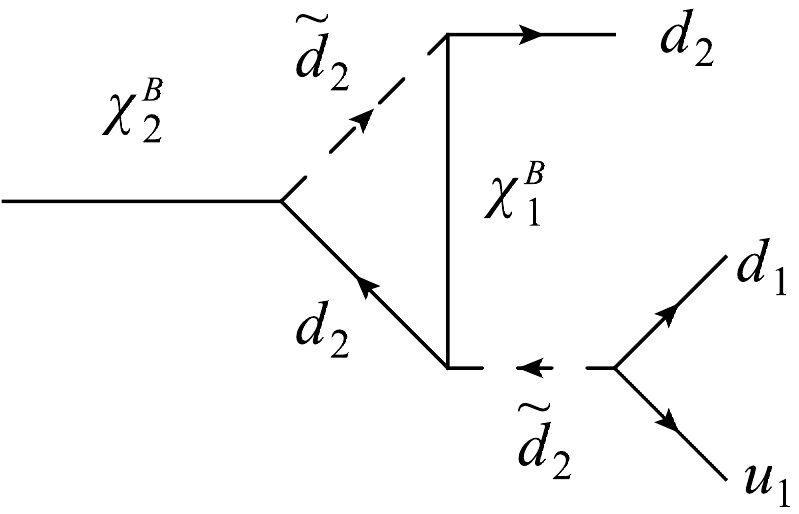}
       \caption{}
       \label{BG:Fig:LoopqqqN}
    \end{subfigure}%
    ~\qquad
    \begin{subfigure}[b]{0.33\textwidth}
       \centering
       \includegraphics[width=\textwidth, bb = 0 0 230 148]{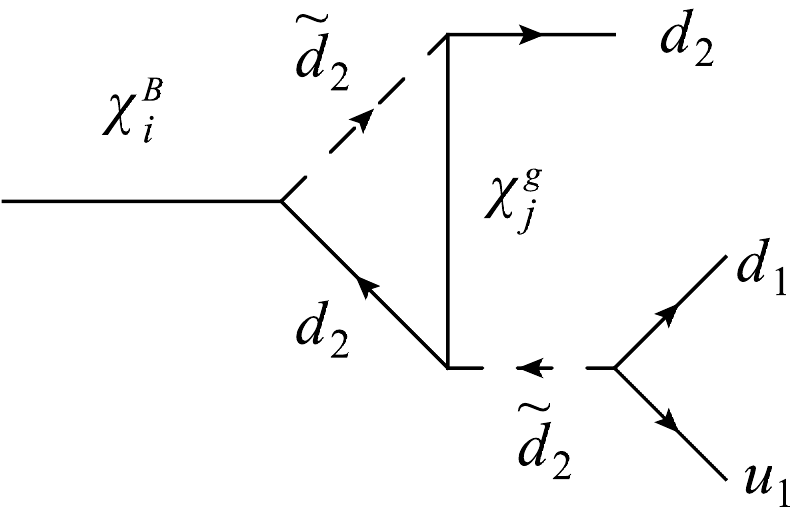}
       \caption{}
       \label{BG:Fig:LoopqqqG}
    \end{subfigure}
  \caption{(a) Loop baryon number breaking decay of $\chi^B_{2}$ via virtual $\chi^B_{1}$. (b) Loop baryon number breaking decay of $\chi^B_{i}$ via virtual $\chi^g_{j}$.}\label{BG:Fig:Loopqqq}
\end{figure}

In the case of heavy gluinos, only $\chi^B_{2}$ will present an asymmetry in its decay to baryons and antibaryons. The asymmetry comes from the interference between the tree-level diagram and the loop diagram of figure \ref{BG:Fig:LoopqqqN}. Other diagrams exist, but either do not lead to any baryon asymmetry or require a dangerous amount of flavour mixing \cite{Cui:2013bta}. In this case, the net decay width to baryons is given by:
\begin{equation}\label{BG:Eq:DecayWidthBiNetBino}
  \Delta \Gamma^B_{\chi_2^B}=\Gamma^B_{\chi^B_{2}\to u_1 d_1 d_2}-\Gamma^B_{\chi^B_{2}\to \overline{u}_1 \overline{d}_1 \overline{d}_2}=\frac{g'^4 Y_d^4| a_1 a_2\lambda''|^2\sin\phi}{2048\pi^4}f\left(\frac{m_1^B}{m_2^B}\right)\left(\frac{m_2^B}{m_{\tilde{d}_2}}\right)^6 m_1^B,
\end{equation}
where the $B$ superscript on $\Gamma$ means that we do not consider contributions from diagrams containing gluinos and:
\begin{equation}\label{BG:Eq:SinPhi}
  \sin\phi=\frac{\text{Im}\{a_1^{*2}a_2^2 \}}{|a_1 a_2|^2}.
\end{equation}
We note that this result illustrates the Nanopoulos-Weinberg theorem. The net width $\Delta \Gamma^B$ would be zero if $m_1^B>m_2^B$ since $f(x)$ is 0 for $x\geq 1$. This also corresponds to the decay of $\chi_2^B$ to $\chi_1^B$ and quarks being forbidden. Obviously, this is not a problem because $m_1^B<m_2^B$ by assumption, but it does show that the decay of $\chi_1^B$ does not lead to baryon asymmetry without at least some new lighter particle.

An interesting property of $\sin\phi$ is that it is zero if at least one of the Majorana masses is zero. This can be understood as follows. First, consider the case of $\rho_1$ set to zero. The interference term between the tree-level diagram and the diagram of figure \ref{BG:Fig:LoopqqqN} can be factorized as a function of the coupling constants times a function depending only on the kinematics. To obtain a net baryon asymmetry, both of these functions must be complex \cite{Kolb:1990vq}.\footnote{It is assumed that the fields have been redefined such that their masses are real.} First, the kinematic function is complex because the loop diagram \ref{BG:Fig:LoopqqqN} can be cut to obtain a tree-level diagram contributing to the decay of $\chi_2^B$ to $\chi_1^B$ and quarks (see \cite{Cutkosky:1960sp}). Second, the phase of $M_1$ can be reabsorbed by a field redefinition of $\tilde{B}$. This effectively makes $g'$ complex. It also transfers a phase to $M_1^D$, which can then be removed by a field redefinition of $\tilde{S}$. Since none of the couplings associated to this field are involved in these diagrams, the phase will only appear in $g'$. Since $g'$ appears as $|g'|^4$ in the interference term, this would not lead to any asymmetry. The procedure obviously breaks down when $\rho_1$ is non-zero, as the field redefinition we did would lead to a complex $\rho_1$. A similar argument holds for $M_1$.

In the case of light gluinos, both binos can potentially contribute to the baryon density:
\begin{equation}\label{BG:Eq:DecayWidthBiNetBGluino}
	\Delta \Gamma^g_{\chi_i^B}=\Gamma^g_{\chi^B_{i}\to u_1 d_1 d_2}-\Gamma^g_{\chi^B_{i}\to \overline{u}_1 \overline{d}_1 \overline{d}_2}=\sum_j\frac{g'^2 g_s^2 Y_d^2| a_i c_j\lambda''|^2\sin\phi'_{ij}}{1536\pi^4}f\left(\frac{m_j^g}{m_i^B}\right)\left(\frac{m_i^B}{m_{\tilde{d}_2}}\right)^6 m_j^g,
\end{equation}
where the $g$ subscript on $\Gamma$ means that only loop diagrams containing gluinos are taken into account and:
\begin{equation}\label{BG:Eq:SinPhiPrimeij}
  \sin\phi'_{ij}=\frac{\text{Im}\{a_i^{*2}c_j^2 \}}{|a_i c_j|^2}.
\end{equation}
We mention that, if binos and gluinos are close in mass, it is possible that certain combinations of binos and gluinos do not lead to any baryon asymmetry. Also, the argument about requiring both Majorana masses to be non-zero for binos does not hold in this case.

Finally, we define bino decay asymmetries as:
\begin{equation}\label{}
	\epsilon^{CP}_{\chi_i^B}=\frac{\Delta \Gamma^B_{\chi_i^B}+\Delta \Gamma^g_{\chi_i^B}}{\Gamma^{\text{Total}}_{\chi_i^B}}.
\end{equation}

\subsection{Annihilation and conversion cross sections}\label{BG:Sec:AnnihilationCS}
We now discuss the annihilation and conversion cross sections of binos. These enter the Boltzmann equations which will be used to calculate the would-be relic density of the binos. The two most important interactions are those with Higgses and quarks.

\subsubsection{Interactions with Higgses}
\begin{figure}
  \centering
    \begin{subfigure}[b]{0.5\textwidth}
      \centering
      \includegraphics[width=0.7\textwidth, bb = 0 0 222 87]{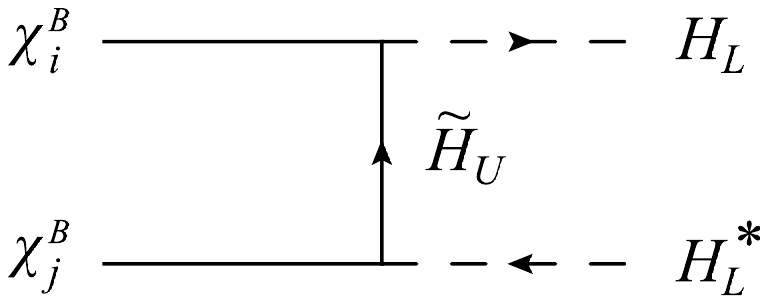}
      \caption{}
      \label{BG:Fig:BBHH}
    \end{subfigure}%
  ~
    \begin{subfigure}[b]{0.5\textwidth}
      \centering    
      \includegraphics[width=0.7\textwidth, bb = 0 0 218 90]{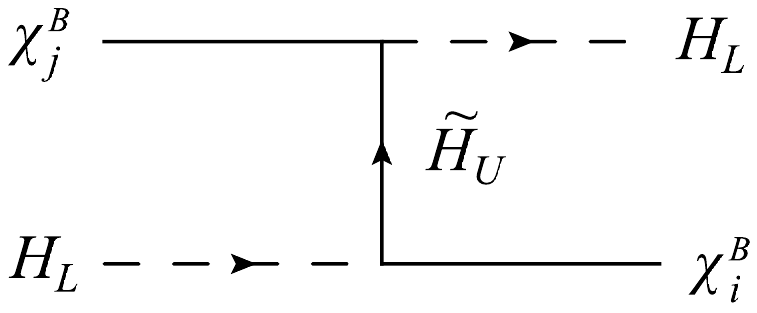}
      \caption{}  
      \label{BG:Fig:BHBH}
    \end{subfigure}
  \caption{(a) Example of diagram contributing to bino annihilation to Higgses. (b) Example of diagram contributing to bino conversion via Higgs scattering.}\label{BG:Fig:2Bscattering}
\end{figure}

An example of annihilation to two Higgses is shown in figure \ref{BG:Fig:BBHH}.  The total cross section is given by:
\begin{equation}\label{BG:Eq:CrossSectionBiBjHLHL}
  \sigma_{\chi^B_{i}\chi^B_{j}\to H_L H_L^*}(s)=\frac{1}{32\pi}\frac{|C_{ij}|^2(s-(m_i^B)^2-(m_j^B)^2)-2\text{Re}\{C_{ij}^2\}m_i^B m_j^B}{\sqrt{(s-(m_i^B+m_j^B)^2)(s-(m_i^B-m_j^B)^2)}},
\end{equation}
where $\sqrt{s}$ is the centre of mass energy and the cross section, like every other in this section, is averaged over all incoming degrees of freedom. As we will be interested in binos of around a TeV or heavier and that decouple at very small $x$, freeze-out will typically take place before the electroweak phase transition. As such, $H_L$ is treated as a complex scalar doublet for calculating relic densities, i.e. no particle has been ``eaten" yet by gauge bosons. 

In addition to annihilation, one must also take into account conversion via scattering. An example is shown in figure \ref{BG:Fig:BHBH}. The associated cross section is:
\begin{equation}\label{BG:Eq:CrossSectionBiHLBjHlsubleading}
  \sigma_{\chi^B_{j}H_L\to \chi^B_{i} H_L}=\frac{1}{64\pi s}\left(\frac{s-(m_i^B)^2}{s-(m_j^B)^2}\right)\left[|C_{ij}|^2\frac{(s+(m_i^B)^2)(s+(m_j^B)^2)}{s}+4 m_i^B m_j^B \text{Re}\{C_{ij}^2\}\right].
\end{equation}

\subsubsection{Interactions with quarks}
Annihilation and conversion can also take place via interactions with quarks. These interactions can either conserve baryon number or break it.

\begin{figure}
  \centering
    \begin{subfigure}[b]{0.5\textwidth}
      \centering
      \includegraphics[width=0.7\textwidth, bb = 0 0 213 126]{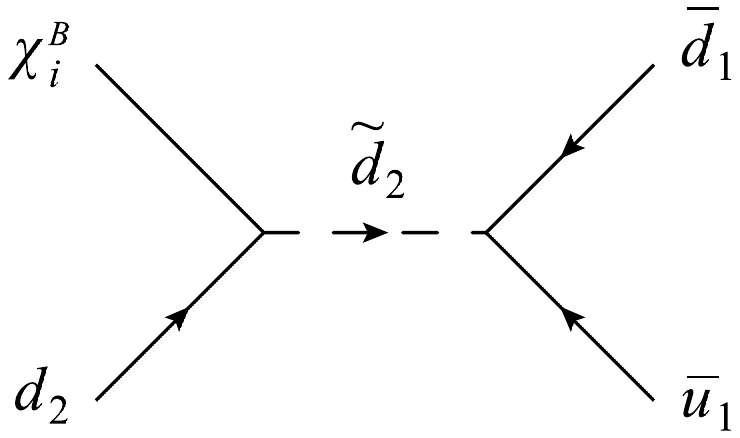}
      \caption{}  
      \label{BG:Fig:Single1}
    \end{subfigure}%
    ~
    \begin{subfigure}[b]{0.5\textwidth}
      \centering    
      \includegraphics[width=0.7\textwidth,bb=0 0 260 208]{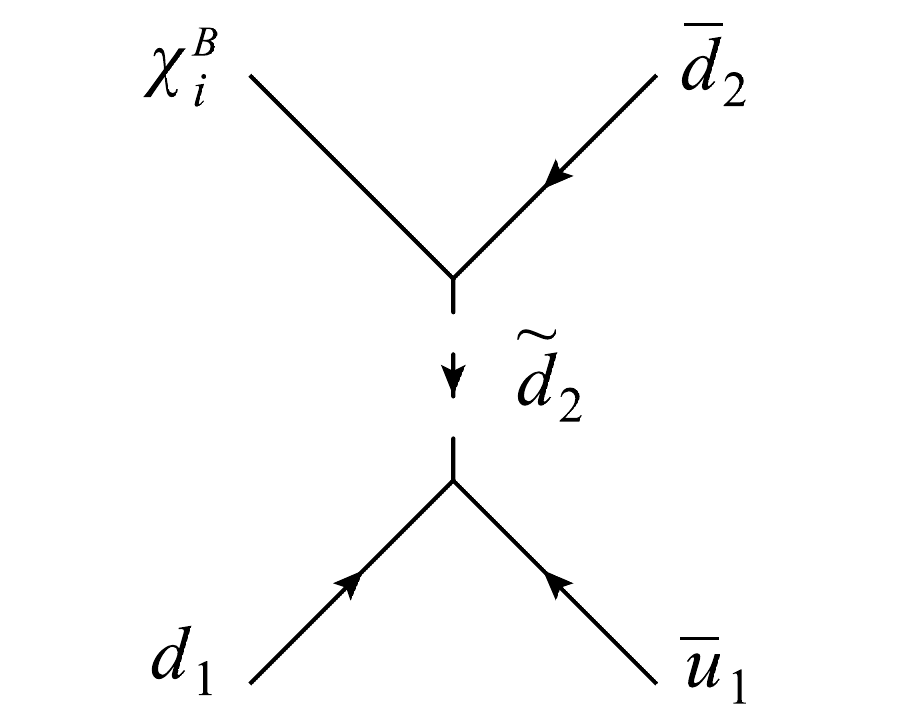}
      \caption{}  
      \label{BG:Fig:Single3}
    \end{subfigure}
  \caption{(a) Example of $s$-channel diagram contributing to annihilation of a single bino. (b) Example of $t$-channel diagram contributing to annihilation of a single bino.}\label{BG:Fig:SingleAnnihilation}
\end{figure}

For $\lambda''$ of $\mathcal{O}(0.1)$ or larger and heavy Higgsinos, baryon number breaking annihilation is expected to dominate because of the large multiplicity and lack of $p$-wave suppression \cite{Cui:2013bta}. Examples of these interactions are shown in figure \ref{BG:Fig:SingleAnnihilation}. The cross section is given by:
\begin{equation}\label{BG:Eq:CrossSectionBiqqq}
	\sigma_{\chi^B_{i}q\to q q}(s)=\frac{g'^2Y_d^2|a_i\lambda''|^2}{48\pi m_{\tilde{d}_2}^4}(5s + (m_i^B)^2).
\end{equation}

\begin{figure}
  \centering
    \begin{subfigure}[b]{0.5\textwidth}
      \centering
      \includegraphics[width=0.7\textwidth, bb = 0 0 214 87]{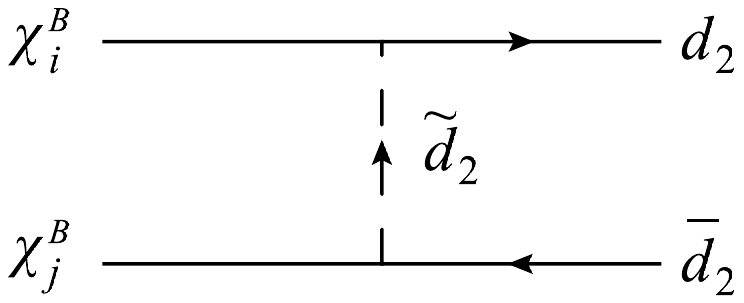}
      \caption{}  
      \label{BG:Fig:BBqq}
    \end{subfigure}%
    ~
    \begin{subfigure}[b]{0.5\textwidth}
      \centering     
      \includegraphics[width=0.7\textwidth, bb = 0 0 220 87]{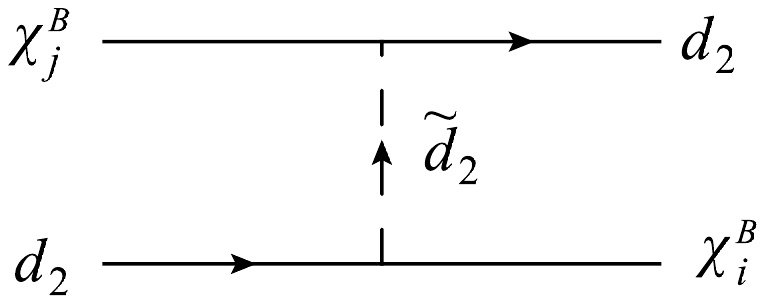}
      \caption{}  
      \label{BG:Fig:BqBq}
    \end{subfigure}
  \caption{(a) Example of diagram contributing to bino pair annihilation to $d_2$ $\overline{d}_2$. (b) Example of a diagram contributing to bino conversion via scattering off quarks.}\label{BG:Fig:2Bscatteringquarks}
\end{figure}

Other subleading interactions with quarks exist that preserve baryon number. These effects are usually subdominant. Annihilation of binos to quarks is shown in figure \ref{BG:Fig:BBqq}. The associated cross section is:
\begin{equation}\label{Bg:Eq:CrossSectionBiBjbbbar}
  \begin{aligned}
  &\sigma_{\chi^B_{i}\chi^B_{j}\to d_2\overline{d}_2}(s)=\frac{g'^4 Y_d^4}{16 \pi m_{\tilde{d}_2}^4}\frac{1}{\sqrt{(s-(m_i^B+m_j^B)^2)(s-(m_i^B-m_j^B)^2)}}\\
  &\hspace{1cm}\left[|a_i a_j|^2(2s^2-((m_i^B)^2+(m_j^B)^2)s-((m_i^B)^2-(m_j^B)^2)^2)-6\text{Re}\{a_i^2 a_j^{*2}\}m_i^B m_j^B s\right].
  \end{aligned}
\end{equation}
In addition, conversion can take place via diagrams like the one of figure \ref{BG:Fig:BqBq}. The associated cross section is:
\begin{equation}\label{BG:Eq:BiqBjq}
  \begin{aligned}
 &\sigma_{\chi^B_{j}d_2\to \chi^B_{i}d_2}(s)=\frac{g'^4 Y_d^4}{96\pi m_{\tilde{d}_2}^4}\frac{(s-(m_i^B)^2)^2}{s^3}\left[|a_i a_j|^2(8s^2+((m_i^B)^2+(m_j^B)^2)s+2(m_i^B)^2 (m_j^B)^2)\right.\\
  &\hspace{6.7cm}\left.+6\text{Re}\{a_i^2 a_j^{*2}\}m_i^B m_j^B s\right].
  \end{aligned}
\end{equation}

\begin{figure}
  \centering
    \begin{subfigure}[b]{0.5\textwidth}
      \centering
      \includegraphics[width=0.7\textwidth, bb = 0 0 214 87]{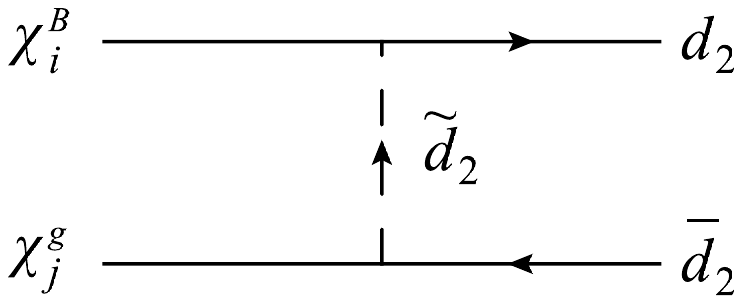}
      \caption{}  
      \label{BG:Fig:Bgqq}
    \end{subfigure}%
    ~
    \begin{subfigure}[b]{0.5\textwidth}
      \centering     
      \includegraphics[width=0.7\textwidth, bb = 0 0 220 87]{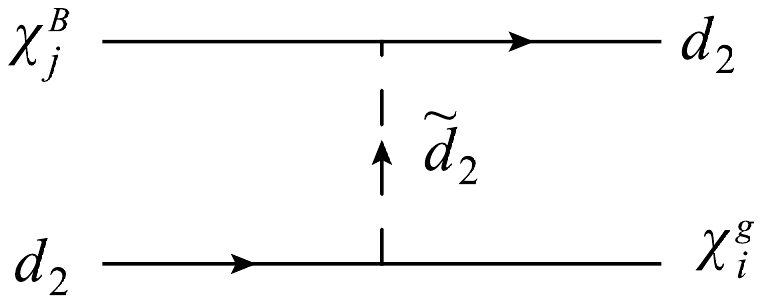}
      \caption{}  
      \label{BG:Fig:Bqgq}
    \end{subfigure}
  \caption{(a) Example of diagram contributing to bino-gluino annihilation to $d_2$ $\overline{d}_2$. (b) Example of diagram contributing to bino conversion to gluino via scattering off quarks.}\label{BG:Fig:Bgscatteringquarks}
\end{figure}

Similar processes exist involving gluinos. Cohannihilation is shown in figure \ref{BG:Fig:Bgqq} and corresponds to a cross section of:
\begin{equation}\label{Bg:Eq:CrossSectionBigjbbbar}
  \begin{aligned}
    &\sigma_{\chi^B_{i}\chi^g_{j}\to d_2\overline{d}_2}(s)=\frac{g'^2g_s^2 Y_d^2}{24 \pi m_{\tilde{d}_2}^4}\frac{1}{\sqrt{(s-(m_i^B+m_j^g)^2)(s-(m_i^B-m_j^g)^2)}}\\
    &\hspace{1cm}\left[|a_i c_j|^2(2s^2-((m_i^B)^2+(m_j^g)^2)s-((m_i^B)^2-(m_j^g)^2)^2)-6\text{Re}\{a_i^2 c_j^{*2}\}m_i^B m_j^g s\right].
  \end{aligned}
\end{equation}
Conversion is shown in figure \ref{BG:Fig:Bqgq} and corresponds to a cross section of:
\begin{equation}\label{BG:Eq:Biqgjq}
  \begin{aligned}
  &\sigma_{\chi^B_{j}d_2\to \chi^g_{i}d_2}(s)=\frac{g'^2g_s^2 Y_d^2}{18\pi m_{\tilde{d}_2}^4}\frac{(s-(m_j^B)^2)^2}{s^3}\left[|a_j c_i|^2(8s^2+((m_j^B)^2+(m_i^g)^2)s+2(m_j^B)^2 (m_i^g)^2)\right.\\
  &\hspace{6.6cm}\left.+6\text{Re}\{a_j^2 c_i^{*2}\}m_j^B m_i^g s\right].
  \end{aligned}
\end{equation}

\subsection{Calculation of the baryon relic density}
To obtain estimates of the baryon relic density $\Omega_{\Delta B}$, we calculate would-be relic densities of binos. For this, we use Boltzmann's equation conveniently rewritten as \cite{Edsjo:1997bg} (see also Ref$.$ \cite{Arcadi:2015ffa}):
\begin{equation}\label{Bg:Eq:Boltzmann}
  \begin{aligned}
    \frac{dY_i}{dx} &=-\sqrt{\frac{ g_*\pi}{45 G}}\frac{m_2^B}{x^2}\left[\langle \sigma_{iX}v_{iX}\rangle(Y_i-Y_i^{\text{eq}})Y_X^{\text{eq}}+\sum_{j=1}^2 \langle \sigma_{ij}v_{ij}\rangle(Y_i Y_j-Y_i^{\text{eq}}Y_j^{\text{eq}})\right.\\
     &\hspace{3.0cm}\left. -\sum_{j=1}^2\langle \sigma_{jX}v_{jX}\rangle\left(Y_j-\frac{Y_j^{\text{eq}}}{Y_i^{\text{eq}}}Y_i\right)Y_X^{\text{eq}}\right].
  \end{aligned}
\end{equation}
The parameter $Y_i$ is given by $Y_i=n_i/s$, where $n_i$ is the number density of particle $i$ and $s$ the entropy per comoving volume (not to be confused with the square root of the centre of mass energy). The parameter $g_*$ corresponds to the number of relativistic degrees of freedom. As we deal with particles with masses of the order of a few TeV and which decouple at small $x$, $g_*$ can safely be approximated by a constant $g_*\approx 106.75$. The parameter $Y_i^{\text{eq}}$ represents the equilibrium value of $Y_i$ and is given by \cite{Edsjo:1997bg}:
\begin{equation}\label{BG:Eq:Yeq}
  Y_i^{\text{eq}}=\frac{45x^2}{4\pi^4 g_*}g_i \left(\frac{m_i}{m_2^B}\right)^2 K_2\left(x\frac{m_i}{m_2^B} \right),
\end{equation}
where $g_i$ is the number of degrees of freedom of the particle $i$ and $K_i(x)$ is a modified Bessel functions of the second kind. The $\langle \sigma_{ij}v_{ij}\rangle$, $\langle \sigma_{iX}v_{iX}\rangle$ and $\langle \sigma_{jX}v_{jX}\rangle$ represent thermally averaged cross sections and can be obtained by combining the results of section \ref{BG:Sec:AnnihilationCS} with the following eq$.$ \cite{Edsjo:1997bg}:
\begin{equation}\label{BG:Eq:ThermallyAveragedCrossSection}
  \langle \sigma_{ij}v_{ij}\rangle=\frac{\int_{(m_i+m_j)^2}^{\infty} \frac{1}{\sqrt{s}} (s-(m_i+m_j)^2)(s-(m_i-m_j)^2)K_1\left(\frac{\sqrt{s}}{T}\right)\sigma_{ij}(s)ds}{8T m_i^2 m_j^2 K_2\left(\frac{m_i}{T}\right)K_2\left(\frac{m_j}{T}\right)},
\end{equation}
where the $i$ and $j$ indices can represent any particle. Annihilation of two binos contributes to $\langle \sigma_{ij}v_{ij}\rangle$, annihilation of a single bino contributes to $\langle \sigma_{iX}v_{iX}\rangle$ and conversion contributes to $\langle \sigma_{jX}v_{jX}\rangle$. 

Since gluinos can annihilate via diagrams that only involve themselves and gluons, they will remain in equilibrium until far later than the binos have decoupled. Therefore, we approximate gluino densities by their equilibrium values when relevant. We also note that neutralinos can potentially decay to the heaviest gluino and that the latter can present an asymmetry in its decay to baryons and antibaryons. We verified that this decay would generally take place long before the gluinos have decoupled and therefore should not contribute to the baryon relic density. We therefore do not consider this contribution.

Once the would-be relic densities of binos are obtained, $\Omega_{\Delta B}$ is approximated by:
\begin{equation}\label{BG:Eq:OmegaDeltaB}
  \Omega_{\Delta B}=\left.\frac{m_p}{(\rho_c/s)_0}\left(\epsilon^{CP}_{\chi_1^B} Y_{\chi_1^B}+\epsilon^{CP}_{\chi_2^B} Y_{\chi_2^B}\right)\right|^{t\to\infty},
\end{equation}
where $m_p$ is the mass of the proton and $(\rho_c/s)_0$ the current ratio of critical density to entropy density. This is a good approximation as long as the decay temperatures of the binos are considerably lower than their freeze-out temperatures. Also, note that baryon number breaking interactions that take place before freeze-out can lead to an additional source of baryon asymmetry. This was studied in Ref$.$ \cite{Arcadi:2015ffa} and found to be negligible because of washout effects. 

\subsection{Results and constraints}\label{BG:Sec:Result}
We now proceed to discuss the relevant constraints and provide a few benchmark plots to illustrate different features. We stress that we do not aim to cover the full parameter space, but to show that it is indeed possible to obtain a baryon density compatible with the observed value of $\Omega_{\Delta B} \sim 0.05$ \cite{Ade:2015xua} .

We first make a few simplifying assumptions out of convenience. We relate parameters by setting $\mu_u=\mu_d\equiv\mu$ and $\lambda_u^s=-\lambda_d^s\equiv\lambda^s$. For decoupled gluinos, we set $\lambda''=0.2$, which is chosen to maximize $\Omega_{\Delta B}$. It is large enough for the $\epsilon^{CP}$'s not to be suppressed by large decay branching ratios to other channels, while not being so large as to make $B$-breaking scattering with quarks too strong. For light gluinos, we instead set it to $\lambda''=1$.

\begin{figure}[t!] 
  \centering 
  \includegraphics[width=0.6\textwidth, bb = 0 0 360 351]{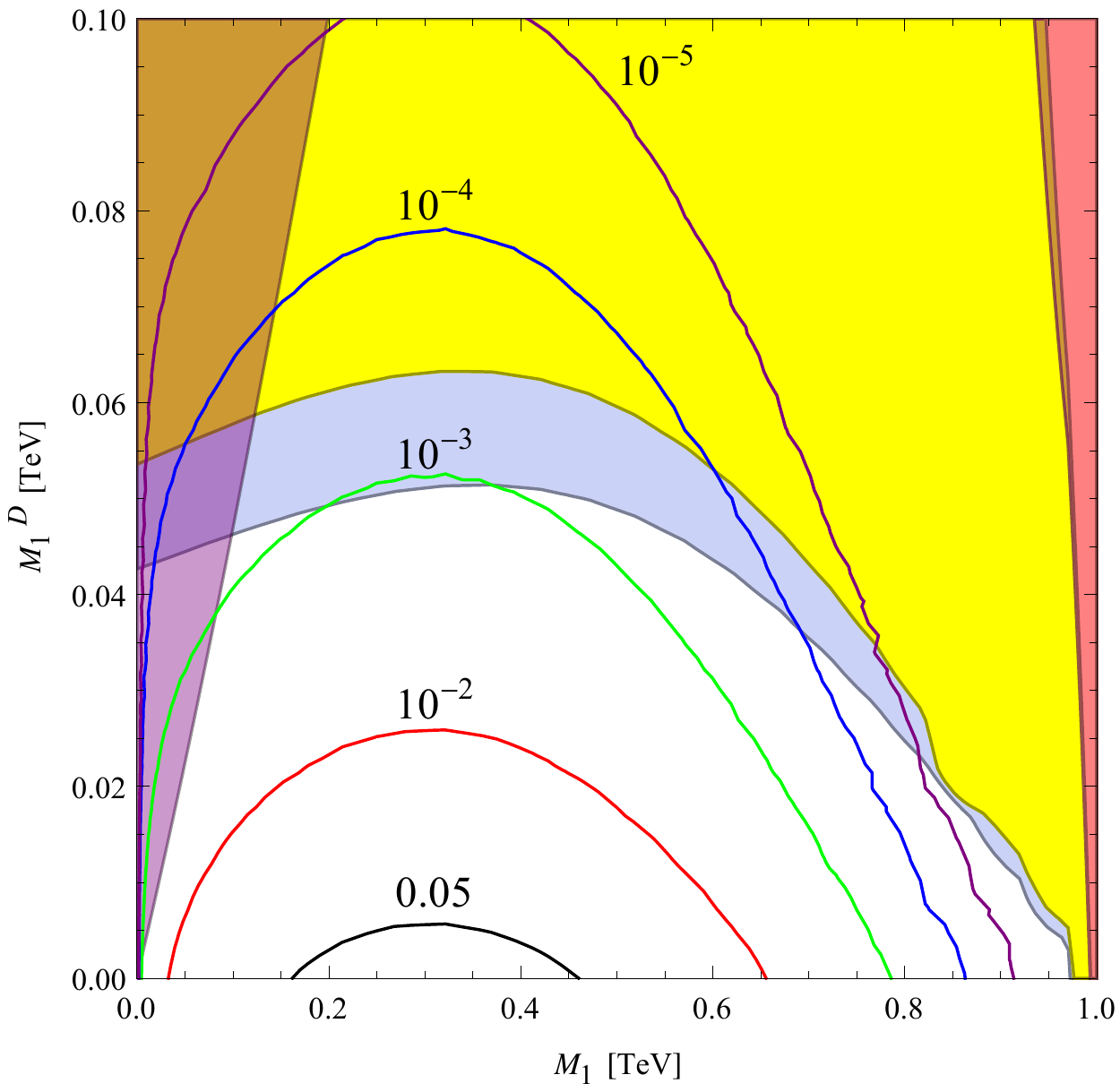} 
  \caption{Contour plots of constant $\Omega_{\Delta B}$ for decoupled Higgsinos and gluinos. The blue region corresponds to $\chi_2^B$ decaying before electroweak phase transition. The yellow region corresponds to $\chi_2^B$ decaying before freeze-out. The purple and pink regions correspond to $\chi_2^B$ having a decay temperature superior to $m_1^B$ and $m_2^B$ respectively.}
  \label{BG:Fig:DeltaBcontours}
\end{figure}

Figure \ref{BG:Fig:DeltaBcontours} shows $\Omega_{\Delta B}$ as a function of $M_1$ and $M_1^D$ for heavy Higgsinos and gluinos. The mass $\rho_1$ is set to $1 \times \text{exp}(3i/4\pi)$ TeV and $m_{\tilde{d}_2}$ to 50 TeV. As can be seen, it is possible to obtain a sufficient baryon relic density, but it requires the $U(1)_R$ breaking to be large. We see that $\Omega_{\Delta B}$ peaks in a region where $\chi_2^B$ is very close to being a pure singlino. In this limit, $\chi_2^B$ can easily decouple early when the Higgsinos are heavy as interactions with quarks are suppressed. It is then not necessary to have the squarks as heavy for it to decouple early and $\epsilon^{CP}$ does not need to be as suppressed. We note that $\Omega_{\Delta B}$ is optimized when $m_1^B/m_2^B$ is close to 0.25. This corresponds to the maximum of $x f(x)$, which controls the asymmetry (see eq$.$ (\ref{BG:Eq:DecayWidthBiNetBino})).

\begin{figure}[t!] 
  \centering 
  \includegraphics[width=0.6\textwidth, bb = 0 0 360 361]{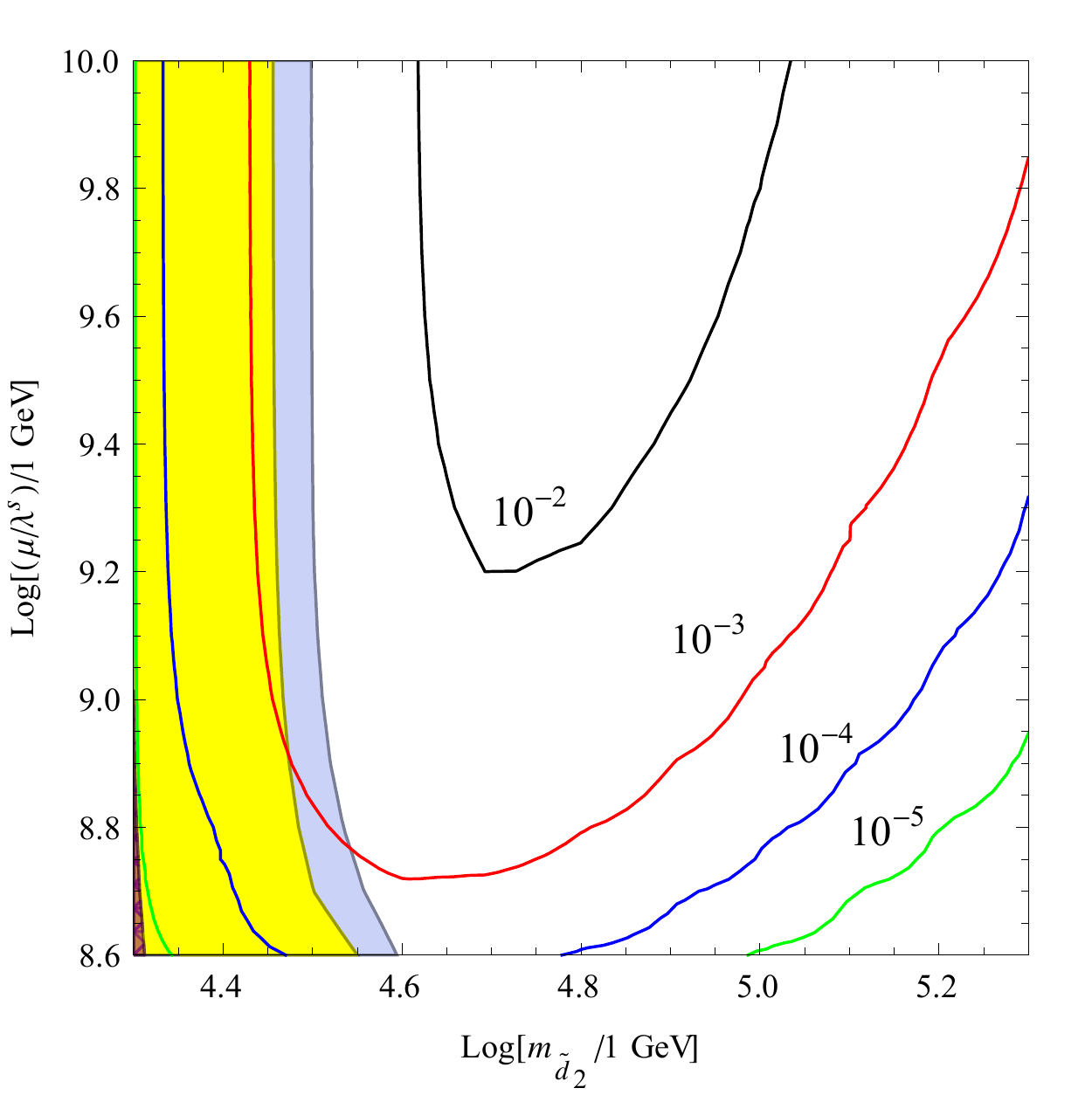} 
  \caption{Contour plots of constant $\Omega_{\Delta B}$ for decoupled gluinos. The blue region corresponds to $\chi_2^B$ decaying before electroweak phase transition. The yellow region corresponds to $\chi_2^B$ decaying before freeze-out. The purple region corresponds to $\chi_2^B$ having a decay temperature inferior to $m_1^B$.}
  \label{BG:Fig:DeltaBcontours2}
\end{figure}

Figure \ref{BG:Fig:DeltaBcontours2} shows $\Omega_{\Delta B}$ as a function of $\mu/\lambda^s$ and $m_{\tilde{d}_2}$ for heavy gluinos. The masses $M_1^D$, $M_1$ and $\rho_1$ are set respectively to 0.02 TeV, 0.25 TeV and $1 \times \text{exp}(3i/4\pi)$ TeV. Obviously, Higgsinos can be made lighter by taking $\lambda^s$ small but eventually the subleading corrections inversely proportional to $\mu^4$ would become non-negligible. In addition, this shows that the requirement of correct $\Omega_{\Delta B}$ indeed leads to a Mini-Split spectrum.

\begin{figure}[t!] 
  \centering 
  \includegraphics[width=0.6\textwidth, bb = 0 0 360 351]{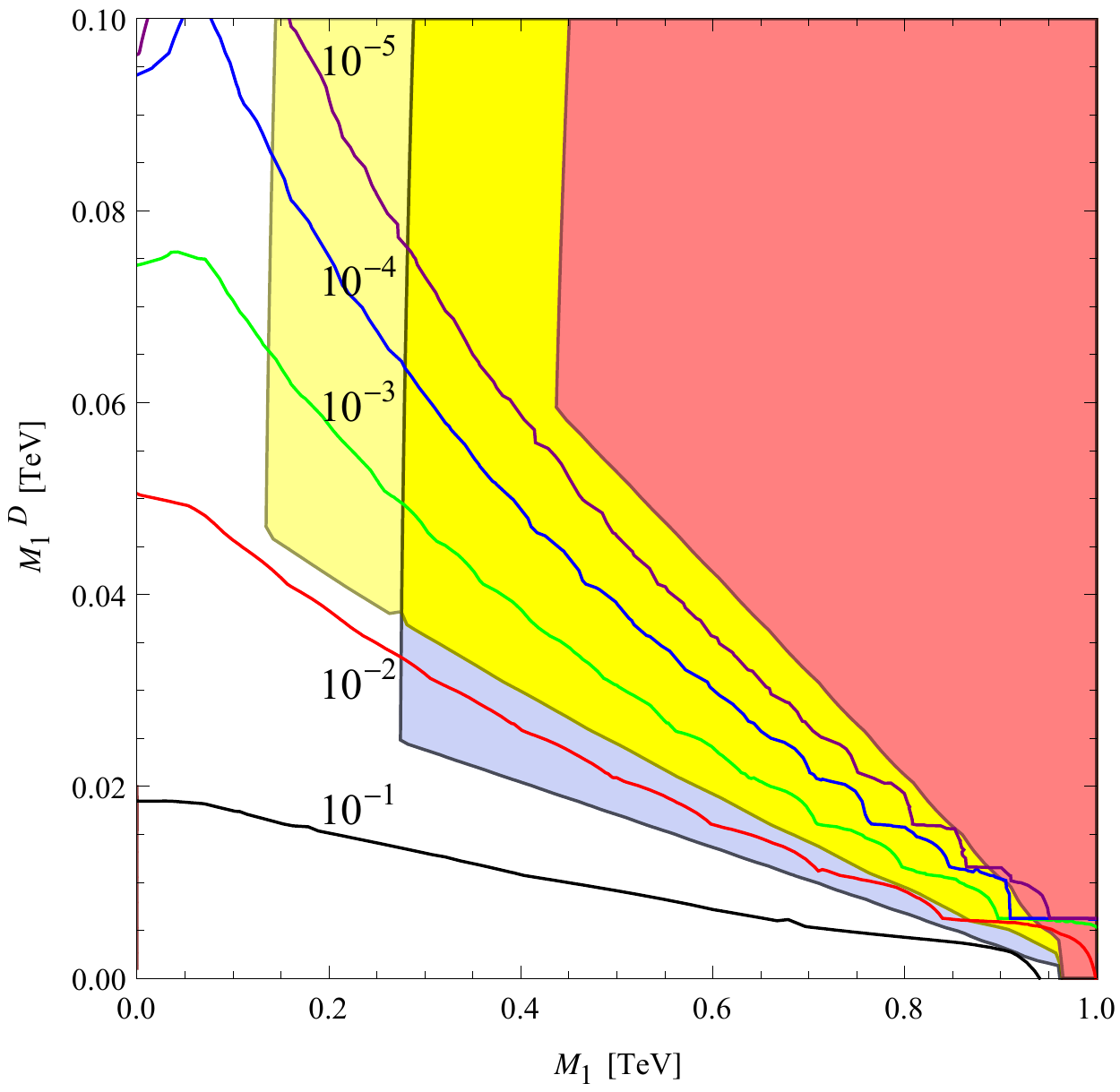} 
  \caption{Contour plots of constant $\Omega_{\Delta B}$ for decoupled Higgsinos. The blue region corresponds to both binos decaying before electroweak phase transition. The yellow region corresponds to both binos decaying before they have decoupled. The pink region corresponds to all binos having a decay temperature superior to at least one  bino or gluino mass.}
	\label{BG:Fig:LightGluinos}
\end{figure}

Figure \ref{BG:Fig:LightGluinos} shows $\Omega_{\Delta B}$ as a function of $M_1$ and $M_1^D$ for decoupled Higgsinos and light gluinos. The mass $\rho_1$ is set to 1 TeV. The gluino Dirac masses is set to 0.5 TeV and the Majorana masses are set to $M_3=0.5 \times \text{exp}(3i/4\pi)$ TeV and $\rho_3=0$ TeV. The mass of $\tilde{d}_2$ is set to 100 TeV. We observe that it is possible to obtain the correct baryon relic density, but that it once again requires the $U(1)_R$ symmetry to be badly broken.

A few additional constraints are also taken into account. The first one concerns washout. The relevant washout processes are carefully discussed in Ref$.$ \cite{Cui:2012jh}. They include inverse decay via on-shell squark and $u_1 d_1 d_2 \to \overline{u}_1 \overline{d}_1 \overline{d}_2$ just to name a few. In the decoupled gluino case, the end result is that these processes are suppressed as long as the decay temperature of $\chi_2^B$ is lower than the masses of the binos. In the case of light gluinos, these can also mediate baryon number breaking interactions and as such we require that the decay temperature of at least one of the binos be smaller than the masses of any of the gluinos and binos.

The second constraint is for the binos to decay after their freeze-out. This constraint is not absolute, as decays that take place slightly before still lead to a relic baryon density, albeit suppressed. We estimate the freeze-out temperature of the binos by taking their relic number density, setting them equal to their equilibrium density and solving for $x$. This defines a freeze-out temperature for each bino. For decoupled gluinos, we then include in the figures the region where $\chi_2^B$ decays before freeze-out. For light gluinos, we include the region where both binos decay before decoupling. Generally speaking, these constraints are far more important than washout.

Finally, if the binos were to decay before the electroweak phase transition, some of the baryon relic density would be converted away by sphaleron effects. This would reduce the baryon relic density by a factor of $28/79$, which is sizable but does not change the qualitative features \cite{Chen:2007fv}. It also corresponds in our plots to a region that does not produce enough baryon relic density. We take the electroweak phase transition to take place at 100 GeV. For decoupled gluinos, we include the regions where the decay of $\chi_2^B$ takes place before the electroweak phase transition. For light gluinos, we show the region where both binos decay before electroweak phase transition.

Another effect to consider is the possibility of entropy dilution. Ref$.$ \cite{Arcadi:2015ffa} studied this and found that it is only relevant for very large scalar masses where the bino decouples while still relativistic. In the case of heavy gluinos, this would lead to a suppression of $\Omega_{\Delta B}$ by a dilution factor of \cite{Arcadi:2015ffa,Baldes:2014rda}:
\begin{equation}\label{BG:Eq:EntropyDilution}
  \xi_s=\text{Max}\left[1,1.8g_*^{1/4}\frac{Y_2(x_{f.o.})m_2^B}{\sqrt{\Gamma_{\chi^B_{2}}^{\text{total}}M_{\text{Pl}}}}\right],
\end{equation}
where $\Gamma_{\chi^B_{2}}^{\text{total}}$ is the value of $Y_2(x_{f.o.})$ when $\chi_2^B$ freezes-out and $M_{\text{Pl}}$ the Planck scale. We have verified that this factor is simply one over all the region shown in our plots. Similar results hold for both binos in the light gluinos case.

\section{Conclusions}
In this paper, we studied the LHC phenomenology of a supersymmetric model with a $U(1)_R$ symmetry which is identified with the baryon number. We also examined how baryogenesis could be realized in this model through the late decay of a neutral gaugino. The model we considered is an extension of the MRSSM with the inclusion of an $R$-parity breaking term of the form $\lambda_{ijk}'' U_i^c D_j^c D_k^c$. Because of the non-standard baryon number assignment of the superpartners, such a term is baryon number conserving in this model. This relaxes the bounds on the $\lambda''$ couplings significantly compared to the RPVMSSM. In particular, the bounds from neutron-antineutron oscillation and from double nucleon decay are considerably loosened. However, they cannot be removed completely as the $U(1)_R$ will be broken by the gravitino mass and communicated to the superpartners of the Standard Model by anomaly mediation. Furthermore, the gravitino must be heavier than the proton to avoid proton decay to a gravitino and a kaon. Flavour physics also puts bounds on products of $\lambda''$ couplings which are the same in our model as in the RPVMSSM. 

The introduction of large $\lambda''$ couplings leads to a collider phenomenology that is significantly different from the MSSM and from the RPVMSSM with very small $\lambda''$ couplings. We examined simplified models where a single one of the $\lambda''$ involving the third generation is large. We looked at both single and pair production of stops and their subsequent decays for bino or Higgsino-up LSP.  When Majorana mass terms are included, a Dirac neutralino splits into two states close in mass. We showed that when this mass splitting is larger than the width of the neutralino, the stop can decay via this neutralino to two same-sign tops. On the other hand when the mass splitting is smaller, the branching ratio of a stop decaying to two same-sign tops is highly suppressed. Because same sign leptons are a powerful tool to reject background, the two cases present different phenomenology and in this work we presented results for both hypotheses. We also presented limits on the masses of the first and second generation of squarks as their production cross section is altered in models with Dirac gluinos.

We note that the structure of these models is quite rich and we did not explore the complete phenomenology of all sectors of the theory. For example, the model has extra scalars as part of the adjoint chiral superfields. Some of these fields could in fact be responsible for obtaining the correct Higgs mass in these models \cite{Fok:2012fb, Bertuzzo:2014bwa, Benakli:2012cy, Diessner:2014ksa}.

The structure of the model also allows for baryogenesis to proceed through late decay of neutralinos. Because of the extra field content, new diagrams contribute to these decays and allow for the conditions of the Nanopolous-Weinberg theorem to be met. The results of the analysis of baryogenesis are presented in section \ref{Sec:Baryogenesis}. We find that with a Mini-Split spectrum, where the scalars are heavier than the gauginos, successful baryogenesis can be achieved. Such mechanism also exists in the RPVMSSM with a split spectrum, but the extended Higgs sector present in our model allows for the higgsinos to be lighter. The mechanism however requires a large breaking of the $U(1)_R$ symmetry. 

\acknowledgments
This work was supported in part by the Natural Sciences and Engineering Research Council of Canada (NSERC). HB acknowledges support from the Ontario Graduate Scholarship (OGS) and from FAPESP. KE acknowledges support from the Alexander Graham Bell Canada Graduate Scholarships Doctoral Program (NSERC CGS D). We would like to thank David London for discussions and collaboration at the early stage of the project. We would also like to thank Alejandro de la Puente for discussions about displaced vertices.

\bibliographystyle{JHEP}
\bibliography{Paper1}

\providecommand{\href}[2]{#2}\begingroup\raggedright\begin{thebibliography}{100}

\bibitem{Fayet:1978qc}
P.~Fayet, {\it {MASSIVE GLUINOS}},  {\em Phys. Lett.} {\bf B78} (1978)
  417--420.

\bibitem{Hall:1990hq}
L.~J. Hall and L.~Randall, {\it {U(1)-R symmetric supersymmetry}},  {\em Nucl.
  Phys.} {\bf B352} (1991) 289--308.

\bibitem{Fox:2002bu}
P.~J. Fox, A.~E. Nelson, and N.~Weiner, {\it {Dirac gaugino masses and
  supersoft supersymmetry breaking}},  {\em JHEP} {\bf 08} (2002) 035,
  [\href{http://arxiv.org/abs/hep-ph/0206096}{{\tt hep-ph/0206096}}].

\bibitem{Nelson:2002ca}
A.~E. Nelson, N.~Rius, V.~Sanz, and M.~Unsal, {\it {The Minimal supersymmetric
  model without a mu term}},  {\em JHEP} {\bf 08} (2002) 039,
  [\href{http://arxiv.org/abs/hep-ph/0206102}{{\tt hep-ph/0206102}}].

\bibitem{Kribs:2007ac}
G.~D. Kribs, E.~Poppitz, and N.~Weiner, {\it {Flavor in supersymmetry with an
  extended R-symmetry}},  {\em Phys. Rev.} {\bf D78} (2008) 055010,
  [\href{http://arxiv.org/abs/0712.2039}{{\tt arXiv:0712.2039}}].

\bibitem{Amigo:2008rc}
S.~D.~L. Amigo, A.~E. Blechman, P.~J. Fox, and E.~Poppitz, {\it {R-symmetric
  gauge mediation}},  {\em JHEP} {\bf 01} (2009) 018,
  [\href{http://arxiv.org/abs/0809.1112}{{\tt arXiv:0809.1112}}].

\bibitem{Benakli:2008pg}
K.~Benakli and M.~D. Goodsell, {\it {Dirac Gauginos in General Gauge
  Mediation}},  {\em Nucl. Phys.} {\bf B816} (2009) 185--203,
  [\href{http://arxiv.org/abs/0811.4409}{{\tt arXiv:0811.4409}}].

\bibitem{Benakli:2010gi}
K.~Benakli and M.~D. Goodsell, {\it {Dirac Gauginos, Gauge Mediation and
  Unification}},  {\em Nucl. Phys.} {\bf B840} (2010) 1--28,
  [\href{http://arxiv.org/abs/1003.4957}{{\tt arXiv:1003.4957}}].

\bibitem{Kribs:2010md}
G.~D. Kribs, T.~Okui, and T.~S. Roy, {\it {Viable Gravity-Mediated
  Supersymmetry Breaking}},  {\em Phys. Rev.} {\bf D82} (2010) 115010,
  [\href{http://arxiv.org/abs/1008.1798}{{\tt arXiv:1008.1798}}].

\bibitem{Abel:2011dc}
S.~Abel and M.~Goodsell, {\it {Easy Dirac Gauginos}},  {\em JHEP} {\bf 06}
  (2011) 064, [\href{http://arxiv.org/abs/1102.0014}{{\tt arXiv:1102.0014}}].

\bibitem{Csaki:2013fla}
C.~Csaki, J.~Goodman, R.~Pavesi, and Y.~Shirman, {\it {The $m_D-b_M$ problem of
  Dirac gauginos and its solutions}},  {\em Phys. Rev.} {\bf D89} (2014), no.~5
  055005, [\href{http://arxiv.org/abs/1310.4504}{{\tt arXiv:1310.4504}}].

\bibitem{Frugiuele:2011mh}
C.~Frugiuele and T.~Gregoire, {\it {Making the Sneutrino a Higgs with a
  $U(1)_R$ Lepton Number}},  {\em Phys. Rev.} {\bf D85} (2012) 015016,
  [\href{http://arxiv.org/abs/1107.4634}{{\tt arXiv:1107.4634}}].

\bibitem{Davies:2011mp}
R.~Davies, J.~March-Russell, and M.~McCullough, {\it {A Supersymmetric One
  Higgs Doublet Model}},  {\em JHEP} {\bf 04} (2011) 108,
  [\href{http://arxiv.org/abs/1103.1647}{{\tt arXiv:1103.1647}}].

\bibitem{Fok:2012fb}
R.~Fok, G.~D. Kribs, A.~Martin, and Y.~Tsai, {\it {Electroweak Baryogenesis in
  R-symmetric Supersymmetry}},  {\em Phys. Rev.} {\bf D87} (2013), no.~5
  055018, [\href{http://arxiv.org/abs/1208.2784}{{\tt arXiv:1208.2784}}].

\bibitem{Frugiuele:2012kp}
C.~Frugiuele, T.~Gregoire, P.~Kumar, and E.~Ponton, {\it {'L=R' -- $U(1)_R$
  Lepton Number at the LHC}},  {\em JHEP} {\bf 05} (2013) 012,
  [\href{http://arxiv.org/abs/1210.5257}{{\tt arXiv:1210.5257}}].

\bibitem{Frugiuele:2012pe}
C.~Frugiuele, T.~Gregoire, P.~Kumar, and E.~Ponton, {\it {'L=R' - $U(1)_R$ as
  the Origin of Leptonic 'RPV'}},  {\em JHEP} {\bf 03} (2013) 156,
  [\href{http://arxiv.org/abs/1210.0541}{{\tt arXiv:1210.0541}}].

\bibitem{Beauchesne:2014pra}
H.~Beauchesne and T.~Gregoire, {\it {Electroweak precision measurements in
  supersymmetric models with a U(1)$_R$ lepton number}},  {\em JHEP} {\bf 05}
  (2014) 051, [\href{http://arxiv.org/abs/1402.5403}{{\tt arXiv:1402.5403}}].

\bibitem{Bertuzzo:2014bwa}
E.~Bertuzzo, C.~Frugiuele, T.~Gregoire, and E.~Ponton, {\it {Dirac gauginos, R
  symmetry and the 125 GeV Higgs}},  {\em JHEP} {\bf 04} (2015) 089,
  [\href{http://arxiv.org/abs/1402.5432}{{\tt arXiv:1402.5432}}].

\bibitem{Carpenter:2015mna}
L.~M. Carpenter and J.~Goodman, {\it {New Calculations in Dirac Gaugino Models:
  Operators, Expansions, and Effects}},  {\em JHEP} {\bf 07} (2015) 107,
  [\href{http://arxiv.org/abs/1501.05653}{{\tt arXiv:1501.05653}}].

\bibitem{Itoyama:2011zi}
H.~Itoyama and N.~Maru, {\it {D-term Dynamical Supersymmetry Breaking
  Generating Split N=2 Gaugino Masses of Mixed Majorana-Dirac Type}},  {\em
  Int. J. Mod. Phys.} {\bf A27} (2012) 1250159,
  [\href{http://arxiv.org/abs/1109.2276}{{\tt arXiv:1109.2276}}].

\bibitem{Itoyama:2013sn}
H.~Itoyama and N.~Maru, {\it {D-term Triggered Dynamical Supersymmetry
  Breaking}},  {\em Phys. Rev.} {\bf D88} (2013), no.~2 025012,
  [\href{http://arxiv.org/abs/1301.7548}{{\tt arXiv:1301.7548}}].

\bibitem{Itoyama:2013vxa}
H.~Itoyama and N.~Maru, {\it {126 GeV Higgs Boson Associated with D-term
  Triggered Dynamical Supersymmetry Breaking}},  {\em Symmetry} {\bf 7} (2015),
  no.~1 193--205, [\href{http://arxiv.org/abs/1312.4157}{{\tt
  arXiv:1312.4157}}].

\bibitem{Heikinheimo:2011fk}
M.~Heikinheimo, M.~Kellerstein, and V.~Sanz, {\it {How Many Supersymmetries?}},
   {\em JHEP} {\bf 04} (2012) 043, [\href{http://arxiv.org/abs/1111.4322}{{\tt
  arXiv:1111.4322}}].

\bibitem{Kribs:2012gx}
G.~D. Kribs and A.~Martin, {\it {Supersoft Supersymmetry is Super-Safe}},  {\em
  Phys. Rev.} {\bf D85} (2012) 115014,
  [\href{http://arxiv.org/abs/1203.4821}{{\tt arXiv:1203.4821}}].

\bibitem{Kribs:2013eua}
G.~D. Kribs and N.~Raj, {\it {Mixed Gauginos Sending Mixed Messages to the
  LHC}},  {\em Phys. Rev.} {\bf D89} (2014), no.~5 055011,
  [\href{http://arxiv.org/abs/1307.7197}{{\tt arXiv:1307.7197}}].

\bibitem{Fok:2010vk}
R.~Fok and G.~D. Kribs, {\it {$\mu$ to e in R-symmetric Supersymmetry}},  {\em
  Phys. Rev.} {\bf D82} (2010) 035010,
  [\href{http://arxiv.org/abs/1004.0556}{{\tt arXiv:1004.0556}}].

\bibitem{Gherghetta:2003he}
T.~Gherghetta and A.~Pomarol, {\it {The Standard model partly supersymmetric}},
   {\em Phys. Rev.} {\bf D67} (2003) 085018,
  [\href{http://arxiv.org/abs/hep-ph/0302001}{{\tt hep-ph/0302001}}].

\bibitem{Riva:2012hz}
F.~Riva, C.~Biggio, and A.~Pomarol, {\it {Is the 125 GeV Higgs the superpartner
  of a neutrino?}},  {\em JHEP} {\bf 02} (2013) 081,
  [\href{http://arxiv.org/abs/1211.4526}{{\tt arXiv:1211.4526}}].

\bibitem{Biggio:2016sdu}
C.~Biggio, J.~A. Dror, Y.~Grossman, and W.~H. Ng, {\it {Probing a slepton Higgs
  on all frontiers}},  {\em JHEP} {\bf 04} (2016) 150,
  [\href{http://arxiv.org/abs/1602.02162}{{\tt arXiv:1602.02162}}].

\bibitem{Brust:2011tb}
C.~Brust, A.~Katz, S.~Lawrence, and R.~Sundrum, {\it {SUSY, the Third
  Generation and the LHC}},  {\em JHEP} {\bf 03} (2012) 103,
  [\href{http://arxiv.org/abs/1110.6670}{{\tt arXiv:1110.6670}}].

\bibitem{Sakharov:1967dj}
A.~D. Sakharov, {\it {Violation of CP Invariance, c Asymmetry, and Baryon
  Asymmetry of the Universe}},  {\em Pisma Zh. Eksp. Teor. Fiz.} {\bf 5} (1967)
  32--35. [Usp. Fiz. Nauk161,61(1991)].

\bibitem{Monteux:2016gag}
A.~Monteux, {\it {New signatures and limits on R-parity violation from resonant
  squark production}},  {\em JHEP} {\bf 03} (2016) 216,
  [\href{http://arxiv.org/abs/1601.03737}{{\tt arXiv:1601.03737}}].

\bibitem{Dreiner:1991pe}
H.~K. Dreiner and G.~G. Ross, {\it {R-parity violation at hadron colliders}},
  {\em Nucl. Phys.} {\bf B365} (1991) 597--613.

\bibitem{Allanach:2012vj}
B.~C. Allanach and B.~Gripaios, {\it {Hide and Seek With Natural Supersymmetry
  at the LHC}},  {\em JHEP} {\bf 05} (2012) 062,
  [\href{http://arxiv.org/abs/1202.6616}{{\tt arXiv:1202.6616}}].

\bibitem{Evans:2012bf}
J.~A. Evans and Y.~Kats, {\it {LHC Coverage of RPV MSSM with Light Stops}},
  {\em JHEP} {\bf 04} (2013) 028, [\href{http://arxiv.org/abs/1209.0764}{{\tt
  arXiv:1209.0764}}].

\bibitem{Bhattacherjee:2013gr}
B.~Bhattacherjee, J.~L. Evans, M.~Ibe, S.~Matsumoto, and T.~T. Yanagida, {\it
  {Natural supersymmetry?s last hope: R-parity violation via UDD operators}},
  {\em Phys. Rev.} {\bf D87} (2013), no.~11 115002,
  [\href{http://arxiv.org/abs/1301.2336}{{\tt arXiv:1301.2336}}].

\bibitem{Graham:2014vya}
P.~W. Graham, S.~Rajendran, and P.~Saraswat, {\it {Supersymmetric crevices:
  Missing signatures of R -parity violation at the LHC}},  {\em Phys. Rev.}
  {\bf D90} (2014), no.~7 075005, [\href{http://arxiv.org/abs/1403.7197}{{\tt
  arXiv:1403.7197}}].

\bibitem{Cui:2013bta}
Y.~Cui, {\it {Natural Baryogenesis from Unnatural Supersymmetry}},  {\em JHEP}
  {\bf 12} (2013) 067, [\href{http://arxiv.org/abs/1309.2952}{{\tt
  arXiv:1309.2952}}].

\bibitem{Cui:2012jh}
Y.~Cui and R.~Sundrum, {\it {Baryogenesis for weakly interacting massive
  particles}},  {\em Phys. Rev.} {\bf D87} (2013), no.~11 116013,
  [\href{http://arxiv.org/abs/1212.2973}{{\tt arXiv:1212.2973}}].

\bibitem{Arcadi:2015ffa}
G.~Arcadi, L.~Covi, and M.~Nardecchia, {\it {Gravitino Dark Matter and
  low-scale Baryogenesis}},  {\em Phys. Rev.} {\bf D92} (2015), no.~11 115006,
  [\href{http://arxiv.org/abs/1507.05584}{{\tt arXiv:1507.05584}}].

\bibitem{Arcadi:2013jza}
G.~Arcadi, L.~Covi, and M.~Nardecchia, {\it {Out-of-equilibrium baryogenesis
  and superweakly interacting massive particle dark matter}},  {\em Phys. Rev.}
  {\bf D89} (2014), no.~9 095020, [\href{http://arxiv.org/abs/1312.5703}{{\tt
  arXiv:1312.5703}}].

\bibitem{Randall:1998uk}
L.~Randall and R.~Sundrum, {\it {Out of this world supersymmetry breaking}},
  {\em Nucl.Phys.} {\bf B557} (1999) 79--118,
  [\href{http://arxiv.org/abs/hep-th/9810155}{{\tt hep-th/9810155}}].

\bibitem{Giudice:1998xp}
G.~F. Giudice, M.~A. Luty, H.~Murayama, and R.~Rattazzi, {\it {Gaugino mass
  without singlets}},  {\em JHEP} {\bf 9812} (1998) 027,
  [\href{http://arxiv.org/abs/hep-ph/9810442}{{\tt hep-ph/9810442}}].

\bibitem{Barbier:2004ez}
R.~Barbier, C.~Berat, M.~Besancon, M.~Chemtob, A.~Deandrea, et~al., {\it
  {R-parity violating supersymmetry}},  {\em Phys.Rept.} {\bf 420} (2005)
  1--202, [\href{http://arxiv.org/abs/hep-ph/0406039}{{\tt hep-ph/0406039}}].

\bibitem{Csaki:2011ge}
C.~Csaki, Y.~Grossman, and B.~Heidenreich, {\it {MFV SUSY: A Natural Theory for
  R-Parity Violation}},  {\em Phys. Rev.} {\bf D85} (2012) 095009,
  [\href{http://arxiv.org/abs/1111.1239}{{\tt arXiv:1111.1239}}].

\bibitem{Abe:2011ky}
{\bf Super-Kamiokande} Collaboration, K.~Abe et~al., {\it {The Search for
  $n-\bar{n}$ oscillation in Super-Kamiokande I}},  {\em Phys. Rev.} {\bf D91}
  (2015) 072006, [\href{http://arxiv.org/abs/1109.4227}{{\tt
  arXiv:1109.4227}}].

\bibitem{Litos:2014fxa}
M.~Litos et~al., {\it {Search for Dinucleon Decay into Kaons in
  Super-Kamiokande}},  {\em Phys. Rev. Lett.} {\bf 112} (2014), no.~13 131803.

\bibitem{Goity:1994dq}
J.~L. Goity and M.~Sher, {\it {Bounds on $\Delta B = 1$ couplings in the
  supersymmetric standard model}},  {\em Phys. Lett.} {\bf B346} (1995) 69--74,
  [\href{http://arxiv.org/abs/hep-ph/9412208}{{\tt hep-ph/9412208}}]. [Erratum:
  Phys. Lett.B385,500(1996)].

\bibitem{Giudice:2011ak}
G.~F. Giudice, B.~Gripaios, and R.~Sundrum, {\it {Flavourful Production at
  Hadron Colliders}},  {\em JHEP} {\bf 08} (2011) 055,
  [\href{http://arxiv.org/abs/1105.3161}{{\tt arXiv:1105.3161}}].

\bibitem{DeSimone:2010tf}
A.~De~Simone, V.~Sanz, and H.~P. Sato, {\it {Pseudo-Dirac Dark Matter Leaves a
  Trace}},  {\em Phys. Rev. Lett.} {\bf 105} (2010) 121802,
  [\href{http://arxiv.org/abs/1004.1567}{{\tt arXiv:1004.1567}}].

\bibitem{Ipek:2016bpf}
S.~Ipek and J.~March-Russell, {\it {Baryogenesis via Particle-Antiparticle
  Oscillations}},  {\em Phys. Rev.} {\bf D93} (2016), no.~12 123528,
  [\href{http://arxiv.org/abs/1604.00009}{{\tt arXiv:1604.00009}}].

\bibitem{Berger:1999zt}
E.~L. Berger, B.~W. Harris, and Z.~Sullivan, {\it {Single top squark production
  via R-parity violating supersymmetric couplings in hadron collisions}},  {\em
  Phys. Rev. Lett.} {\bf 83} (1999) 4472--4475,
  [\href{http://arxiv.org/abs/hep-ph/9903549}{{\tt hep-ph/9903549}}].

\bibitem{Alwall:2014hca}
J.~Alwall, R.~Frederix, S.~Frixione, V.~Hirschi, F.~Maltoni, O.~Mattelaer,
  H.~S. Shao, T.~Stelzer, P.~Torrielli, and M.~Zaro, {\it {The automated
  computation of tree-level and next-to-leading order differential cross
  sections, and their matching to parton shower simulations}},  {\em JHEP} {\bf
  07} (2014) 079, [\href{http://arxiv.org/abs/1405.0301}{{\tt
  arXiv:1405.0301}}].

\bibitem{Alloul:2013bka}
A.~Alloul, N.~D. Christensen, C.~Degrande, C.~Duhr, and B.~Fuks, {\it
  {FeynRules 2.0 - A complete toolbox for tree-level phenomenology}},  {\em
  Comput. Phys. Commun.} {\bf 185} (2014) 2250--2300,
  [\href{http://arxiv.org/abs/1310.1921}{{\tt arXiv:1310.1921}}].

\bibitem{Plehn:2000be}
T.~Plehn, {\it {Single stop production at hadron colliders}},  {\em Phys.
  Lett.} {\bf B488} (2000) 359--366,
  [\href{http://arxiv.org/abs/hep-ph/0006182}{{\tt hep-ph/0006182}}].

\bibitem{Beenakker:1997ut}
W.~Beenakker, M.~Kramer, T.~Plehn, M.~Spira, and P.~M. Zerwas, {\it {Stop
  production at hadron colliders}},  {\em Nucl. Phys.} {\bf B515} (1998) 3--14,
  [\href{http://arxiv.org/abs/hep-ph/9710451}{{\tt hep-ph/9710451}}].

\bibitem{Beenakker:2010nq}
W.~Beenakker, S.~Brensing, M.~Kramer, A.~Kulesza, E.~Laenen, and I.~Niessen,
  {\it {Supersymmetric top and bottom squark production at hadron colliders}},
  {\em JHEP} {\bf 08} (2010) 098, [\href{http://arxiv.org/abs/1006.4771}{{\tt
  arXiv:1006.4771}}].

\bibitem{Beenakker:2011fu}
W.~Beenakker, S.~Brensing, M.~n. Kramer, A.~Kulesza, E.~Laenen, L.~Motyka, and
  I.~Niessen, {\it {Squark and Gluino Hadroproduction}},  {\em Int. J. Mod.
  Phys.} {\bf A26} (2011) 2637--2664,
  [\href{http://arxiv.org/abs/1105.1110}{{\tt arXiv:1105.1110}}].

\bibitem{Beenakker:2016lwe}
W.~Beenakker, C.~Borschensky, M.~Krämer, A.~Kulesza, and E.~Laenen, {\it
  {NNLL-fast: predictions for coloured supersymmetric particle production at
  the LHC with threshold and Coulomb resummation}},
  \href{http://arxiv.org/abs/1607.07741}{{\tt arXiv:1607.07741}}.

\bibitem{Beenakker:2016gmf}
W.~Beenakker, C.~Borschensky, R.~Heger, M.~Krämer, A.~Kulesza, and E.~Laenen,
  {\it {NNLL resummation for stop pair-production at the LHC}},  {\em JHEP}
  {\bf 05} (2016) 153, [\href{http://arxiv.org/abs/1601.02954}{{\tt
  arXiv:1601.02954}}].

\bibitem{Beenakker:1996ed}
W.~Beenakker, R.~Hopker, and M.~Spira, {\it {PROSPINO: A Program for the
  production of supersymmetric particles in next-to-leading order QCD}},
  \href{http://arxiv.org/abs/hep-ph/9611232}{{\tt hep-ph/9611232}}.

\bibitem{Aad:2014aqa}
{\bf ATLAS} Collaboration, G.~Aad et~al., {\it {Search for new phenomena in the
  dijet mass distribution using $p-p$ collision data at $\sqrt{s}=8$ TeV with
  the ATLAS detector}},  {\em Phys. Rev.} {\bf D91} (2015), no.~5 052007,
  [\href{http://arxiv.org/abs/1407.1376}{{\tt arXiv:1407.1376}}].

\bibitem{ATLAS-CONF-2016-030}
{\bf ATLAS Collaboration} Collaboration, {\it {Search for light dijet
  resonances with the ATLAS detector using a Trigger-Level Analysis in LHC pp
  collisions at $\sqrt{s}=13$~TeV}},  Tech. Rep. ATLAS-CONF-2016-030, CERN,
  Geneva, Jun, 2016.

\bibitem{ATLAS-CONF-2016-069}
{\bf ATLAS Collaboration} Collaboration, {\it {Search for New Phenomena in
  Dijet Events with the ATLAS Detector at $\sqrt{s}$=13 TeV with 2015 and 2016
  data}},  Tech. Rep. ATLAS-CONF-2016-069, CERN, Geneva, Aug, 2016.

\bibitem{Khachatryan:2016ecr}
{\bf CMS} Collaboration, V.~Khachatryan et~al., {\it {Search for narrow
  resonances in dijet final states at $\sqrt(s)=$ 8 TeV with the novel CMS
  technique of data scouting}},  {\em Phys. Rev. Lett.} {\bf 117} (2016), no.~3
  031802, [\href{http://arxiv.org/abs/1604.08907}{{\tt arXiv:1604.08907}}].

\bibitem{CMS-PAS-EXO-16-032}
{\bf CMS Collaboration} Collaboration, {\it {Searches for narrow resonances
  decaying to dijets in proton-proton collisions at 13 TeV using 12.9 inverse
  femtobarns.}},  Tech. Rep. CMS-PAS-EXO-16-032, CERN, Geneva, 2016.

\bibitem{ATLAS-CONF-2016-060}
{\bf ATLAS Collaboration} Collaboration, {\it {Search for resonances in the
  mass distribution of jet pairs with one or two jets identified as $b$-jets
  with the ATLAS detector with 2015 and 2016 data}},  Tech. Rep.
  ATLAS-CONF-2016-060, CERN, Geneva, Aug, 2016.

\bibitem{Sjostrand:2014zea}
T.~Sjöstrand, S.~Ask, J.~R. Christiansen, R.~Corke, N.~Desai, P.~Ilten,
  S.~Mrenna, S.~Prestel, C.~O. Rasmussen, and P.~Z. Skands, {\it {An
  Introduction to PYTHIA 8.2}},  {\em Comput. Phys. Commun.} {\bf 191} (2015)
  159--177, [\href{http://arxiv.org/abs/1410.3012}{{\tt arXiv:1410.3012}}].

\bibitem{deFavereau:2013fsa}
{\bf DELPHES 3} Collaboration, J.~de~Favereau, C.~Delaere, P.~Demin,
  A.~Giammanco, V.~Lemaître, A.~Mertens, and M.~Selvaggi, {\it {DELPHES 3, A
  modular framework for fast simulation of a generic collider experiment}},
  {\em JHEP} {\bf 02} (2014) 057, [\href{http://arxiv.org/abs/1307.6346}{{\tt
  arXiv:1307.6346}}].

\bibitem{Dobbs:2001ck}
M.~Dobbs and J.~B. Hansen, {\it {The HepMC C++ Monte Carlo event record for
  High Energy Physics}},  {\em Comput. Phys. Commun.} {\bf 134} (2001) 41--46.

\bibitem{Aad:2016kww}
{\bf ATLAS} Collaboration, G.~Aad et~al., {\it {A search for top squarks with
  R-parity-violating decays to all-hadronic final states with the ATLAS
  detector in $\sqrt{s}$ = 8 TeV proton-proton collisions}},  {\em JHEP} {\bf
  06} (2016) 067, [\href{http://arxiv.org/abs/1601.07453}{{\tt
  arXiv:1601.07453}}].

\bibitem{ATLAS-CONF-2016-022}
{\bf ATLAS Collaboration} Collaboration, {\it {A search for R-parity violating
  decays of the top squark in four jet final states with the ATLAS detector at
  $\sqrt{s}=13$ TeV}},  Tech. Rep. ATLAS-CONF-2016-022, CERN, Geneva, May,
  2016.

\bibitem{ATLAS-CONF-2016-084}
{\bf ATLAS Collaboration} Collaboration, {\it {A search for pair produced
  resonances in four jets final states in proton-proton collisions at
  $\sqrt{s}$=13 TeV with the ATLAS experiment}},  Tech. Rep.
  ATLAS-CONF-2016-084, CERN, Geneva, Aug, 2016.

\bibitem{Khachatryan:2014lpa}
{\bf CMS} Collaboration, V.~Khachatryan et~al., {\it {Search for pair-produced
  resonances decaying to jet pairs in proton–proton collisions at $\sqrt{s}
  =$ 8 TeV}},  {\em Phys. Lett.} {\bf B747} (2015) 98--119,
  [\href{http://arxiv.org/abs/1412.7706}{{\tt arXiv:1412.7706}}].

\bibitem{ATLAS-CONF-2016-037}
{\bf ATLAS Collaboration} Collaboration, {\it {Search for supersymmetry with
  two same-sign leptons or three leptons using 13.2 fb$^{−1}$ of $\sqrt{s} =
  13$ TeV $pp$ collision data collected by the ATLAS detector}},  Tech. Rep.
  ATLAS-CONF-2016-037, CERN, Geneva, Aug, 2016.

\bibitem{ATLAS-CONF-2016-057}
{\bf ATLAS Collaboration} Collaboration, {\it {Search for massive
  supersymmetric particles in multi-jet final states produced in pp collisions
  at $\sqrt{s}~=$~13 TeV using the ATLAS detector at the LHC}},  Tech. Rep.
  ATLAS-CONF-2016-057, CERN, Geneva, Aug, 2016.

\bibitem{ATLAS-CONF-2016-094}
{\bf ATLAS Collaboration} Collaboration, {\it {Search for new physics in a
  lepton plus high jet multiplicity final state with the ATLAS experiment using
  sqrt(s) = 13 TeV proton-proton collision data}},  Tech. Rep.
  ATLAS-CONF-2016-094, CERN, Geneva, Aug, 2016.

\bibitem{ATLAS-CONF-2016-013}
{\it {Search for production of vector-like top quark pairs and of four top
  quarks in the lepton-plus-jets final state in $pp$ collisions at
  $\sqrt{s}=13$ TeV with the ATLAS detector}},  Tech. Rep. ATLAS-CONF-2016-013,
  CERN, Geneva, Mar, 2016.

\bibitem{ATLAS-CONF-2016-032}
{\bf ATLAS Collaboration} Collaboration, {\it {Search for new physics using
  events with $b$-jets and a pair of same charge leptons in 3.2 fb$^{-1}$ of
  $pp$ collisions at $\sqrt{s}=13$ TeV with the ATLAS detector}},  Tech. Rep.
  ATLAS-CONF-2016-032, CERN, Geneva, Jun, 2016.

\bibitem{CMS-PAS-SUS-16-020}
{\bf CMS Collaboration} Collaboration, {\it {Search for SUSY in same-sign
  dilepton events at 13 TeV}},  Tech. Rep. CMS-PAS-SUS-16-020, CERN, Geneva,
  2016.

\bibitem{Cacciari:2011ma}
M.~Cacciari, G.~P. Salam, and G.~Soyez, {\it {FastJet User Manual}},  {\em Eur.
  Phys. J.} {\bf C72} (2012) 1896, [\href{http://arxiv.org/abs/1111.6097}{{\tt
  arXiv:1111.6097}}].

\bibitem{Cacciari:2005hq}
M.~Cacciari and G.~P. Salam, {\it {Dispelling the $N^{3}$ myth for the $k_t$
  jet-finder}},  {\em Phys. Lett.} {\bf B641} (2006) 57--61,
  [\href{http://arxiv.org/abs/hep-ph/0512210}{{\tt hep-ph/0512210}}].

\bibitem{Read:2002hq}
A.~L. Read, {\it {Presentation of search results: The CL(s) technique}},  {\em
  J. Phys.} {\bf G28} (2002) 2693--2704. [,11(2002)].

\bibitem{Junk:1999kv}
T.~Junk, {\it {Confidence level computation for combining searches with small
  statistics}},  {\em Nucl. Instrum. Meth.} {\bf A434} (1999) 435--443,
  [\href{http://arxiv.org/abs/hep-ex/9902006}{{\tt hep-ex/9902006}}].

\bibitem{Cui:2014twa}
Y.~Cui and B.~Shuve, {\it {Probing Baryogenesis with Displaced Vertices at the
  LHC}},  {\em JHEP} {\bf 02} (2015) 049,
  [\href{http://arxiv.org/abs/1409.6729}{{\tt arXiv:1409.6729}}].

\bibitem{CMS:2014wda}
{\bf CMS} Collaboration, V.~Khachatryan et~al., {\it {Search for Long-Lived
  Neutral Particles Decaying to Quark-Antiquark Pairs in Proton-Proton
  Collisions at $\sqrt{s} =$ 8 TeV}},  {\em Phys. Rev.} {\bf D91} (2015), no.~1
  012007, [\href{http://arxiv.org/abs/1411.6530}{{\tt arXiv:1411.6530}}].

\bibitem{Liu:2015bma}
Z.~Liu and B.~Tweedie, {\it {The Fate of Long-Lived Superparticles with
  Hadronic Decays after LHC Run 1}},  {\em JHEP} {\bf 06} (2015) 042,
  [\href{http://arxiv.org/abs/1503.05923}{{\tt arXiv:1503.05923}}].

\bibitem{Beenakker:1996ch}
W.~Beenakker, R.~Hopker, M.~Spira, and P.~M. Zerwas, {\it {Squark and gluino
  production at hadron colliders}},  {\em Nucl. Phys.} {\bf B492} (1997)
  51--103, [\href{http://arxiv.org/abs/hep-ph/9610490}{{\tt hep-ph/9610490}}].

\bibitem{Kulesza:2008jb}
A.~Kulesza and L.~Motyka, {\it {Threshold resummation for squark-antisquark and
  gluino-pair production at the LHC}},  {\em Phys. Rev. Lett.} {\bf 102} (2009)
  111802, [\href{http://arxiv.org/abs/0807.2405}{{\tt arXiv:0807.2405}}].

\bibitem{Kulesza:2009kq}
A.~Kulesza and L.~Motyka, {\it {Soft gluon resummation for the production of
  gluino-gluino and squark-antisquark pairs at the LHC}},  {\em Phys. Rev.}
  {\bf D80} (2009) 095004, [\href{http://arxiv.org/abs/0905.4749}{{\tt
  arXiv:0905.4749}}].

\bibitem{Beenakker:2009ha}
W.~Beenakker, S.~Brensing, M.~Kramer, A.~Kulesza, E.~Laenen, and I.~Niessen,
  {\it {Soft-gluon resummation for squark and gluino hadroproduction}},  {\em
  JHEP} {\bf 12} (2009) 041, [\href{http://arxiv.org/abs/0909.4418}{{\tt
  arXiv:0909.4418}}].

\bibitem{Beenakker:2011sf}
W.~Beenakker, S.~Brensing, M.~Kramer, A.~Kulesza, E.~Laenen, and I.~Niessen,
  {\it {NNLL resummation for squark-antisquark pair production at the LHC}},
  {\em JHEP} {\bf 01} (2012) 076, [\href{http://arxiv.org/abs/1110.2446}{{\tt
  arXiv:1110.2446}}].

\bibitem{Beenakker:2013mva}
W.~Beenakker, T.~Janssen, S.~Lepoeter, M.~Krämer, A.~Kulesza, E.~Laenen,
  I.~Niessen, S.~Thewes, and T.~Van~Daal, {\it {Towards NNLL resummation: hard
  matching coefficients for squark and gluino hadroproduction}},  {\em JHEP}
  {\bf 10} (2013) 120, [\href{http://arxiv.org/abs/1304.6354}{{\tt
  arXiv:1304.6354}}].

\bibitem{Beenakker:2014sma}
W.~Beenakker, C.~Borschensky, M.~Krämer, A.~Kulesza, E.~Laenen, V.~Theeuwes,
  and S.~Thewes, {\it {NNLL resummation for squark and gluino production at the
  LHC}},  {\em JHEP} {\bf 12} (2014) 023,
  [\href{http://arxiv.org/abs/1404.3134}{{\tt arXiv:1404.3134}}].

\bibitem{ATLAS-CONF-2016-078}
{\bf ATLAS Collaboration} Collaboration, {\it {Further searches for squarks and
  gluinos in final states with jets and missing transverse momentum at
  $\sqrt{s}$ =13 TeV with the ATLAS detector}},  Tech. Rep.
  ATLAS-CONF-2016-078, CERN, Geneva, Aug, 2016.

\bibitem{Arvanitaki:2012ps}
A.~Arvanitaki, N.~Craig, S.~Dimopoulos, and G.~Villadoro, {\it {Mini-Split}},
  {\em JHEP} {\bf 1302} (2013) 126, [\href{http://arxiv.org/abs/1210.0555}{{\tt
  arXiv:1210.0555}}].

\bibitem{Cui:2016rqt}
Y.~Cui, T.~Okui, and A.~Yunesi, {\it {LHC Signatures of WIMP-triggered
  Baryogenesis}},  {\em Phys. Rev.} {\bf D94} (2016), no.~11 115022,
  [\href{http://arxiv.org/abs/1605.08736}{{\tt arXiv:1605.08736}}].

\bibitem{Nanopoulos:1979gx}
D.~V. Nanopoulos and S.~Weinberg, {\it {Mechanisms for Cosmological Baryon
  Production}},  {\em Phys. Rev.} {\bf D20} (1979) 2484.

\bibitem{Kolb:1990vq}
E.~W. Kolb and M.~S. Turner, {\it {The Early Universe}},  {\em Front. Phys.}
  {\bf 69} (1990) 1--547.

\bibitem{Cutkosky:1960sp}
R.~E. Cutkosky, {\it {Singularities and discontinuities of Feynman
  amplitudes}},  {\em J. Math. Phys.} {\bf 1} (1960) 429--433.

\bibitem{Edsjo:1997bg}
J.~Edsjo and P.~Gondolo, {\it {Neutralino relic density including
  coannihilations}},  {\em Phys. Rev.} {\bf D56} (1997) 1879--1894,
  [\href{http://arxiv.org/abs/hep-ph/9704361}{{\tt hep-ph/9704361}}].

\bibitem{Ade:2015xua}
{\bf Planck} Collaboration, P.~A.~R. Ade et~al., {\it {Planck 2015 results.
  XIII. Cosmological parameters}},  {\em Astron. Astrophys.} {\bf 594} (2016)
  A13, [\href{http://arxiv.org/abs/1502.01589}{{\tt arXiv:1502.01589}}].

\bibitem{Chen:2007fv}
M.-C. Chen, {\it {TASI 2006 Lectures on Leptogenesis}},  in {\em {Proceedings
  of Theoretical Advanced Study Institute in Elementary Particle Physics :
  Exploring New Frontiers Using Colliders and Neutrinos (TASI 2006)}},
  pp.~123--176, 2007.
\newblock \href{http://arxiv.org/abs/hep-ph/0703087}{{\tt hep-ph/0703087}}.

\bibitem{Baldes:2014rda}
I.~Baldes, N.~F. Bell, A.~Millar, K.~Petraki, and R.~R. Volkas, {\it {The role
  of CP violating scatterings in baryogenesis - case study of the neutron
  portal}},  {\em JCAP} {\bf 1411} (2014), no.~11 041,
  [\href{http://arxiv.org/abs/1410.0108}{{\tt arXiv:1410.0108}}].

\bibitem{Benakli:2012cy}
K.~Benakli, M.~D. Goodsell, and F.~Staub, {\it {Dirac Gauginos and the 125 GeV
  Higgs}},  {\em JHEP} {\bf 06} (2013) 073,
  [\href{http://arxiv.org/abs/1211.0552}{{\tt arXiv:1211.0552}}].

\bibitem{Diessner:2014ksa}
P.~Dießner, J.~Kalinowski, W.~Kotlarski, and D.~Stöckinger, {\it {Higgs boson
  mass and electroweak observables in the MRSSM}},  {\em JHEP} {\bf 12} (2014)
  124, [\href{http://arxiv.org/abs/1410.4791}{{\tt arXiv:1410.4791}}].

\end{thebibliography}\endgroup

\end{document}